\documentclass{article}

\PassOptionsToPackage{table}{xcolor}
\usepackage{arxiv}
\usepackage[utf8]{inputenc} % allow utf-8 input
\usepackage[T1]{fontenc}    % use 8-bit T1 fonts
\usepackage{cite}
\usepackage{hyperref}
\usepackage{url}            % simple URL typesetting
\usepackage{booktabs}       % professional-quality tables
\usepackage{tabularx}
\usepackage{siunitx}
\usepackage{array}
\usepackage{bm}
\usepackage{amsfonts}       % blackboard math symbols
\usepackage{amsmath}
\usepackage{physics}
\usepackage{amssymb}
\usepackage{nicefrac}       % compact symbols for 1/2, etc.
\usepackage{microtype}
\usepackage{cleveref}
\usepackage{xcolor}
\usepackage{authblk}
\usepackage{algorithm}
\usepackage{algpseudocode}
\usepackage{color,soul}
\usepackage{caption}
\usepackage{subcaption}
\usepackage{threeparttable}
\usepackage{lipsum}
\usepackage{fancyhdr}       % header
\usepackage{graphicx}       % graphics
\usepackage{svg}            % include .svg figures via \includesvg
\usepackage{enumitem}
\usepackage{textgreek}
\usepackage{float}
\usepackage[section]{placeins}
\usepackage{colortbl}
\graphicspath{{media/}}     % organize your images and other figures under media/ folder


\AtBeginDocument{\setlength{\headheight}{12pt}}
\newcommand{\tens}[1]{\bm{#1}}

\newcommand{\standardtablewidth}{0.88\textwidth}
\newcolumntype{Y}{>{\centering\arraybackslash}X}
\makeatletter
\newcommand{\printfnsymbol}[1]{%
  \textsuperscript{\@fnsymbol{#1}}%
}

\makeatother

\title{JAX-FEM-ANISO: Differentiable GPU-Accelerated Finite Element Framework for Inverse Identification of Finite-Strain Anisotropic Plasticity}

\author[a]{Deepak Sharma}
\author[a]{Itzel Salgado}
\author[b]{Lu Huang}
\author[b]{Hui-Ping Wang}
\author[a,$\ast$]{Jian Cao}

\affil[a]{Department of Mechanical Engineering, \protect\\
Northwestern University, Evanston, 60208, IL, USA}
\affil[b]{General Motors, Warren, 48090, Michigan, USA}
\affil[$\ast$]{Corresponding author e-mail: \href{mailto:jcao@northwestern.edu}{jcao@northwestern.edu}}

\date{}

\begin{document}

\maketitle

\begin{abstract}
Calibrating advanced anisotropic plasticity models is essential for high-fidelity simulations of metal forming, welding, and additive manufacturing. Conventional workflows, however, often require extensive multi-test experimental campaigns, computationally intensive forward simulations, and finite-difference sensitivity calculations that scale poorly with the number of material parameters. We present a fully differentiable, GPU-accelerated finite element framework JAX-FEM-ANISO, for forward simulation and inverse parameter identification of finite-strain anisotropic plasticity. Built on JAX-FEM, the framework exploits modern accelerator architectures by parallelizing the three major computational bottlenecks in nonlinear FEM: elemental weak-form and tangent-stiffness evaluation, global sparse matrix assembly, and sparse linear solution. For a large-scale forward problem with 3 million degrees of freedom, JAX-FEM-ANISO on a single NVIDIA H100 GPU achieves up to 9.4$\times$ speed-up over a 24-core CPU Abaqus baseline. Automatic differentiation is applied through the constitutive update and solver workflow, providing consistent Jacobians for complex constitutive models without manual derivation and accurate gradients for PDE-constrained inverse analysis. Compared with finite differences, the JAX-AD gradients avoid step-size sensitivity and provide the required sensitivities at substantially lower computational cost. For inverse characterization, we combine information-rich, topology-optimized heterogeneous specimens with full-field displacement data to identify advanced constitutive model parameters from a single test, replacing what would otherwise require many conventional experiments. We demonstrate accurate recovery of anisotropic yield and hardening parameters in progressively challenging settings, including uniform and spatially varying material properties. The resulting AD-based formulation enables efficient optimization in high-dimensional parameter spaces where finite-difference approaches are computationally infeasible. These results establish differentiable, GPU-accelerated FEM as a practical high-throughput engine for simulation, characterization, and optimization workflows in advanced manufacturing.
\end{abstract}


\section{Introduction}
Accurate simulation of metal forming, welding, and additive manufacturing requires computational models that capture the geometric and material nonlinearities inherent in finite-strain plasticity~\cite{SimoHughes1998,Belytschko2000}. Anisotropic plasticity models, such as Hill--48~\cite{Hill1948}, Barlat's Yld2000-2d~\cite{Barlat2003}, and Yld2004-3D~\cite{Barlat2005}, are widely used in automotive and aerospace applications to simulate complex sheet-metal forming processes. These constitutive models account for directional variations in yield stress and plastic flow that develop during manufacturing, enabling more accurate predictions of springback, formability limits, and final part geometry. Their implementation and calibration remain challenging because finite-strain anisotropic plasticity couples geometric and material nonlinearities~\cite{deSouzaNeto2008}. Practical finite element implementations therefore require iterative Newton--Raphson schemes at both local material-point and global structural levels, together with consistent tangent operators and robust return-mapping algorithms for history-dependent internal variables~\cite{Michaleris1994,Miehe2002,deSouzaNeto2008,Aravas1987}.
	
    Calibrating anisotropic plasticity models traditionally requires extensive experimental campaigns. For example, Hill--48 in plane stress and Yld2000-2d require at least four distinct tests: three uniaxial tests in different orientations (0°, 45°, and 90°) and one biaxial tension test~\cite{Guner2012,Kim2014}. More complex constitutive models, such as Yld2004-3D, provide greater flexibility and accuracy but require additional experiments for parameter identification. Conventional testing also relies on specialized equipment, time-intensive procedures, and limited spatial information from single-point strain gauges. Digital image correlation (DIC) has changed this setting by providing high-resolution full-field displacement and strain data across specimen surfaces~\cite{Sutton2009}, enabling more information-rich material characterization. Consequently, inverse identification methods have been developed to extract constitutive parameters from heterogeneous strain and displacement fields using reduced experimental datasets~\cite{Kim2014,Bertin2017,Coppieters2018,Zhang2023}. This integrated paradigm, which combines tailored specimen geometries, full-field DIC measurements, and numerical inverse identification for calibrating advanced constitutive models, is commonly referred to as `Material Testing 2.0' and has been widely adopted~\cite{PierronGrediac2021}. Within this framework, two prominent inverse identification strategies have emerged: the Virtual Fields Method (VFM) and finite element model updating (FEMU)~\cite{Martins2018,Marek2019}. VFM is valued for computational efficiency, whereas FEMU offers greater flexibility for nonlinear constitutive models and complex specimen geometries by formulating parameter identification as a finite-element-constrained optimization problem~\cite{Martins2018,PierronGrediac2021}. In FEMU, constitutive parameters are updated iteratively by minimizing the discrepancy between measured and predicted responses, typically using gradient-based optimization. Recent studies~\cite{Coppieters2018,Zhang2022,Zhang2023,Conde2023,Goncalves2025} have shown that, when combined with carefully designed heterogeneous experiments, FEMU can identify anisotropic plasticity parameters using one or only a few information-rich tests, thereby reducing the need for extensive conventional testing campaigns. These studies demonstrate the feasibility of near one-shot discovery of advanced yield criteria while also highlighting persistent challenges associated with parameter coupling and identifiability in highly flexible constitutive models. FEMU-based optimization, however, depends critically on accurate gradients of the objective function with respect to constitutive parameters. These gradients are commonly computed using finite-difference (FD) approximations~\cite{Zhang2023,Conde2023}, which are sensitive to step-size selection and scale poorly with the number of parameters because each gradient component requires an additional forward simulation~\cite{SeidlGranzow2022,Kumar2025}.
	
    The limitations of finite-difference sensitivity analysis have motivated the use of automatic differentiation (AD), which computes derivatives by systematically applying the chain rule to the computational graph of a simulation code~\cite{Margossian2019}. AD can simplify the implementation of complex nonlinear solvers by eliminating manual derivations of consistent tangent operators for path-dependent elastoplastic constitutive models~\cite{RotheHartmann2015,SeidlGranzow2022,Dummer2024}. It also supports gradient-based optimization workflows by providing accurate gradients for inverse identification, topology optimization, and machine learning applications~\cite{Jia2025,JiaZhang2026}. In their CPU-based implementation, Seidl and Granzow~\cite{SeidlGranzow2022} explored AD for inverse parameter identification in finite-strain isotropic and anisotropic plasticity models. They computed objective-function gradients using both forward- and reverse-mode AD and highlighted the efficiency of adjoint methods for problems with many parameters. Differentiable programming frameworks such as JAX~\cite{Bradbury2018}, PyTorch~\cite{Paszke2019}, and NVIDIA Warp~\cite{Macklin2022} provide high-level interfaces for derivative computation and facilitate rapid prototyping of forward and inverse problems across scientific domains.
	
    As industrial applications move toward higher-fidelity simulations with millions of degrees of freedom and increasingly complex material behavior, traditional CPU-based finite element implementations face substantial computational bottlenecks. This trend has motivated high-performance computing (HPC) approaches based on graphics processing units (GPUs). The rapid maturation of GPU architectures and parallel programming models has expanded the scope of large-scale solid mechanics simulations. For example, the NVIDIA H100 SXM datacenter GPU delivers a theoretical peak of approximately 34 teraFLOPS (TFLOPS) in double precision (FP64) and 67 TFLOPS in single precision (FP32), along with extremely high on-package memory bandwidth (3.35 TB/s) enabled by 80 GB of HBM3~\cite{NvidiaH100Datasheet}. In contrast, the Intel Xeon 6952P is a high-end 96-core, 192-thread Granite Rapids server CPU. It has a theoretical FP64 peak of approximately 9 TFLOPS and a FP32 peak of 19 TFLOPS at its 3.2 GHz all-core turbo frequency. Its memory bandwidth ranges from 0.61 TB/s to 0.84 TB/s across 12 channels~\cite{IntelXeon6952P}. In practice, memory bandwidth rather than floating-point capability often limits CPU performance because scientific computing workloads typically exhibit low arithmetic intensity~\cite{Williams2009}. The superior computing speed and memory bandwidth of GPUs have drawn significant attention for computationally intensive scientific computing tasks~\cite{MartinezFrutos2015,Aissa2017,Stavroulakis2017,GhyselsSynk2022,Li2023,Kiran2024MatrixFree,Hong2022,Liao2023,Traff2023,Kiran2024SpMV,Karatarakis2014}. More recently, differentiable programming frameworks have coupled GPU acceleration with AD, allowing end-to-end gradient computation for inverse design and optimization. JAX-FEM provides a general-purpose, GPU-accelerated differentiable finite element framework for three-dimensional nonlinear mechanics and topology optimization~\cite{Xue2023}, while JAX-CPFEM extends these ideas to crystal plasticity, achieving significant speedups over CPU-based MPI implementations and enabling gradient-based microstructure design without manual tangent derivations~\cite{Hu2025}. These capabilities have also enabled GPU-accelerated, AD-driven calibration of crystal plasticity models using multi-objective optimization~\cite{Hu2026}. They have been used for differentiable discovery and updating of history-dependent constitutive models~\cite{FerreiraBessa2025}. Related work has also demonstrated the co-design of geometry and spatially varying, temperature-dependent material distributions under coupled thermoelastic constraints~\cite{Knapik2025}. Complementary advances in GPU-optimized multiphysics solvers, such as multilevel thermal simulations for laser powder bed fusion, further demonstrate the scalability of these approaches to manufacturing-scale problems~\cite{LeonorWagner2024}. Together, these developments point to a growing convergence between high-performance computing and differentiable simulation: GPUs provide throughput for high-fidelity forward analyses, while AD enables efficient and accurate gradient computation for optimization-driven workflows.

Despite these advances, the use of GPU acceleration and AD-based sensitivities for large-scale forward and inverse analyses of finite-strain anisotropic plasticity remains largly unexplored, particularly for advanced yield criteria such as Barlat's Yld2004-3D. This gap is important because the difficulty is not confined to a single part of the simulation. At the material-point level, advanced anisotropic models require nonlinear return-mapping algorithms and consistent algorithmic tangents whose manual derivation is lengthy, model-specific, and error-prone, especially when material and geometric nonlinearities are coupled. At the structural level, each Newton iteration requires assembly and solution of large sparse systems, so forward simulations become expensive as the mesh is refined to resolve heterogeneous deformation fields. In inverse identification, these costs are multiplied across many optimization iterations, and finite-difference sensitivities further multiply the cost by requiring additional forward solves for each material parameter. This scaling becomes prohibitive for complex yield functions with many parameters and is effectively impractical when parameters are represented as spatial fields rather than as a small set of scalar constants. These challenges motivate a framework in which AD provides both local consistent tangents and global objective-function gradients through the full computational graph, while GPU acceleration reduces the cost of constitutive updates, sparse assembly, and large linear solves. Such a combination directly targets the two central bottlenecks in FEMU for advanced finite-strain plasticity: derivative complexity and forward-simulation cost.

 In this paper, we implement the general three-step modular framework for finite-strain plasticity proposed by Miehe~\cite{Miehe2002} within the GPU-accelerated Python library JAX-FEM~\cite{Xue2023}. The implementation uses JAX's automatic vectorization, automatic differentiation, and just-in-time (JIT) compilation to maintain a compact code structure while efficiently using high-end GPU hardware. We apply GPU acceleration to the three computationally demanding stages of a FEM forward simulation: (1) constitutive updates and elemental stiffness-matrix calculation, (2) sparse assembly, and (3) linear solution. For the solution of large-scale linear systems, we integrate NVIDIA's AmgX library with JAX-FEM. AmgX is a GPU-native algebraic multigrid (AMG) solver that provides up to 10$\times$ acceleration over CPU-based implementations for the computationally intensive linear-solver portion of simulations~\cite{AmgX}. To our knowledge, this paper presents first JAX-based differentiable, GPU-accelerated implementation of finite-strain anisotropic plasticity for both forward and inverse problems. We verify the implementation through representative benchmark tests and evaluate its computational efficiency by comparing execution times with commercial finite element software. Specifically, we benchmark the framework on datacenter-grade GPUs (NVIDIA A100 and H100) that are well suited for FP64 scientific computing workloads and compare performance against both CPU-based and GPU-accelerated implementations in Abaqus.

The primary contributions of this paper are summarized as follows:
\begin{enumerate}[label=(\alph*)]
\item We develop a fully differentiable, GPU-accelerated FEM framework JAX-FEM-ANISO, for finite-strain anisotropic elasto-plasticity, with automatic differentiation through constitutive updates, global equilibrium solution, and PDE-constrained inverse analysis.
\item We design an end-to-end inverse identification methodology that combines information-rich, topology-optimized heterogeneous specimen geometries, full-field displacement data, and AD-based sensitivities to efficiently identify complex constitutive models with many material parameters.
\item The proposed approach enables the identification of spatially varying material parameters, allowing the characterization of heterogeneities induced by manufacturing processes such as forming, welding, and additive manufacturing.
\item The framework is verified through representative benchmark problems, demonstrating its accuracy, scalability, and extensibility for practical engineering applications.
\end{enumerate}

\section{Methodology}

\subsection{Overview}
\begin{figure}[!htbp]
\centering
\includegraphics[width=\textwidth]{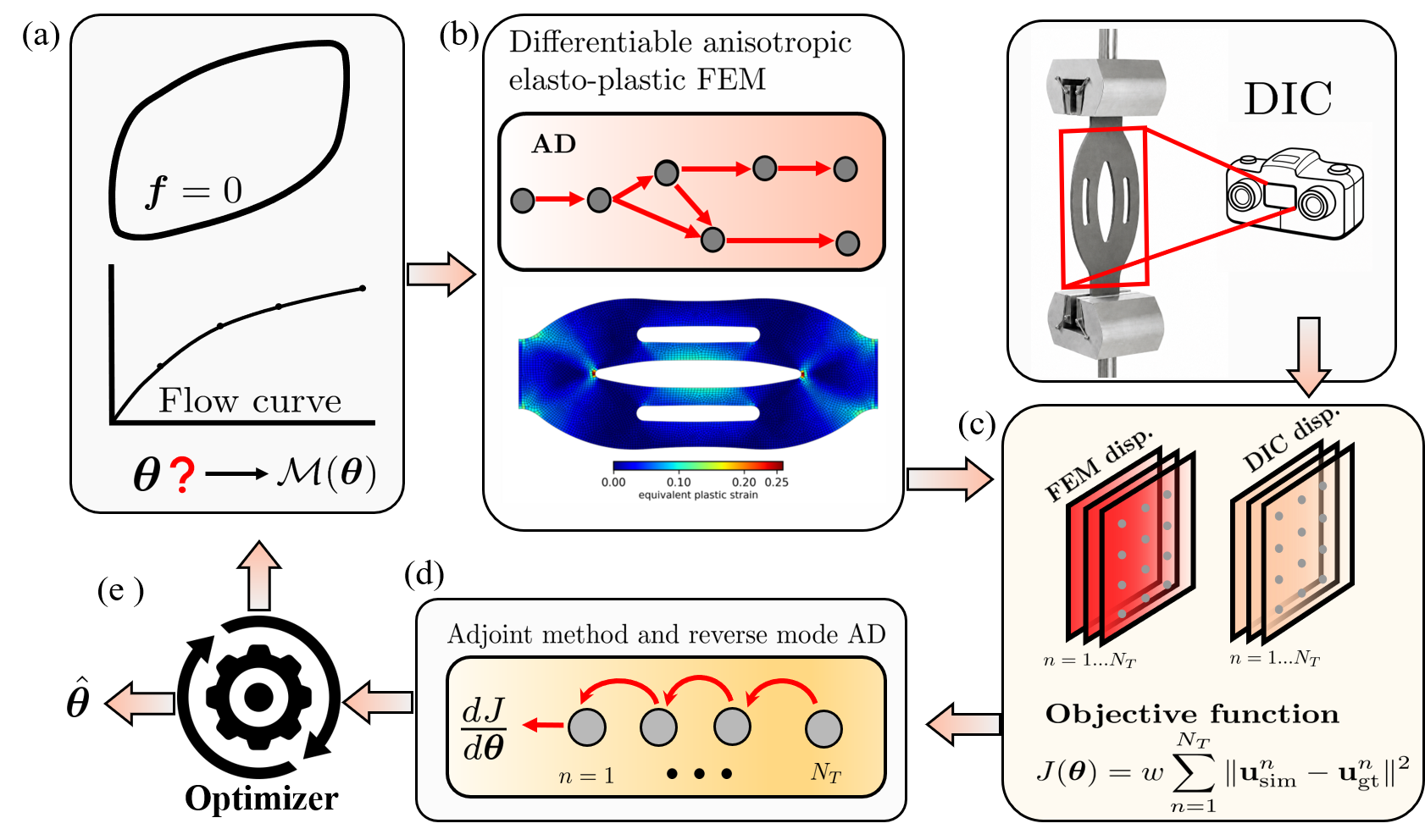}
\caption{Overview of the proposed differentiable GPU-accelerated framework for gradient-based parameter identification of finite-strain anisotropic plasticity: (a) parameterization of the constitutive model ($\mathcal{M}$) by a set of material parameters ($\bm{\theta}$), (b) forward simulation using differentiable FEM, (c) objective function evaluation by comparing the mismatch of the simulated and ground truth displacement fields, (d) gradient computation using the adjoint method with reverse-mode AD, and (e) optimization of material parameters.}
\label{fig:overview}
\end{figure}

For gradient based parameters identification of finite-strain anisotropic plasticity, the proposed framework uses differentiable GPU accelerated FEM along with the shape or topology optimized specimen geometries. The overall workflow of the framework is illustrated in Fig.~\ref{fig:overview}. It consists of the following steps,
\begin{enumerate}
\item [(a)] The constitutive model ($\mathcal{M}$) of the material is parameterized by a set of material parameters ($\bm{\theta}$), which includes the parameters of anisotropic yield surface ($f$) and hardening curve of the material. The material parameters $\bm{\theta}$ are the unknowns to be identified through the optimization process.
\item [(b)] For a specified specimen geometry, constitutive model ($\mathcal{M}$), material parameters ($\bm{\theta}$) and loading conditions, the forward problem is solved using a GPU-accelerated differentiable FEM to obtain the predicted displacement field $\bm{u}_{sim}^{n}$ at required time steps $n=1,2,...,N_T$. During forward solve with large deformation geometric and material nonlinearities, the consistent stiffness materix required for Newton-Raphson iterations is automatically computed uisng JAX forward-mode AD with automatic vectorization over elements and quadrature points. Automatic vectorization provides substantial efficiency gains for both runtime and memory usage by avoiding explicit for-loops and enabling batch processing of the local constitutive updates.
\item [(c)] The objective function ($J$) is computed by comparing the mismatch of the simulated displacement field $\bm{u}_{sim}^{n}$ with the ground truth displacement field $\bm{u}_{gt}^{n}$ at the same time steps. The $\bm{u}_{gt}^{n}$ can be obtained directly from the  experimental DIC measurements or synthetically from high-fidelity simulations. 
\item [(d)] The adjoint method is used with reverse-mode AD to compute the gradient of the objective function ($\mathrm{d}J/\mathrm{d}\boldsymbol{\theta}$) with respect to the material parameters in a memory-efficient and fully automatic manner. The gradient calculation for history dependent constitutive models is nontrivial due to the path-dependence of the internal variables and the need to backpropagate through the entire loading history. This poses significant computational and memeory demands. Our JAX implementation with automatic vectorization and checkpointing strategies allows efficient gradient computation even for complex constitutive models and long loading histories.
\item [(e)] The material parameters $\bm{\theta}$ are updated using a gradient-based optimization algorithm, such as L-BFGS, to minimize the objective function. The optimization iterates through steps (a) to (e) until convergence is achieved, resulting in the identified material parameters ($\hat{\bm{\theta}}$) that best match the observed deformation behavior.
\end{enumerate}
The details of our differentiable simulation framework are discussed in the following sections.

\subsection{Forward Problem}
This section describes the additive-decomposition-based finite-strain plasticity formulation used to model metal plasticity. Following Miehe et al.~\cite{Miehe2002}, the constitutive equations are formulated in logarithmic strain space. The framework is rate-consistent and suitable for large deformations in anisotropic metals.

We first summarize the governing equations and their weak form in the reference configuration, then introduce the constitutive model and its implicit time discretization, and finally describe the finite element discretization of the weak form.
\subsubsection{Governing Equations}

\begin{figure}[!htbp]
\centering
\includegraphics[width=0.4\textwidth]{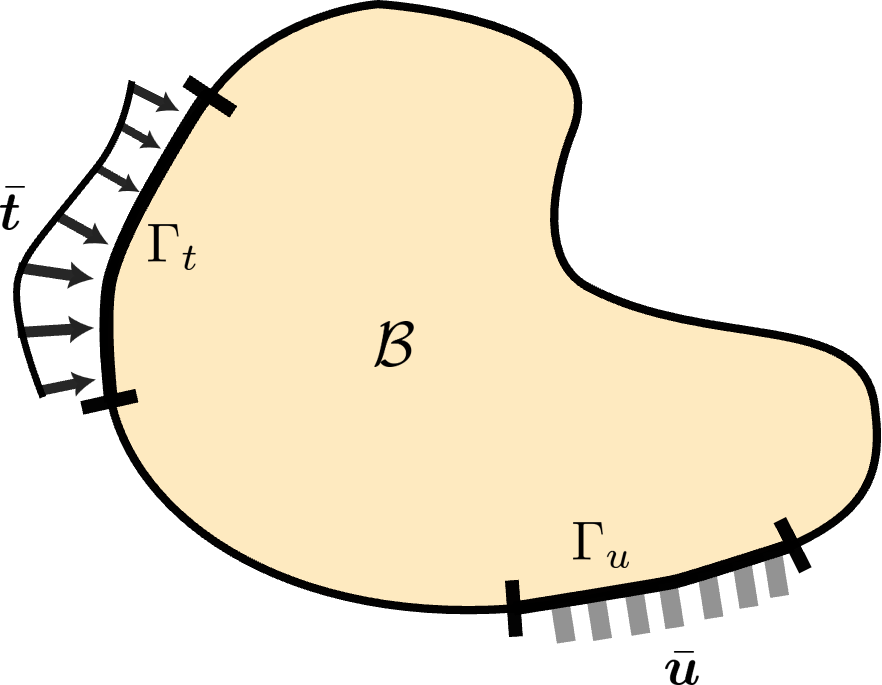}
\caption{Schematic of an elastoplastic solid body with displacement and traction boundary conditions.}
\label{fig:Continuum_body}
\end{figure}

We consider an elasto-plastic solid body occupying a reference configuration 
$\mathcal{B} \subset \mathbb{R}^3$ with boundary $\partial \mathcal{B}$ (Fig.~\ref{fig:Continuum_body}). 
The boundary is decomposed into two parts: the displacement boundary 
$\Gamma_u$ and the traction boundary $\Gamma_t$, such that 
$\partial \mathcal{B} = \Gamma_u \cup \Gamma_t$ and $\Gamma_u \cap \Gamma_t = \emptyset$. 
Neglecting inertial effects and body forces, the governing equations of the elastoplastic solid in the reference configuration are~\cite{Miehe2002},
\begin{equation}
\begin{cases}
-\nabla_0 \cdot \bm{P} = \bm{0}, & \text{in } \mathcal{B} \\[6pt]
\bm{u} = \bar{\bm{u}}, & \text{on } \Gamma_u \\[6pt]
\bm{P} \cdot \bm{n} = \bar{\bm{t}}, & \text{on } \Gamma_t
\end{cases}
\label{eq:equilibrium}
\end{equation}
where $\bm{u} : \mathcal{B} \to \mathbb{R}^3$ is the displacement field, 
$\bm{P}$ is the first Piola--Kirchhoff stress tensor, $\bar{\bm{u}}$ is the prescribed displacement on $\Gamma_u$, and $\bar{\bm{t}}$ is the prescribed traction vector on $\Gamma_t$. Here $\bm{n}$ denotes the outward unit normal to the boundary in the reference configuration, and $\nabla_0(\cdot)$ is the gradient with respect to the reference coordinates.
The weak form of equilibrium follows from multiplying Eq.~\eqref{eq:equilibrium} by an admissible virtual displacement $\delta \bm{u}$, integrating over $\mathcal{B}$, and applying the divergence theorem. The resulting weak form in the reference configuration is: find 
$\bm{u} \in \mathcal{U}$ such that for all virtual displacements 
$\delta \bm{u} \in \mathcal{V}$,
\begin{equation}
    \int_{\mathcal{B}} \bm{P} : \nabla_0 \delta \bm{u} \, dV 
    = \int_{\Gamma_t} \bar{\bm{t}} \cdot \delta \bm{u} \, dA,
    \label{eq:weak_form}
\end{equation}
where $\mathcal{U}$ is the space of kinematically admissible displacements satisfying the essential boundary conditions and $\mathcal{V}$ is the associated space of virtual displacements.

\subsubsection{Additive Lagrangian Modular Approach}\label{subsubsec:constitutive_model}

Finite-strain plasticity with material anisotropy is essential for modeling sheet metals subjected to complex loading paths. Several algorithmic challenges arise at large strains, including the choice of strain measure, objective stress mapping, robust return mapping for path-dependent internal variables, and numerical integration schemes that avoid volumetric locking. To address these issues, we adopt the modular formulation of Miehe~\cite{Miehe2002}. The framework proceeds in three stages. First, a \textit{geometric preprocessor} transforms the finite-strain kinematics into logarithmic strain space by computing the total logarithmic strain from the right Cauchy--Green tensor. Second, the \textit{constitutive model} operates entirely in logarithmic strain space, where elastic and plastic strains combine additively and standard small-strain plasticity algorithms, including anisotropic yield criteria, associative flow rules, and hardening laws, apply without modification. Third, a \textit{geometric postprocessor} maps the resulting logarithmic stresses back to their finite-strain Lagrangian counterparts through transformation tensors derived from the sensitivity of logarithmic strain with respect to the deformation gradient. Separating geometric pre- and postprocessing from the constitutive model allows classical small-strain formulations to be embedded directly into finite-strain computations, simplifying implementation while maintaining thermodynamic consistency and numerical robustness.

\noindent\textit{(a) Geometric Preprocessor}: The deformation gradient is defined as,
\begin{equation}
    \bm{F} = \bm{I} + \nabla_0 \bm{u}
\end{equation}
where $\bm{I}$ is the second-order identity tensor. The right Cauchy--Green tensor is computed as,
\begin{equation}
    \bm{C} = \bm{F}^\top \bm{F}
\end{equation}
The total logarithmic strain $\bm{\varepsilon}$ is obtained from the spectral decomposition of $\bm{C}$ as,
\begin{equation}
    \label{eq:c_decomp}
    \bm{\varepsilon} = \tfrac{1}{2}\ln \bm{C} = \sum_{a=1}^{3} \ln(\lambda_a) \, \bm{n}_a \otimes \bm{n}_a
\end{equation}
where $\lambda_a$ and $\bm{n}_a$ are the eigenvalues and eigenvectors of the right Cauchy--Green tensor, respectively. Similarly, the plastic deformation is characterized by a plastic metric $\bm{G}^p$ from which the logarithmic plastic strain is computed as,
\begin{equation}
    \bm{\varepsilon}_\mathrm{p} = \tfrac{1}{2}\ln \bm{G}^p
\end{equation}
In practice, the logarithmic plastic strain $\bm{\varepsilon}_\mathrm{p}$ appears as an internal state variable in the constitutive model. This preprocessing step transforms the nonlinear finite-strain kinematics into an additive strain space where small-strain constitutive models apply directly.

\noindent\textit{(b) Constitutive Model}: In logarithmic strain space, the additive decomposition of elastic and plastic strains is,
\begin{equation}
    \bm{\varepsilon} = \bm{\varepsilon}_\mathrm{e} + \bm{\varepsilon}_\mathrm{p}
\end{equation}
where $\bm{\varepsilon}_\mathrm{e}$ denotes the recoverable logarithmic elastic strain. This construction mirrors the classical small-strain theory and extends naturally to finite strains. In logarithmic strain space, the elastic response is assumed linear,
\begin{equation}
    \bm{T} = \mathbb{E} : \bm{\varepsilon}_\mathrm{e}
    \label{eq:T}
\end{equation}
where $\bm{T}$ is the work-conjugate logarithmic stress tensor and $\mathbb{E}$ is the fourth-order elasticity tensor. $\mathbb{E}$ is specified compactly by nine scalar stiffness coefficients $\{d_i\}_{i=1}^9$ in a local orthotropy frame and rotated into the global coordinate system using structural tensors, thereby enabling an efficient representation of general orthotropic elastic behavior.

\vspace{0.5em}

Plastic flow is governed by an anisotropic yield function ($f$) with an isotropic Voce hardening law,
\begin{equation}
    f := \phi(\bm{T}) - \sigma_y(\alpha)
\end{equation}
where $\sigma_y$ denotes the yield stress, $\alpha$ is the equivalent plastic strain, and $\phi(\bm{T})$ denotes the effective stress. The yield stress evolves with the isotropic hardening variable $\alpha$ according to the exponential Voce law,
\begin{equation}
    \sigma_y(\alpha) = \sigma_0 + \sqrt{\frac{2}{3}}\,Q\bigl(1 - e^{-b\alpha}\bigr)
    \label{eq:Hard}
\end{equation}
where $\sigma_0$, $Q$, and $b$ are material parameters.

% \paragraph{Hill--48 model}:
\noindent\textit{Hill--48 model}: For the Hill--48 yield criterion, the effective stress $\phi(\bm{T})$ is obtained using a fourth-order Hill tensor $\mathbb{H}$,
\begin{equation}
\label{eq:Phi_hill}
    \phi(\bm{T}) = \sqrt{\bm{T} : \mathbb{H} : \bm{T}}
\end{equation}
The tensor $\mathbb{H}$ encodes the orthotropic yield-stress ratios and shear-yield ratios in the material frame and is rotated into the global frame using the material orientation. The parameterization used in the implementation is given in Appendix~\ref{sec:appendix_A}.

\noindent\textit{Barlat model}: The Barlat Yld2004-18p yield criterion is implemented as an alternative anisotropic yield function, particularly suited for sheet-metal forming applications. Its effective stress is computed from the principal values of two linearly transformed deviatoric stress tensors,
\begin{equation}
\begin{aligned}
    \label{eq:Phi_barlat}
    \phi(\bm{T}) = \left[\frac{1}{4}\Big(
    |s'_1 - s''_1|^m + |s'_1 - s''_2|^m + |s'_1 - s''_3|^m \right.
    +\, |s'_2 - s''_1|^m + |s'_2 - s''_2|^m + |s'_2 - s''_3|^m \\
    \left. +\, |s'_3 - s''_1|^m + |s'_3 - s''_2|^m + |s'_3 - s''_3|^m
    \Big)\right]^{1/m}
\end{aligned}
\end{equation}
where $s'_i$ and $s''_j$ ($i,j = 1,2,3$) are the eigenvalues of the transformed stress tensors $\bm{s}'$ and $\bm{s}''$, and $m$ is a material exponent. The two transformations are parameterized by 18 anisotropy coefficients and are summarized in Appendix~\ref{sec:appendix_A}.

For both yield criteria, the evolution of the plastic strain and the internal variable follows an associative flow rule,
\begin{align}
    \dot{\bm{\varepsilon}}^p &= \dot{\gamma} \frac{\partial \phi}{\partial \bm{T}} = \dot{\gamma} \,\bm{N}(\bm{T})
    \label{eq:ep_evol} \\[3pt]
    \dot{\alpha} &= \dot{\gamma}\,\mathcal{W}
\end{align}
where $\dot{\gamma}$ is the plastic multiplier, $\bm{N}$ is the plastic flow direction, and $\mathcal{W}$ is a model-dependent scalar. We use $\mathcal{W}=\sqrt{2/3}$ for the Hill--48 model and $\mathcal{W}=1$ for the Barlat Yld2004-18p model. The plastic multiplier $\dot{\gamma}\geq 0$ is determined by the Kuhn--Tucker conditions,
\begin{equation}
    \dot{\gamma} \;\geq\; 0, 
    \quad f(\bm{T}, \alpha) \;\leq\; 0, 
    \quad \dot{\gamma}\,f(\bm{T}, \alpha) \;=\; 0
    \label{eq:KT}
\end{equation}

\noindent\textit{(c) Geometric Postprocessor}: Once the constitutive model provides the updated stress and internal variables in logarithmic strain space, these quantities are mapped back to finite-strain stress measures for the global equilibrium problem. This mapping uses transformation tensors obtained from the chain rule applied to the logarithmic strain measure. The Lagrangian (second Piola--Kirchhoff) stress $\bm{S}$ is recovered through the fourth-order transformation tensor $\mathbb{P}_L = 2\,\partial\bm{\varepsilon}/\partial\bm{C}$, which relates the rate of logarithmic strain to the material time derivative of the right Cauchy--Green tensor. The first Piola--Kirchhoff stress then follows from $\bm{P} = \bm{F}\bm{S}$. This postprocessing step is purely geometric and independent of the constitutive response; it relies only on function evaluations and spectral decompositions of the current metric tensor. The consistent elastoplastic tangent moduli in the Lagrangian configuration are obtained by applying the fourth-order tensor $\mathbb{P}_L$ and the sixth-order tensor $\mathbb{L}_L = 4\,\partial^2\bm{\varepsilon}/\partial\bm{C}^2$ to transform the tangent operator from logarithmic strain space to the configuration space.

\subsubsection{Implicit Time Discretization} \label{subsubsec:implicit_time_discretization}

We next outline the backward-Euler time discretization of the constitutive equations over a time step $[t_n, t_{n+1}]$. In logarithmic strain space, the return-mapping algorithm reduces to the classical small-strain plasticity procedure. Let $\bm{T}_n$, $\bm{F}_n$, $\bm{\varepsilon}_n^{\,p}$, and $\alpha_n$ be known at time $t_n \in [0,T]$ and location $\bm{X} \in \mathcal{B}$. The total logarithmic strain $\bm{\varepsilon}_{n+1}$ and $\bm{C}_{n+1}$ are assumed to be known from the current deformation state at time $t_{n+1}$. We seek the updated quantities $\bm{T}_{n+1}$, $\bm{\varepsilon}_{n+1}^{\,p}$, $\alpha_{n+1}$, and the incremental plastic multiplier $\Delta \gamma$ at time $t_{n+1}$ that satisfy the constitutive relations and consistency conditions Eq.~\eqref{eq:T} and Eq.~\eqref{eq:Hard}--\eqref{eq:KT}, where $\Delta\gamma = \dot{\gamma}\,\Delta t$.

The time discretization leads to the standard elastic-predictor/plastic-corrector return-mapping algorithm. Assuming that the entire deformation increment is elastic, we first compute the trial stress,
\begin{equation}
    \bm{T}_{n+1}^{\text{tr}}
    = \mathbb{E} : \bigl(\bm{\varepsilon}_{n+1} - \bm{\varepsilon}_{n}^{\,p}\bigr) \label{eq:Tr_stress}
\end{equation}
The yield condition evaluated at the trial state is,
\begin{equation}
    f^{\text{tr}} = \phi(\bm{T}_{n+1}^{\text{tr}}) - \sigma_y(\alpha_n)
\end{equation}
The elastic prediction is accepted if $f^{\text{tr}} \leq 0$,
\begin{equation}
\begin{aligned}
    \Delta \gamma &= 0, \qquad
    \bm{T}_{n+1} = \bm{T}_{n+1}^{\text{tr}} \\[3pt]
    \bm{\varepsilon}_{n+1}^{\,p} &= \bm{\varepsilon}_{n}^{\,p}, \qquad
    \alpha_{n+1} = \alpha_n
\end{aligned}
\end{equation}
If $f^{\text{tr}} > 0$, a plastic corrector step is performed by solving the implicit nonlinear evolution equations using the closest-point projection (CPP) algorithm~\cite{deSouzaNeto2008}. The discretized evolution equations [Eq.~\eqref{eq:ep_evol}--Eq.~\eqref{eq:KT}] are,

\begin{align}
    \bm{\varepsilon}_{n+1}^{\,p} &= \bm{\varepsilon}_{n}^{\,p} + \Delta\gamma\,\bm{N}_{n+1} \label{eq:Ep_update} \\
    \alpha_{n+1} &= \alpha_n + \mathcal{W}\,\Delta\gamma \label{eq:alpha_update} \\
    f_{n+1} &= \phi(\bm{T}_{n+1}) - \sigma_y(\alpha_{n+1}) = 0 \label{eq:yield_final}
\end{align}
Substituting Eq.~\eqref{eq:Ep_update} into Eq.~\eqref{eq:Tr_stress} gives,
\begin{equation}
        \bm{T}_{n+1} - \mathbb{E} : \bigl(\bm{\varepsilon}_{n+1} - \bm{\varepsilon}_{n}^{\,p} - \Delta\gamma\,\bm{N}_{n+1}\bigr)=\bm{0} 
\end{equation}
The CPP solves for the unknowns $\bm{y} = [\bm{T}_{n+1} \;\;\alpha_{n+1}\;\; \Delta\gamma  ]^\top$ by forming a nonlinear residual system and solving it with a Newton--Raphson scheme. The residual vector $\bm{G}(\bm{y})$ is,
\begin{equation}
    \bm{G}(\bm{y}) = \begin{Bmatrix}
        \bm{T}_{n+1} - \mathbb{E} : \bigl(\bm{\varepsilon}_{n+1} - \bm{\varepsilon}_{n}^{\,p} - \Delta\gamma\,\bm{N}_{n+1}\bigr) \\
        \alpha_{n+1} - \alpha_n - \mathcal{W}\,\Delta\gamma \\
        \phi(\bm{T}_{n+1}) - \sigma_y(\alpha_{n+1}) 
    \end{Bmatrix} = \begin{Bmatrix}\bm{0} \\ 0 \\ 0\end{Bmatrix}
    \label{eq:local_cpp}
\end{equation}
The fully implicit Newton--Raphson iterations are performed at each integration point as,
\begin{equation}
    \label{eq:local_solver}
    \bm{J}^{G}(\bm{y}^k)\,\delta\bm{y}^k = -\bm{G}(\bm{y}^k), 
    \qquad \bm{y}^{k+1} = \bm{y}^k + \delta\bm{y}^k
\end{equation}
where $\bm{J}^{G} = \partial\bm{G}/\partial\bm{y}$ is the Jacobian of the local system of equations (Eq.~\eqref{eq:local_cpp}).

To improve robustness when the full Newton correction is too aggressive for highly anisotropic nonquadratic yield surfaces, we augment the local update with a safeguarded line search along the Newton direction, following Scherzinger~\cite{Scherzinger2017}. The local Newton update is written as $\bm{y}^{k+1}=\bm{y}^k+\zeta^k\delta\bm{y}^k$, where $\zeta^k \in (0,1]$ is selected to reduce a scalar residual merit function. The Goldstein sufficient-decrease test, quadratic step-size update, and safeguard used in the implementation are given in Appendix~\ref{sec:appendix_B}. The overall stress update algorithm is summarized in Appendix~\ref{sec:appendix_C}.

This implicit scheme applies to return mapping for general constitutive models and provides quadratic convergence when the full Newton step is accepted. For complex material models, manual derivation and implementation of the Jacobian can be tedious and error-prone~\cite{ScaletAuricchio2018,RotheHartmann2015}, motivating approximate tangent strategies such as secant or quasi-Newton procedures~\cite{Alfano1999}. In the present implementation, $\bm{J}^{G}$ is obtained through automatic differentiation (AD) in JAX, which ensures consistency between the residual and its linearization without manual derivation. This approach retains the convergence properties of the full Newton--Raphson method while avoiding manual algebraic manipulation of the tangent operator.

Once the corrected stress $\bm{T}_{n+1}$ is found, the second Piola--Kirchhoff stress $\bm{S}_{n+1}$ is recovered using the projection operator,
\begin{equation}
\begin{aligned}
\label{eq:proj_lagrange}
\bm{S}_{n+1} = \mathbb{P}_L : \bm{T}_{n+1}
\end{aligned}
\end{equation}
The first Piola--Kirchhoff stress is finally recovered as,
\begin{equation}
    \label{eq:pk_stress}
    \bm{P}_{n+1} = \bm{F}_{n+1}\,\bm{S}_{n+1}
\end{equation}
$\bm{P}_{n+1}$ is used in the weak form (Eq.~\eqref{eq:weak_form}) to automatically compute the consistent tangent stiffness.

\subsubsection{Finite Element Formulation}
\label{subsec:fe_formulation}

The weak form in Eq.~\eqref{eq:weak_form} is discretized using a standard displacement-based finite element approximation in the reference configuration. The body $\mathcal{B}$ is partitioned into finite elements $\mathcal{B}^e$ such that,
\begin{equation}
    \mathcal{B} = \bigcup_{e=1}^{n_\mathrm{el}} \mathcal{B}^e
\end{equation}
where $n_\mathrm{el}$ denotes the number of elements. Within each element, the displacement field is approximated as,
\begin{equation}
    \bm{u}^h(\bm{X}) = \sum_{I=1}^{n_\mathrm{en}} N_I(\bm{X})\,\bm{d}_I
\end{equation}
where $N_I$ are the standard shape functions, $n_\mathrm{en}$ is the number of nodes per element, and $\bm{d}_I$ are the nodal displacement vectors. The corresponding virtual displacement is interpolated in the same way,
\begin{equation}
    \delta \bm{u}^h(\bm{X}) = \sum_{I=1}^{n_\mathrm{en}} N_I(\bm{X})\,\delta \bm{d}_I
\end{equation}
The deformation gradient at quadrature points is,
\begin{equation}
    \bm{F}^h(\bm{X}) 
    = \bm{I} + \nabla_0 \bm{u}^h(\bm{X})
    = \bm{I} + \sum_{I=1}^{n_\mathrm{en}} \bm{B}_I(\bm{X})\,\bm{d}_I
\end{equation}
where $\bm{B}_I = \nabla_0 N_I \otimes \bm{I}$ denotes the gradient of the shape functions with respect to the reference coordinates.

In nearly incompressible regimes, a purely displacement-based formulation with standard interpolation for $\bm{u}$ may exhibit severe volumetric locking. To alleviate this pathology while retaining the same kinematic approximation, we adopt the $F$-bar methodology of Neto \emph{et al.}~\cite{deSouzaNeto1996LowOrder,deSouzaNeto2005FBar}. In this approach, the deformation gradient at each integration point is modified so that the incompressibility constraint is enforced in an average sense at the element level rather than pointwise.

The $F$-bar method is based on a multiplicative decomposition of the deformation gradient into isochoric and volumetric contributions,
\begin{equation}
    \bm{F}^{h} = \bm{F}^{h}_{\text{iso}} \,\bm{F}^{h}_{\text{vol}},
    \qquad 
    \bm{F}^{h}_{\text{iso}} = J^{-1/3}\bm{F}^{h},
    \qquad 
    \bm{F}^{h}_{\text{vol}} = J^{1/3}\bm{I},
    \qquad 
    J = \det \bm{F}^{h}
\end{equation}
where $\bm{F}^{h}_{\text{iso}}$ and $\bm{F}^{h}_{\text{vol}}$ denote the isochoric and volumetric parts of the deformation gradient, respectively. For each element, an averaged Jacobian $\bar{J}$ is then computed and used to construct a modified deformation gradient $\bar{\bm{F}}^h$ that is passed to the constitutive model,
\begin{equation}
    \bar{\bm{F}}^h = \bar{J}^{1/3} J^{-1/3} \bm{F}^h,
    \qquad 
    \bar{J} = \frac{1}{V_e} \int_{\mathcal{B}^e} J \, dV
\end{equation}
with $V_e$ the element volume in the reference configuration. In this way, the volumetric contribution is replaced by an element-wise projected counterpart while the isochoric part retains the standard Gauss-point resolution. For materials with decoupled volumetric--deviatoric response, this construction yields a constant pressure field within each element and effectively suppresses volumetric locking.

Substituting the finite element approximations into Eq.~\eqref{eq:weak_form} gives the nonlinear algebraic equilibrium equation,
\begin{equation}
    \bm{R}(\bm{d}) = \bm{0}
\end{equation}
The global residual is assembled from element residuals as,
\begin{equation}
    \bm{R}(\bm{d}) =
    \mathop{\mathcal{A}}_{e=1}^{n_\mathrm{el}}
    \bm{R}^e(\bm{d}^e)
\end{equation}
where $\mathcal{A}$ denotes the standard finite element assembly operator and $\bm{d}^e \in \mathbb{R}^{3n_\mathrm{en}}$ is the local displacement vector for element $e$. The residual component associated with node $I$ of element $e$ is,
\begin{equation}
    \bm{R}^e_I(\bm{d}^e)
    =
    \int_{\mathcal{B}^e}
    \bm{P}(\bar{\bm{F}}^h,\bm{h}_n)
    \cdot
    \nabla_0 N_I \, dV
    -
    \int_{\Gamma_t^e}
    \bar{\bm{t}}\,N_I \, dA
    \qquad I=1,\ldots,n_\mathrm{en}
\end{equation}
where $\bm{h}_n$ denotes the history variables from the previous converged increment, such as $\bm{\varepsilon}^{p}_n$ and $\alpha_n$ at the quadrature points. Thus, each element residual is a local map,
\begin{equation}
    \bm{R}^e : \mathbb{R}^{3n_\mathrm{en}}
    \rightarrow
    \mathbb{R}^{3n_\mathrm{en}}
    \qquad
    \bm{d}^e \mapsto \bm{R}^e(\bm{d}^e)
\end{equation}
The corresponding element stiffness matrix is the Jacobian of this local residual,
\begin{equation}
    \bm{K}^e(\bm{d}^e)
    =
    \frac{\partial \bm{R}^e}{\partial \bm{d}^e}
    \in
    \mathbb{R}^{3n_\mathrm{en}\times 3n_\mathrm{en}}
\end{equation}
In classical finite element implementations, this linearization often requires manual derivation of the consistent material tangent and associated geometric transformations. In the present JAX-based implementation, $\bm{K}^e$ is computed directly by automatic differentiation of the element residual kernel. This keeps the element stiffness matrix consistent with the implemented residual, including the $F$-bar kinematics, logarithmic-strain constitutive update, and implicit return-mapping algorithm.

The global tangent stiffness matrix follows from assembly of the element stiffness matrices,
\begin{equation}
    \bm{K}(\bm{d})
    =
    \frac{\partial \bm{R}}{\partial \bm{d}}
    =
    \mathop{\mathcal{A}}_{e=1}^{n_\mathrm{el}}
    \bm{K}^e(\bm{d}^e)
\end{equation}
The nonlinear system is solved using a global Newton--Raphson scheme. At iteration $k$, the linearized system is,
\begin{equation}
\label{eq:Stiffness_mat}
    \bm{K}(\bm{d}^k)\Delta\bm{d}^k
    =
    -\bm{R}(\bm{d}^k)
    \qquad
    \bm{d}^{k+1}
    =
    \bm{d}^k+\Delta\bm{d}^k
\end{equation}

\subsubsection{Implementation Aspects} \label{subsec:Implementation_sec}
\paragraph{Differentiable spectral decomposition and projection operator.}

The geometric pre- and postprocessors defined above require the spectral map $\bm{C}\mapsto\bm{\varepsilon}(\bm{C})$ (Eq.~\eqref{eq:c_decomp}) and its first- and second-order projection tensors $\mathbb{P}_L$ and $\mathbb{L}_L$. However, in JAX's automatic-differentiation framework, the standard symmetric eigensolver \texttt{jax.numpy.linalg.eigh} is not differentiable in the presence of repeated eigenvalues. In such cases, the eigenvectors are not uniquely defined, the derivative of the eigendecomposition becomes ill-posed, and JAX explicitly disables higher-order autodiff through \texttt{eigh}.

We use two complementary treatments for this issue. The first is a purely numerical regularization: before calling \texttt{eigh}, we replace $\bm{C}$ by a slightly perturbed tensor $\tilde{\bm{C}} = \bm{C} + \delta\bm{I}$ with a very small diagonal perturbation $\delta \ll 1$. This breaks exact eigenvalue multiplicities and allows JAX to differentiate through the spectral map. The resulting gradients are approximate but sufficiently accurate and robust for the forward and inverse simulations considered here.

The second approach is an analytic custom Jacobian--vector product (JVP) treatment based on the Seth--Hill spectral projection formulas of Miehe and Lambrecht~\cite{MieheLambrecht2001}, specialized to the logarithmic strain measure. We treat the spectral strain map and projection operation as custom primitives and provide Jacobian--vector products using closed-form expressions for $\mathbb{P}_L$ and $\mathbb{L}_L$. These formulas are written in terms of eigenprojectors, divided differences, and their repeated-eigenvalue limits, so the directional derivatives remain well-defined without perturbing $\bm{C}$ or differentiating through the eigensolver itself. This preserves stable evaluation of the logarithmic strain, stress pull-back, and consistent tangent contributions while retaining compatibility with higher-order automatic differentiation in the JAX framework.

\paragraph{Implicit Differentiation of the Constitutive Update Solver.}

The plastic corrector at each integration point is implemented as a nonlinear solver (Eq.~\eqref{eq:local_solver}) that runs Newton iterations on the local residual equations. Rather than differentiating through every iteration and loop in this solver, we expose the constitutive update as a custom primitive in JAX and attach a custom JVP rule. This JVP is based on the implicit function theorem and uses the local residual Jacobian together with one automatic-differentiation call with respect to all inputs. As a result, the implementation obtains exact directional derivatives of the updated stress and internal variables with respect to the strain and material parameters without recording the Newton iterations in the computational graph. This treatment provides stable, consistent algorithmic tangents for the global Newton solver while keeping the code differentiable and efficient for gradient-based inverse design.Details of the custom JVP implementation are provided in Appendix~\ref{sec:appendix_D}. The overall stress-update procedure and the local return-mapping algorithm are summarized in Appendix~\ref{sec:appendix_C}.

\begin{figure}[!htbp]
\centering
\includegraphics[width=0.8\textwidth]{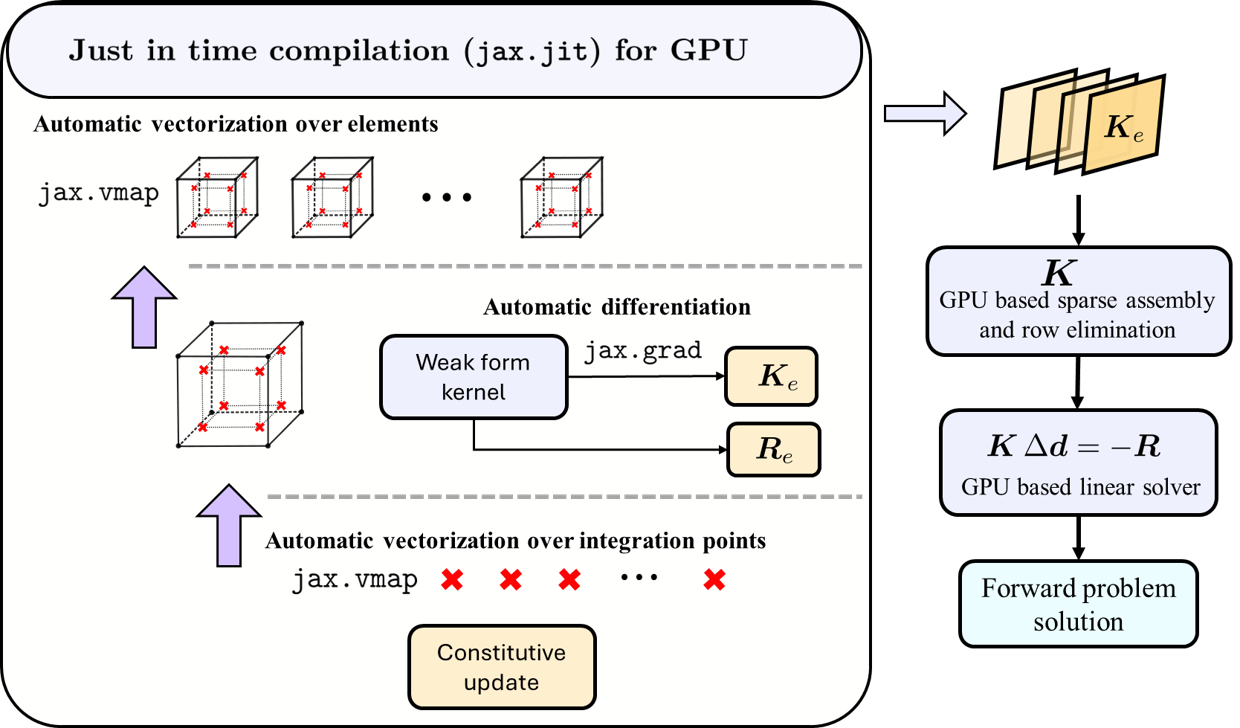}
\caption{Overview of the differentiable GPU-accelerated forward-solve implementation.}
\label{fig:Forward_overview}
\end{figure}

\paragraph{Parallel Element-wise Residual and Stiffness Calculation.}
The implementation follows the residual-based formulation described in Section~\ref{subsec:fe_formulation}. A weak-form kernel maps the local displacement vector $\bm{d}^e$ and element history variables to the element residual $\bm{R}^e(\bm{d}^e)$. The kernel evaluates $\bm{F}^h$, applies the $F$-bar projection, calls the constitutive update at quadrature points, and contracts the returned first Piola--Kirchhoff stress with the reference shape-function gradients. The element stiffness matrix is obtained as,
\begin{equation}
    \bm{K}^e = \frac{\partial \bm{R}^e}{\partial \bm{d}^e}
\end{equation}
using JAX automatic differentiation. Both residual and stiffness kernels are vectorized with \texttt{jax.vmap}, with one level over quadrature points and another over elements. The batched kernels are compiled with \texttt{jax.jit}, allowing XLA to optimize GPU execution. The resulting element Jacobians are converted into sparse triplets and assembled into the global tangent matrix $\bm{K}$. The overall implementation is illustrated in Fig.~\ref{fig:Forward_overview}.

\paragraph{GPU-accelerated global sparse matrix assembly.} Sparse finite element assembly on GPUs is challenging because many element contributions target shared global degrees of freedom, producing irregular memory access and potential write conflicts. Specialized data layouts, communication-aware implementations, and coloring or warp-based strategies have therefore been widely studied~\cite{Fu2014GPUPipeline,Meng2014GPUEM,Kiran2019GPUColoring}. In our implementation, the element-level Jacobians are first represented as triplets of values and row/column indices and are then assembled on the GPU using \texttt{CuPy} sparse operations. To control peak memory use, the triplets are processed in chunks: each chunk is converted to COO format and accumulated into a global CSR matrix. Dirichlet rows are eliminated directly on the GPU by batched CUDA kernels that zero off-diagonal entries and reset the diagonal to unity. Thus, both assembly and boundary-condition enforcement remain accelerator resident, yielding a scalable construction of the global tangent matrix. The procedure is detailed in Appendix~\ref{sec:appendix_E}, where it is summarized in Algorithm~\ref{alg:gpu_chunked_assembly}, and GPU/CPU assembly timings are reported in Table~\ref{tab:assembly_computation_times}.

\paragraph{GPU-accelerated linear solvers.} To solve the linear system arising from the global Newton--Raphson iteration (Eq.~\eqref{eq:Stiffness_mat}), the framework provides several GPU-accelerated sparse linear solver options. For iterative solves, it supports native JAX routines such as CG and BiCGSTAB. In addition, JAX-FEM is interfaced with NVIDIA's AMGX library, which provides optimized GPU implementations of iterative solvers for large sparse systems. The AMGX interface allows users to select solver configurations, including algebraic multigrid preconditioners and Krylov subspace methods, for a given problem size and matrix structure. Custom external linear solvers are invoked through \texttt{jax.pure\_callback} to preserve differentiability within JAX. For problems where iterative methods are ineffective or fail to converge robustly, the framework also provides access to the NVIDIA CuDSS sparse direct solver. CPU-based direct solvers available through SciPy and \texttt{pypardiso} remain available for smaller problems or hardware configurations without suitable GPU support. In the present implementation, AMGX BiCGSTAB is used as the default option, while CuDSS is employed as a fallback when the iterative solve does not converge within a prescribed number of iterations. A comparative assessment of these solver choices in terms of computational performance is provided in Appendix~\ref{sec:appendix_F}.

\paragraph{Automatic time-stepping scheme.}
To improve robustness in strongly nonlinear loading regimes, the forward solver employs an automatic adaptive time-stepping strategy. At each increment, a trial step size ($\Delta t_{initial}$) is proposed within prescribed minimum ($\Delta t_{\min}$) and maximum ($\Delta t_{\max}$) bounds. The step size is clipped when necessary so that the solver exactly hits user-defined time markers. The initial guess for the Newton solve of a given step is obtained by linear or quadratic extrapolation from the two most recent converged solutions. If the global Newton solve fails to converge or produces non-finite solutions, the increment is rejected and the step size is reduced by a fixed cutback factor before retrying. Conversely, after repeated successful steps, the time increment is increased up to a prescribed maximum, thereby reducing the total number of load steps while retaining stability near difficult regions of the response.

\FloatBarrier

\subsection{Inverse Problem}
We formulate constitutive-parameter identification as a PDE-constrained optimization problem. Gradient-based optimization requires derivatives of the objective with respect to all model parameters. Direct forward sensitivity analysis therefore becomes prohibitively expensive as the parameter dimension grows. We instead adopt a discrete adjoint formulation, which is equivalent to reverse-mode differentiation of the time-discrete forward problem and has a computational cost that is nearly independent of the number of parameters~\cite{SeidlGranzow2022}.
The objective function is defined as the discrepancy between simulated and ground-truth displacement fields over all time steps,
\begin{equation}
J(\boldsymbol{\theta}) = \sum_{n=1}^{N_T} J^n
= w \sum_{n=1}^{N_T} \| \mathbf{u}_{\text{sim}}^n - \mathbf{u}_{\text{gt}}^n \|^2
\end{equation}
where $\boldsymbol{\theta}$ denotes the vector of constitutive material parameters, and $\mathbf{u}_{\text{sim}}^n$ and $\mathbf{u}_{\text{gt}}^n$ are the simulated and ground-truth displacement fields at time step $n$, respectively. Here, $w$ is a scalar weighting factor and $N_T$ denotes the total number of time steps. 

The inverse problem is then posed as the constrained minimization,
\begin{equation}
\begin{aligned}
\min_{\boldsymbol{\theta}} \quad & J(\boldsymbol{\theta}) \\
\text{s.t.} \quad &
\mathbf{R}^n(\mathbf{u}^n, \mathbf{u}^{n-1}, \boldsymbol{\theta}) = \mathbf{0}
\quad n = 1, \ldots, N_T
\end{aligned}
\end{equation}
In the reduced formulation adopted here, internal variables are not treated as independent optimization variables. Their evolution is embedded in the constitutive update and is therefore implicitly accounted for through the global residual $\mathbf{R}^n$.
Solving the inverse problem with gradient-based optimizers requires the total derivative $\mathrm{d}J/\mathrm{d}\boldsymbol{\theta}$. The adjoint method computes this sensitivity efficiently by introducing Lagrange multipliers that enforce the equilibrium constraints. The resulting adjoint equations eliminate the need for expensive forward sensitivity solves~\cite{SeidlGranzow2022,giles2000introduction}. The following sections describe the adjoint formulation, its implementation through custom reverse-mode automatic differentiation in JAX, and its verification against finite-difference approximations.

\subsubsection{Adjoint Formulation}

To derive the adjoint equations, we augment the objective with the equilibrium constraints using the method of Lagrange multipliers. The Lagrangian functional is defined as,
\begin{equation}
\label{eq:Lagrangian}
\mathcal{L}(\mathbf{u}^n,\mathbf{u}^{n-1},\boldsymbol{\theta})
=
\sum_{n=1}^{N_T}
\left[
J^n(\mathbf{u}^n,\boldsymbol{\theta})+
(\boldsymbol{\lambda}^n)^T
\mathbf{R}^n(\mathbf{u}^n,\mathbf{u}^{n-1},\boldsymbol{\theta})
\right]
\end{equation}
where $\boldsymbol{\lambda}^n \in \mathbb{R}^{N_{\text{dof}}}$ are the Lagrange multipliers enforcing the global equilibrium constraint at time step $n$. Because the forward simulation satisfies $\mathbf{R}^n=\mathbf{0}$ for all $n$, the augmented terms vanish identically,
\begin{equation}
J(\boldsymbol{\theta}) = \mathcal{L}(\mathbf{u}^n,\mathbf{u}^{n-1},\boldsymbol{\theta})
\end{equation}
Consequently, the total derivatives are equal as well,
\begin{equation}
\frac{\mathrm{d}J}{\mathrm{d}\boldsymbol{\theta}}
=
\frac{\mathrm{d}\mathcal{L}}{\mathrm{d}\boldsymbol{\theta}}
\end{equation}
Expanding the right-hand side via the chain rule gives,
\begin{equation}
\label{eq:total_derivative_chain}
\frac{\mathrm{d}J}{\mathrm{d}\boldsymbol{\theta}}
=
\frac{\mathrm{d}\mathcal{L}}{\mathrm{d}\boldsymbol{\theta}}
=
\frac{\partial \mathcal{L}}{\partial \boldsymbol{\theta}}
+
\sum_{n=1}^{N_T}
\frac{\partial \mathcal{L}}{\partial \mathbf{u}^n}
\frac{\mathrm{d}\mathbf{u}^n}{\mathrm{d}\boldsymbol{\theta}}
\end{equation}
The matrices $\mathrm{d}\mathbf{u}^n/\mathrm{d}\boldsymbol{\theta}$ are the forward sensitivity matrices. Computing and storing these matrices becomes prohibitively expensive as the parameter dimension grows. The adjoint method avoids this cost by choosing the multipliers $\boldsymbol{\lambda}^n$ so that the sensitivity-dependent terms in Eq.~\eqref{eq:total_derivative_chain} vanish exactly,
\begin{equation}
\frac{\partial \mathcal{L}}{\partial \mathbf{u}^n} = \mathbf{0}
\quad n = 1, \ldots, N_T
\label{eq:adj_u}
\end{equation}
Eq.~\eqref{eq:adj_u} constitutes the \emph{adjoint equation}. Once this equation is satisfied, the sensitivity-dependent term in Eq.~\eqref{eq:total_derivative_chain} drops out and the gradient reduces to,
\begin{equation}
\frac{\mathrm{d}J}{\mathrm{d}\boldsymbol{\theta}}
=
\frac{\partial \mathcal{L}}{\partial \boldsymbol{\theta}}
\end{equation}
Differentiating the Lagrangian in Eq.~\eqref{eq:Lagrangian} with respect to $\boldsymbol{\theta}$ while holding all state and adjoint variables fixed yields the explicit gradient formula,
\begin{equation}
\label{eq:main_grad}
\frac{\mathrm{d}J}{\mathrm{d}\boldsymbol{\theta}}
=
\frac{\partial \mathcal{L}}{\partial \boldsymbol{\theta}}
=
\sum_{n=1}^{N_T}
\left[
\frac{\partial J^n}{\partial \boldsymbol{\theta}}
+
(\boldsymbol{\lambda}^n)^T
\frac{\partial \mathbf{R}^n}{\partial \boldsymbol{\theta}}
\right]
\end{equation}
It remains to determine the Lagrange multipliers $\boldsymbol{\lambda}^n$ from the adjoint equation~\eqref{eq:adj_u}. Because $\mathbf{u}^n$ appears in $J^n$, $\mathbf{R}^n$, and $\mathbf{R}^{n+1}$, differentiating the Lagrangian with respect to $\mathbf{u}^n$ and setting the result to zero gives,
\begin{equation}
\frac{\partial \mathcal{L}}{\partial \mathbf{u}^n}
=
\frac{\partial J^n}{\partial \mathbf{u}^n}
+
(\boldsymbol{\lambda}^n)^T \frac{\partial \mathbf{R}^n}{\partial \mathbf{u}^n}
+
(\boldsymbol{\lambda}^{n+1})^T \frac{\partial \mathbf{R}^{n+1}}{\partial \mathbf{u}^n}
=
\mathbf{0}
\end{equation}
Rearranging and transposing yields the adjoint linear system at time step $n$,
\begin{equation}
\label{eq:dR_u}
\left(
\frac{\partial \mathbf{R}^n}{\partial \mathbf{u}^n}
\right)^T
\boldsymbol{\lambda}^n
=
-
\left[
\left(
\frac{\partial J^n}{\partial \mathbf{u}^n}
\right)^T
+
\left(
\frac{\partial \mathbf{R}^{n+1}}{\partial \mathbf{u}^n}
\right)^T
\boldsymbol{\lambda}^{n+1}
\right]
\end{equation}
where $\partial \mathbf{R}^n/\partial \mathbf{u}^n$ is the tangent stiffness matrix (Eq.~\eqref{eq:Stiffness_mat}) assembled during the forward Newton solve at step $n$. Eq.~\eqref{eq:dR_u} is integrated backward in time from $n=N_T$ to $n=1$, with the terminal condition $\boldsymbol{\lambda}^{N_T+1} = \mathbf{0}$.
This backward sweep yields the complete set of adjoint variables $\{\boldsymbol{\lambda}^n\}_{n=1}^{N_T}$, after which the gradient is assembled via Eq.~\eqref{eq:main_grad}. The implementation of this procedure within JAX's automatic differentiation framework is described in the following section.

\subsubsection{Implementation via Custom Reverse-Mode AD}

We implement the adjoint method using JAX's \texttt{custom\_vjp} (vector-Jacobian product) mechanism~\cite{xu2020adcme}. This mechanism avoids differentiating directly through the iterative Newton--Raphson solver at each time step. Instead, a custom backward pass solves the linear adjoint systems in Eq.~\eqref{eq:dR_u}. All partial derivative terms required for adjoint assembly are computed automatically using \texttt{jax.vjp}, without manual derivation of linearization kernels. The execution flow is presented in Algorithm~\ref{alg:adjoint_ad}. This approach maintains gradient accuracy while reducing computational cost.

\newcommand{\KsubT}{\mathbf{K}_T}
\newcommand{\fadj}{\mathbf{f}_{\mathrm{adj}}}

\begin{algorithm}[!htbp]
\caption{Adjoint-Based Gradient Computation with JAX-AD}
\label{alg:adjoint_ad}
\begin{algorithmic}
\Require Material parameters $\boldsymbol{\theta}$
\Ensure Total gradient $\frac{\mathrm{d}J}{\mathrm{d}\boldsymbol{\theta}}$

\State \textbf{Forward Pass (Primal Solve and Store History)}
\For{$n=1$ to $N_T$}
    \State Retrieve previous state $\mathbf{u}^{n-1}$ from $\mathcal{H}[n-1]$
    \State Solve $\mathbf{R}^n(\mathbf{u}^n, \mathbf{u}^{n-1}; \boldsymbol{\theta}) = \mathbf{0}$ for $\mathbf{u}^n$
    \Comment{Implicit equilibrium solve}
    \State Store $\mathcal{H}[n] \gets (\mathbf{u}^n, \mathbf{u}^{n-1}, \boldsymbol{\theta})$
    \Comment{Store step inputs for AD}
\EndFor

\Statex
\State \textbf{Backward Pass (Adjoint Solve and Gradient Accumulation)}
\State Initialize total gradient: $\frac{\mathrm{d}J}{\mathrm{d}\boldsymbol{\theta}} \gets \mathbf{0}$
\State Initialize terminal adjoint: $\boldsymbol{\lambda}^{N_T+1} \gets \mathbf{0}$

\For{$n=N_T$ down to $1$}
    \State $(\mathbf{u}^n, \mathbf{u}^{n-1}, \boldsymbol{\theta}) \gets \mathcal{H}[n]$
    \Comment{Retrieve stored quantities at step $n$}

    \State 
    $\mathbf{v}^n
    \gets
  -
    \left[
    \left(
    \frac{\partial J^n}{\partial \mathbf{u}^n}
    \right)^T
    +
    \left(
    \frac{\partial \mathbf{R}^{n+1}}{\partial \mathbf{u}^n}
    \right)^T
    \boldsymbol{\lambda}^{n+1}
    \right]$
    \Comment{RHS of Eq.~\eqref{eq:dR_u}}

    \State Evaluate tangent stiffness $\frac{\partial \mathbf{R}^n}{\partial \mathbf{u}^n}$
    \Comment{Compute using JAX-AD}

    \State Solve
    $\left(
    \frac{\partial \mathbf{R}^n}{\partial \mathbf{u}^n}
    \right)^T
    \boldsymbol{\lambda}^n
    =
    \mathbf{v}^n $
    \Comment{Adjoint linear solve}

    \State Evaluate VJP: 
     $\mathbf{v}_{\mathbf{R},\boldsymbol{\theta}}^n
    \gets
    \left(
    \frac{\partial \mathbf{R}^n}{\partial \boldsymbol{\theta}}
    \right)^T
    \boldsymbol{\lambda}^n$
    \Comment{Obtain using \texttt{jax.vjp}}
    
    $\frac{\mathrm{d}J}{\mathrm{d}\boldsymbol{\theta}}
    \gets
    \frac{\mathrm{d}J}{\mathrm{d}\boldsymbol{\theta}}
    +
    \left(
    \frac{\partial J^n}{\partial \boldsymbol{\theta}}
    \right)^T
    +
    \mathbf{v}_{\mathbf{R},\boldsymbol{\theta}}^n$
    \Comment{Accumulate gradient Eq.~\eqref{eq:main_grad}}
\EndFor

\State \Return $\frac{\mathrm{d}J}{\mathrm{d}\boldsymbol{\theta}}$
\end{algorithmic}
\end{algorithm}

\subsubsection{Finite Difference Approximation}

The finite-difference method provides a straightforward alternative for computing gradients because it requires only objective-function evaluations. We use central finite differences to verify the adjoint gradients. For each parameter component $i=1,\ldots,N_p$, the gradient is approximated as,
\begin{equation}
\left[
\frac{\mathrm{d}J}{\mathrm{d}\boldsymbol{\theta}}
\right]_i
\approx
\frac{1}{2h}
\left(
J(\boldsymbol{\theta} + h\mathbf{e}_i)
-
J(\boldsymbol{\theta} - h\mathbf{e}_i)
\right)
\end{equation}
where $\mathbf{e}_i$ is the $i$-th standard basis vector and $h>0$ is the finite-difference step size. This procedure requires $2N_p$ complete forward simulations, giving the computational cost $\mathrm{Cost}_{\mathrm{FD}} = 2N_p \times \mathrm{Cost}_{\mathrm{forward}}$,
which scales linearly with parameter dimension. In contrast, the adjoint method computes the gradient at a cost comparable to at most two forward solves regardless of $N_p$. Algorithm~\ref{alg:finite_difference} summarizes the finite-difference procedure used as an independent verification tool for the JAX-AD-based adjoint gradients.

\begin{algorithm}[!htbp]
\caption{Finite Difference Gradient}
\label{alg:finite_difference}
\begin{algorithmic}
\Require Material parameters $\boldsymbol{\theta} \in \mathbb{R}^{N_p}$, step size $h > 0$
\Ensure Gradient approximation $\dfrac{\mathrm{d}J}{\mathrm{d}\boldsymbol{\theta}} \in \mathbb{R}^{N_p}$

\For{$i = 1$ to $N_p$} \Comment{Loop over all parameters}
  \State $\boldsymbol{\theta}^{+} \gets \boldsymbol{\theta} + h \mathbf{e}_i$
  \Comment{Perturb $i$-th parameter forward}
  \State $\mathbf{u}^{n,+} \gets$ $\mathbf{R}^n(\mathbf{u}^n, \mathbf{u}^{n-1}, \boldsymbol{\theta}^{+}) = \mathbf{0}$ for $n = 1, \ldots, N_T$
    \Comment{Forward solve}
  \State Evaluate $J^{+}_i \gets J(\boldsymbol{\theta}^{+})$
  
  \State $\boldsymbol{\theta}^{-} \gets \boldsymbol{\theta} - h \mathbf{e}_i$
  \Comment{Perturb $i$-th parameter backward}
  \State $\mathbf{u}^{n,-} \gets$ $\mathbf{R}^n(\mathbf{u}^n, \mathbf{u}^{n-1}, \boldsymbol{\theta}^{-}) = \mathbf{0}$ for $n = 1, \ldots, N_T$
  \Comment{Forward solve}
  \State Evaluate $J^{-}_i \gets J(\boldsymbol{\theta}^{-})$
  
  \State Compute $\left[ \dfrac{\mathrm{d}J}{\mathrm{d}\boldsymbol{\theta}} \right]_i \gets \dfrac{1}{2h} \left( J^{+}_i - J^{-}_i \right)$
  \Comment{$i$-th gradient component}
\EndFor

\State \Return $\dfrac{\mathrm{d}J}{\mathrm{d}\boldsymbol{\theta}}$

\end{algorithmic}
\end{algorithm}

\section{Results}
\label{sec:results}

This section presents verification studies and performance benchmarks for the proposed differentiable finite element framework.
We begin with the forward problem by assessing the accuracy and computational performance of the JAX-FEM-ANISO solver on representative plasticity benchmarks.
We then verify gradients computed by the adjoint-based AD method against second-order finite-difference approximations.
Finally, we demonstrate inverse material-parameter identification on geometries of varying complexity.

\subsection[Performance of Forward Simulations with JAX-FEM-ANISO]{Performance of Forward Simulations with JAX-FEM-ANISO}
We evaluate the forward solver through benchmark examples covering large-deformation anisotropic plasticity. For selected cases, GPU wall-clock times are compared with CPU and GPU runs in the commercial solver Abaqus to quantify the speed-up achieved by the JAX-FEM-ANISO implementation.

\subsubsection{Hill--48 Plasticity Model Verification}
\begin{figure}[!htbp]
    \centering
    \includegraphics[width=0.3\textwidth]{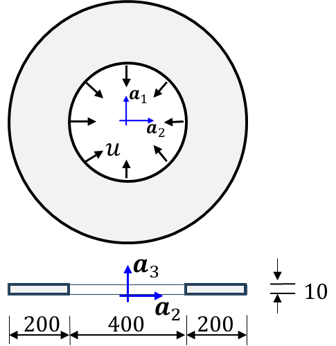}
    \caption{Schematic diagram of the thin circular flange considered for the drawing simulations. All dimensions are in mm.}
    \label{fig:CircularDraw}
\end{figure}

In this section, we verify JAX-FEM-ANISO finite-strain anisotropic plasticity simulations with the Hill--48 plasticity model against Abaqus results using a circular sheet drawing benchmark. This example idealizes the deformation of the outer portion of a flat circular blank during deep drawing. The problem is a standard benchmark for anisotropic plasticity simulations and has been used extensively in foundational studies~\cite{Miehe2002,PapadopoulosLu1998}.
The schematic diagram of the hollow circular sheet geometry, including dimensions and loading conditions, is shown in Fig.~\ref{fig:CircularDraw}. The geometry is discretized using 940 hexahedral elements and 2960 nodes, with a single element layer through the thickness. The bottom face is constrained against out-of-plane ($a_3$) displacement, while a total radial displacement of 75 mm is applied to all nodes on the inner boundary. The simulation employs isotropic elastic properties and anisotropic plastic behavior, with material orientations aligned with the axes shown in Fig.~\ref{fig:CircularDraw}. The Hill plasticity model is parameterized by the initial axial yield stresses $\sigma^y_{ii}$ and shear yield stresses $\tau^y_{ij}$ along the principal axes of orthotropy. In this benchmark, we set $\sigma^y_{11} = \sigma^y_{22} = \sigma^y_{33} = \sigma_0$ and $\tau^y_{12} = \tau^y_{23} = \tau^y_{13}$, so that the anisotropy ratio $\delta = \sigma^y_{11}/\tau^y_{12}$ controls the degree of plastic anisotropy. To assess the robustness of the implementation, simulations are performed for two distinct anisotropy ratios: (a) $\delta = 3.4641$ and (b) $\delta = 0.86603$. Isotropic hardening following the Voce hardening law is adopted for all simulations, with complete material parameters provided in Table~\ref{tab:material_parameters}.

\begin{figure}[!htbp]
    \centering
    \begin{tikzpicture}
        % Node for the image
        \node[anchor=south west, inner sep=0] (image) at (0,0) {\includegraphics[width=0.8\textwidth]{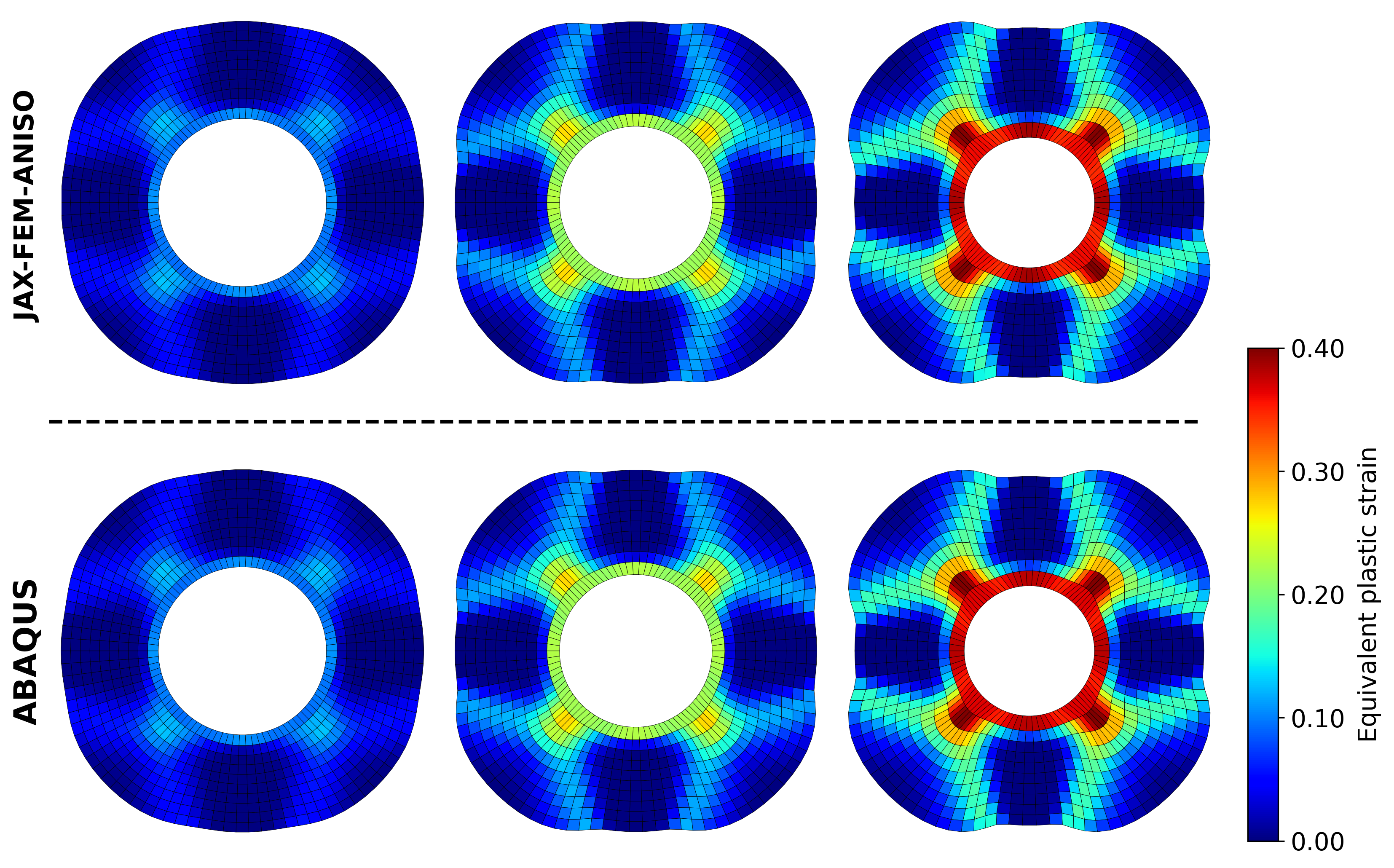}};
        % Scope to define coordinates relative to image size (0 to 1)
        \begin{scope}[x={(image.south east)},y={(image.north west)}]
            \node at (0.02, 1.00) {(a)};
            % --- TOP LABELS (Displacement) ---
            \node[above] at (0.18, 0.98) {$u = 25$ mm};
            \node[above] at (0.46, 0.98) {$u = 50$ mm};
            \node[above] at (0.75, 0.98) {$u = 75$ mm};
        \end{scope}
    \end{tikzpicture}

    \vspace{0.1cm} % 
    
    \begin{tikzpicture}
        % Node for the image
        \node[anchor=south west, inner sep=0] (image) at (0,0) {\includegraphics[width=0.8\textwidth]{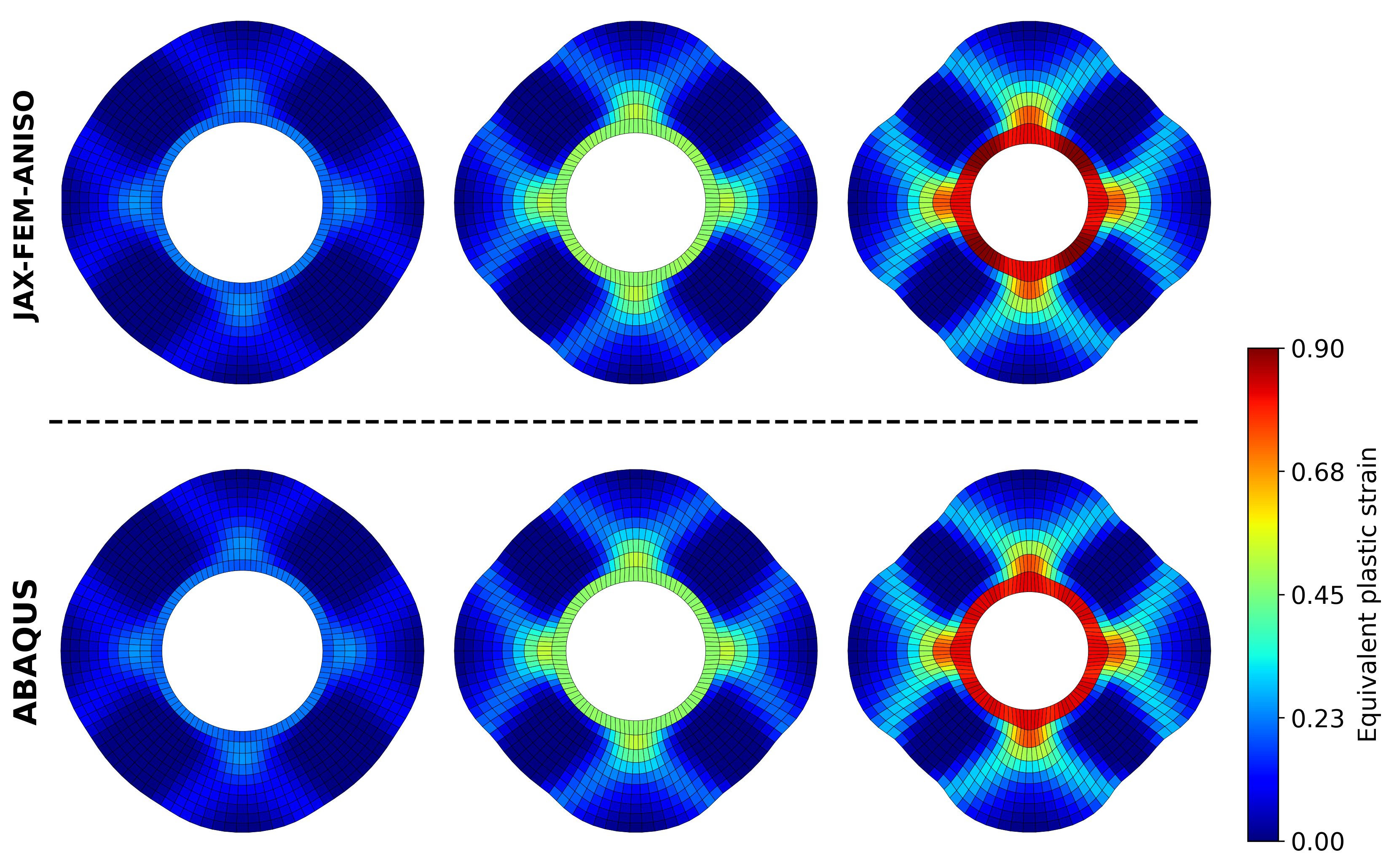}};
        
        % Scope to define coordinates relative to image size (0 to 1)
        \begin{scope}[x={(image.south east)},y={(image.north west)}]
            \node at (0.02, 0.95) {(b)};
            
        \end{scope}
    \end{tikzpicture}
    
    \caption[Comparison of equivalent plastic strain in flange drawing simulations between JAX-FEM-ANISO and Abaqus]{Comparison of equivalent plastic strain in flange drawing simulations between JAX-FEM-ANISO and Abaqus. The deformed meshes are shown at radial displacements of $u = 25$, $50$, and $75$ mm. Results correspond to anisotropic responses characterized by (a) $\delta = 3.4641$ and (b) $\delta = 0.86603$.}
    \makeatletter\def\@currentlabelname{Comparison of equivalent plastic strain in flange drawing simulations between JAX-FEM-ANISO and Abaqus}\makeatother
    \label{fig:CircularResult2}
\end{figure}
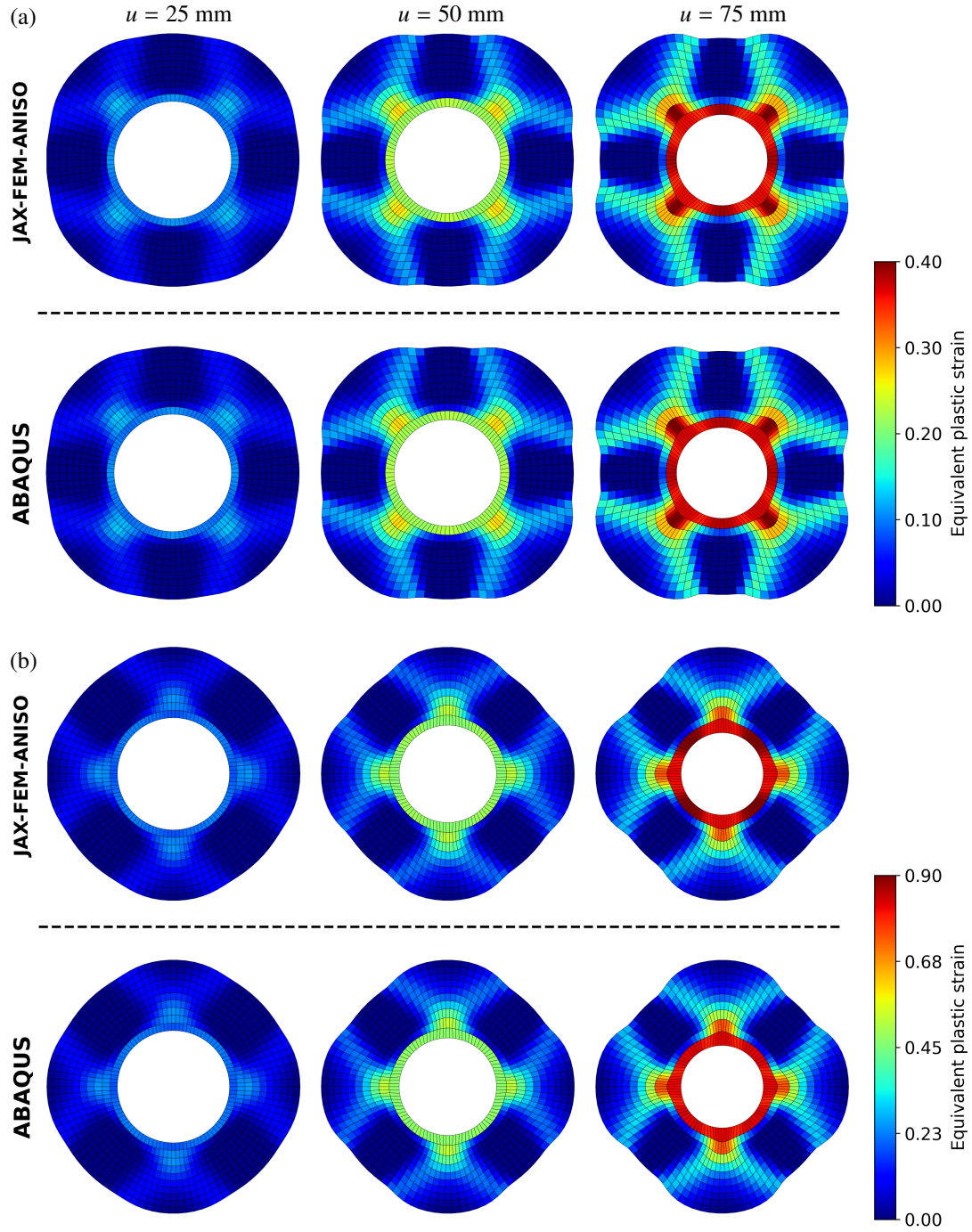

The JAX-FEM-ANISO simulation results for both parameter sets are compared with Abaqus results in Fig.~\ref{fig:CircularResult2}(a) and Fig.~\ref{fig:CircularResult2}(b). These figures show the equivalent plastic strain distribution on the deformed mesh at three displacement levels: $u = 25$ mm, $u=50$ mm, and $u=75$ mm. For both material parameter sets, the outer ring exhibits a wavelike deformation pattern, which is documented in the sheet-metal forming literature as earing. The JAX-FEM-ANISO and Abaqus results show excellent agreement for this large-deformation plasticity benchmark.
% Table
\begin{table}[!htbp]
\centering
\begin{threeparttable}
\begin{minipage}{\standardtablewidth}
\caption{Material parameters used in the circular sheet drawing benchmark.}
\label{tab:material_parameters}
\centering
\begin{tabular*}{\textwidth}{@{\extracolsep{\fill}}ccccc@{}}
\toprule
E & $\sigma_0$ & $\nu$ & Q & b \\
\midrule
200 GPa & 450 MPa & 0.3 & 400 MPa & 0.2 \\
\bottomrule
\end{tabular*}
\end{minipage}
\end{threeparttable}
\end{table}

% Figure for resutls

\FloatBarrier

\subsubsection{Barlat Yld2004-18p Plasticity Model Verification} 

\begin{figure}[!htbp]
    \centering
    \begin{minipage}[t]{0.48\textwidth}
        \centering
        \begin{tikzpicture}
            \node[anchor=south west, inner sep=0] (image) at (0,0) {\includegraphics[width=\textwidth]{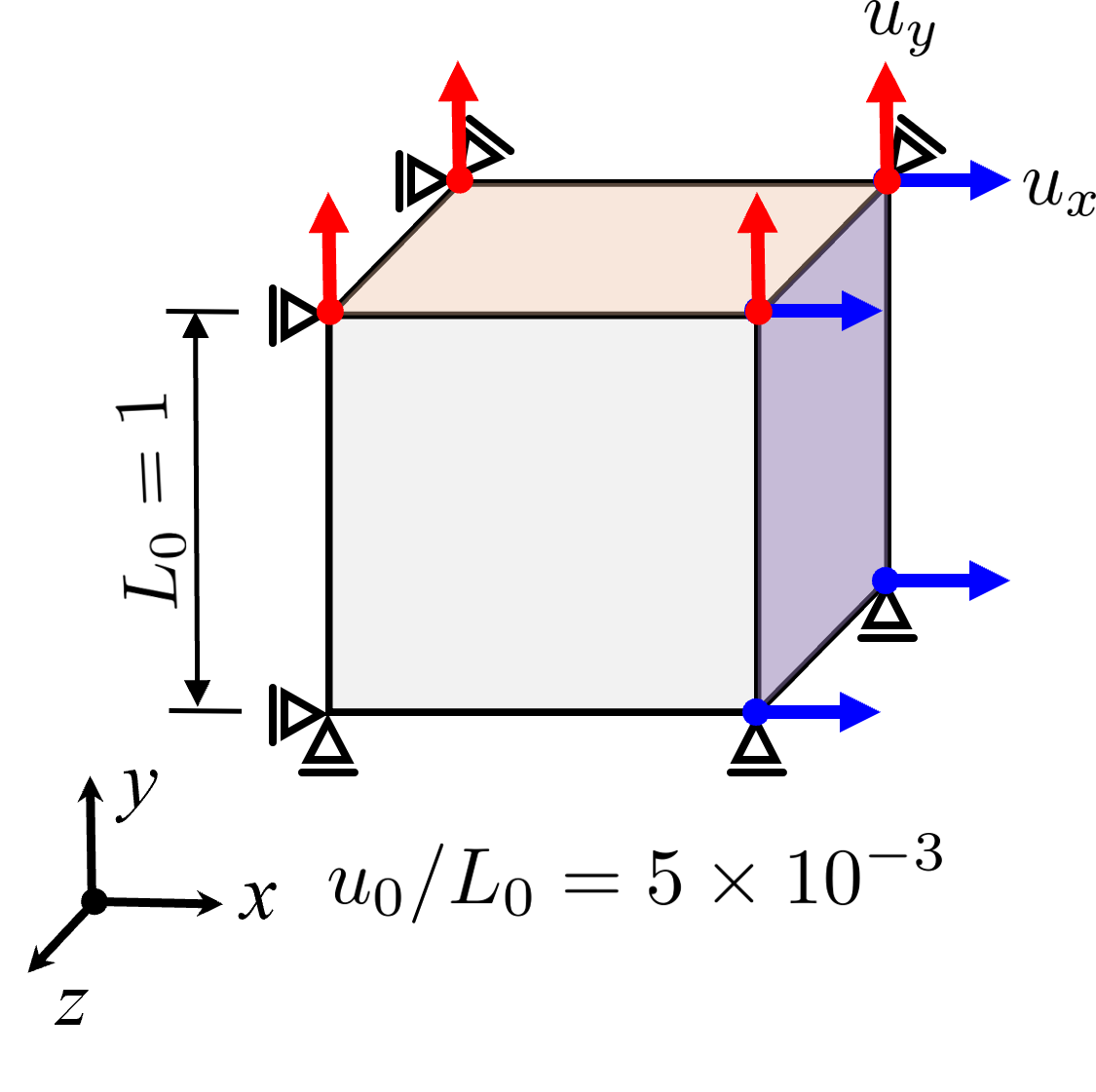}};
            \begin{scope}[x={(image.south east)},y={(image.north west)}]
                \node[anchor=north west] at (0.02,0.98) {(a)};
            \end{scope}
        \end{tikzpicture}
    \end{minipage}
    \hfill
    \begin{minipage}[t]{0.48\textwidth}
        \centering
        \begin{tikzpicture}
            \node[anchor=south west, inner sep=0] (image) at (0,0) {\includegraphics[width=\textwidth]{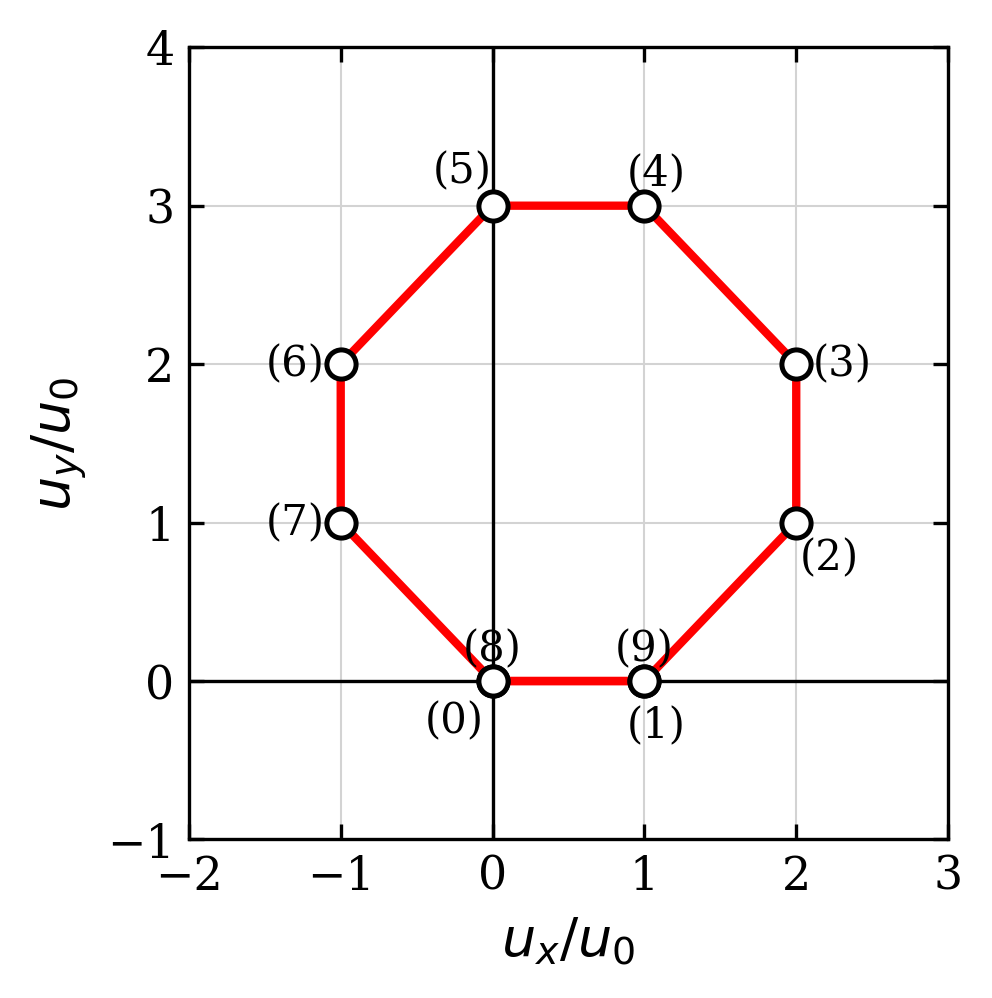}};
            \begin{scope}[x={(image.south east)},y={(image.north west)}]
                \node[anchor=north west] at (0.02,0.98) {(b)};
            \end{scope}
        \end{tikzpicture}
    \end{minipage}
    \caption{Verification setup for the Barlat Yld2004-18p model. (a) Unit-cube single-element benchmark with the applied boundary conditions. (b) Prescribed displacement history imposed on the highlighted nodes.}
    \label{fig:BarlatSetup}
\end{figure}

The JAX-FEM-ANISO implementation of the Yld2004-18p model is verified against reference solutions generated in Abaqus with the Unified Material Model Driver for Plasticity (UMMDp) UMAT of Takizawa et al.~\cite{Takizawa2018}. The benchmark consists of a unit cube discretized with a single HEX-8 element. The boundary conditions and prescribed loading history are shown in Fig.~\ref{fig:BarlatSetup}(a) and (b), respectively. Rigid-body motion is suppressed by constraining the faces at $x=0$, $y=0$, and $z=0$ against displacement in the $x$, $y$, and $z$-directions, respectively, while leaving the remaining degrees of freedom unconstrained to preserve the intended deformation path. The rolling direction (RD) and transverse direction (TD) are aligned with the $x$ and $y$-axes, respectively. The material parameters~\cite{Barlat2005} used in this benchmark are listed in Table~\ref{tab:barlat_verification_material_parameters}. Perfect plasticity is assumed for this verification example.

\begin{figure}[!htbp]
    \centering
    \includegraphics[width=0.5\textwidth]{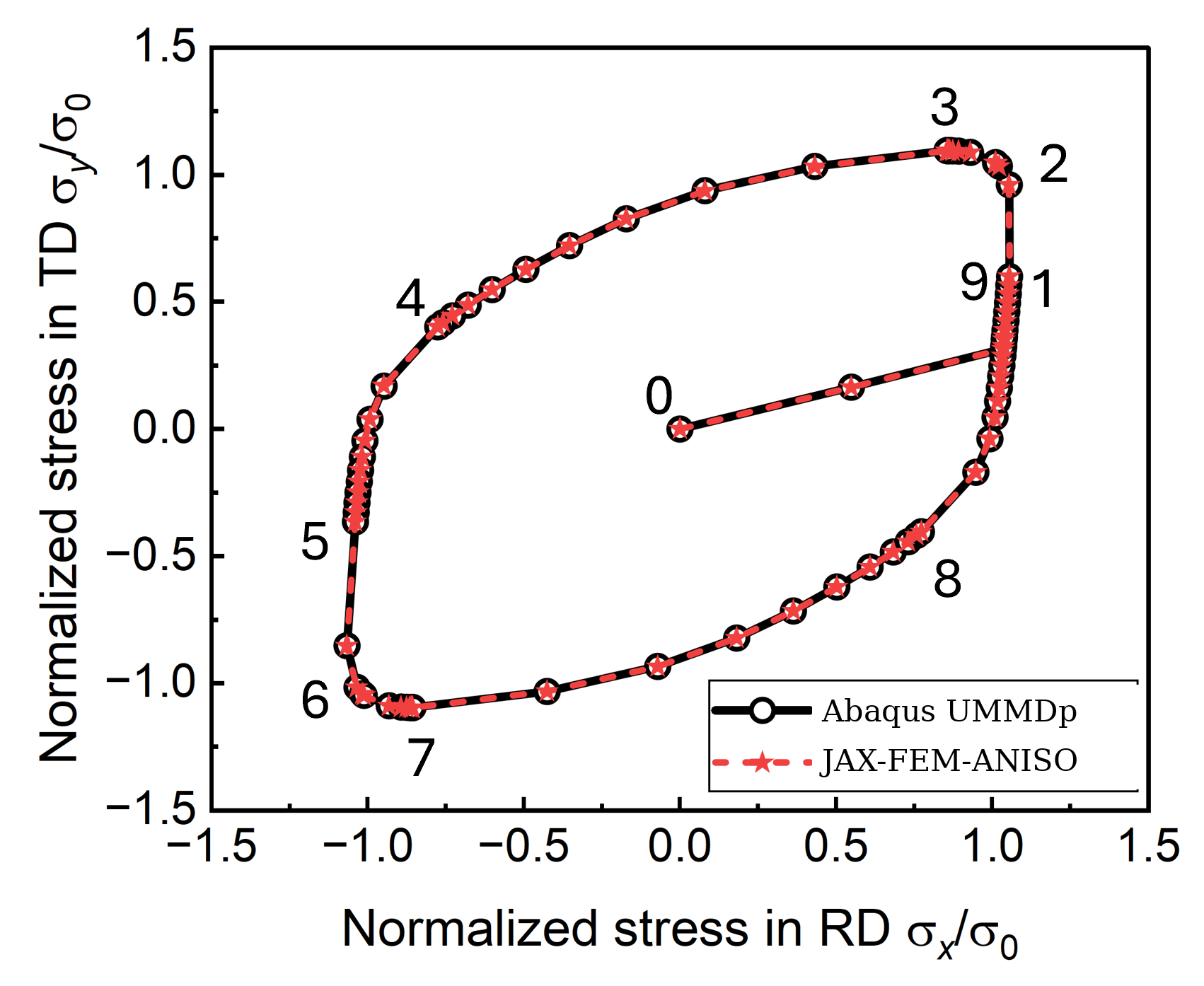}
    \caption[Comparison of the stress plots obtained using JAX-FEM-ANISO and Abaqus for the Barlat Yld2004-18p verification test]{Comparison of the stress plots obtained using JAX-FEM-ANISO and Abaqus for the Barlat Yld2004-18p verification test.}
    \label{fig:BarlatVerification}
\end{figure}

Fig.~\ref{fig:BarlatVerification} compares the stress plots obtained from JAX-FEM-ANISO and Abaqus. The two results are in excellent agreement, with the JAX-FEM-ANISO response overlapping the Abaqus reference solution throughout the verification test. This agreement verifies the present Yld2004-18p implementation and shows that the consistent tangent obtained through JAX automatic differentiation reproduces the expected constitutive response.

\begin{table}[!htbp]
    \centering
\begin{threeparttable}
\begin{minipage}{\standardtablewidth}
    \caption{Material parameters used in the Barlat Yld2004-18p verification test~\cite{Barlat2005}.}
    \label{tab:barlat_verification_material_parameters}
    \setlength{\tabcolsep}{2pt}
\centering
    \begin{tabular*}{\textwidth}{@{\extracolsep{\fill}}ccccccccc@{}}
        \toprule
        $c'_{12}$ & $c'_{13}$ & $c'_{21}$ & $c'_{23}$ & $c'_{31}$ & $c'_{32}$ & $c'_{44}$ & $c'_{55}$ & $c'_{66}$ \\
        \midrule
        -0.069888 & 0.936408 & 0.079143 & 1.003060 & 0.524741 & 1.363180 & 1.023770 & 1.069060 & 0.954322 \\
        \midrule
        $c''_{12}$ & $c''_{13}$ & $c''_{21}$ & $c''_{23}$ & $c''_{31}$ & $c''_{32}$ & $c''_{44}$ & $c''_{55}$ & $c''_{66}$ \\
        \midrule
        0.981171 & 0.476741 & 0.575316 & 0.866827 & 1.145010 & -0.079294 & 1.051660 & 1.147100 & 1.404620 \\
        \midrule
        $m$ & $E$ & $\nu$ & $\sigma_0$ &  &  &  &  &  \\
        \midrule
        8.0 & 1000 & 0.3 & 1.0 &  &  &  &  &  \\
        \bottomrule
    \end{tabular*}
\end{minipage}
\end{threeparttable}
\end{table}

\FloatBarrier

\subsubsection{Computational Performance Benchmarks}

\begin{figure}[!htbp]
    \centering
    \includegraphics[width=0.5\textwidth]{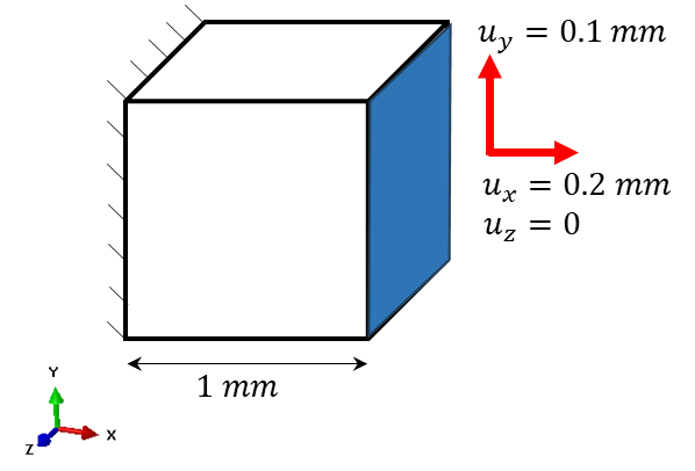}
    \caption{Schematic of the unit-cube example with loading and boundary conditions used for the forward-simulation benchmarks.}
    \label{fig:UnitCubePerformance}
\end{figure}

To assess the computational performance and scalability of JAX-FEM-ANISO, we consider the unit-cube geometry shown in Fig.~\ref{fig:UnitCubePerformance}. The left face of the cube is fixed, while the right face is displaced by 0.2 mm in the positive $x$-direction and 0.1 mm in the positive $y$-direction, creating a nonuniform elastoplastic deformation. The constitutive response combines isotropic elasticity, Hill--48 anisotropic plasticity, and Voce isotropic hardening. The material parameters used in this benchmark are listed in Table~\ref{tab:performance_material_parameters}. The linear system is solved using the AMGX BiCGSTAB solver on the GPU. Because of the strong geometric and material nonlinearities, the loading is applied through the automatic time-stepping strategy described in Section~\ref{subsec:Implementation_sec}, with $\Delta t_{initial}=10^{-3}$, $\Delta t_{\min}=10^{-5}$, and $\Delta t_{\max}=5\times10^{-2}$.

\begin{figure}[!htbp]
    \centering
    \includegraphics[width=0.8\textwidth]{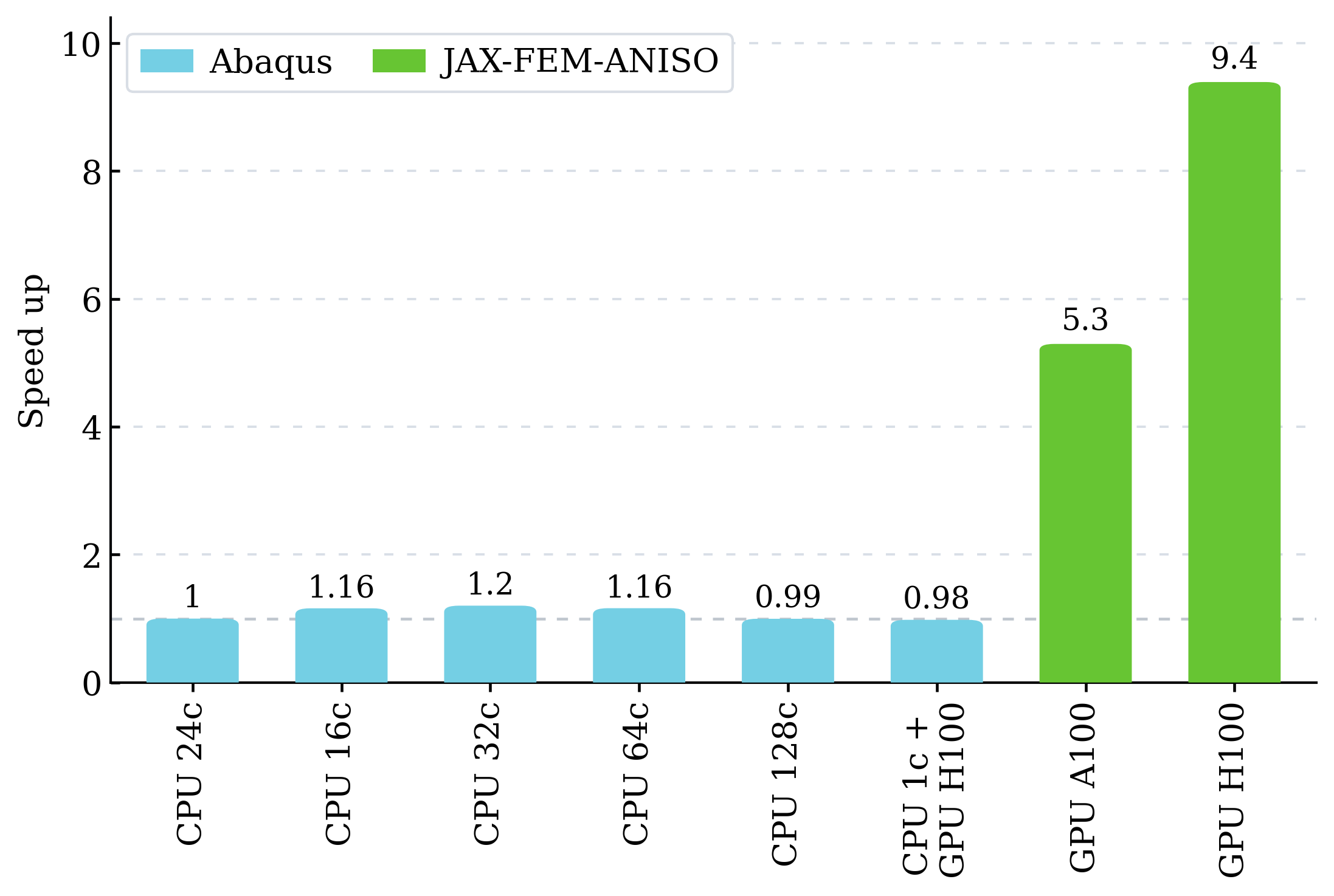}
    \caption[Speed-up comparison between JAX-FEM-ANISO and Abaqus for the unit-cube benchmark]{Speed-up comparison between JAX-FEM-ANISO and Abaqus for the unit-cube benchmark with $100\times100\times100$ elements, using the Abaqus CPU (24-core) case as the reference configuration. The Abaqus simulations are performed on different CPU configurations ranging from 1 core to 128 cores, together with an Abaqus GPU-enabled run.}
    \label{fig:ForwardGPU}
\end{figure}

For the first benchmark, the cube is discretized with $100\times100\times100$ hexahedral elements. This simulation compares JAX-FEM-ANISO with several Abaqus CPU and GPU configurations, including 16, 24, 32, 64, and 128-core CPU runs, as well as an Abaqus run using an H100 GPU. All simulations use the same loading protocol and adaptive stepping settings. In JAX-FEM-ANISO, the computation completes in 31 load steps and 111 nonlinear iterations. The H100 and A100 GPU executions require 2030 s and 3598 s, respectively. The computation times are listed in Table~\ref{tab:forward_gpu_time}, and the corresponding speed-up comparison is shown in Fig.~\ref{fig:ForwardGPU}. Relative to the Abaqus 24-core CPU workstation, which is used here as the reference configuration, JAX-FEM-ANISO on the H100 GPU achieves a speed-up of 9.4$\times$. In contrast, the Abaqus GPU-enabled run yields a speed-up of 1.6$\times$, and increasing the number of Abaqus CPU cores beyond 32 does not produce a meaningful reduction in wall-clock time.

\begin{figure}[!htbp]
    \centering
    \includegraphics[width=0.7\textwidth]{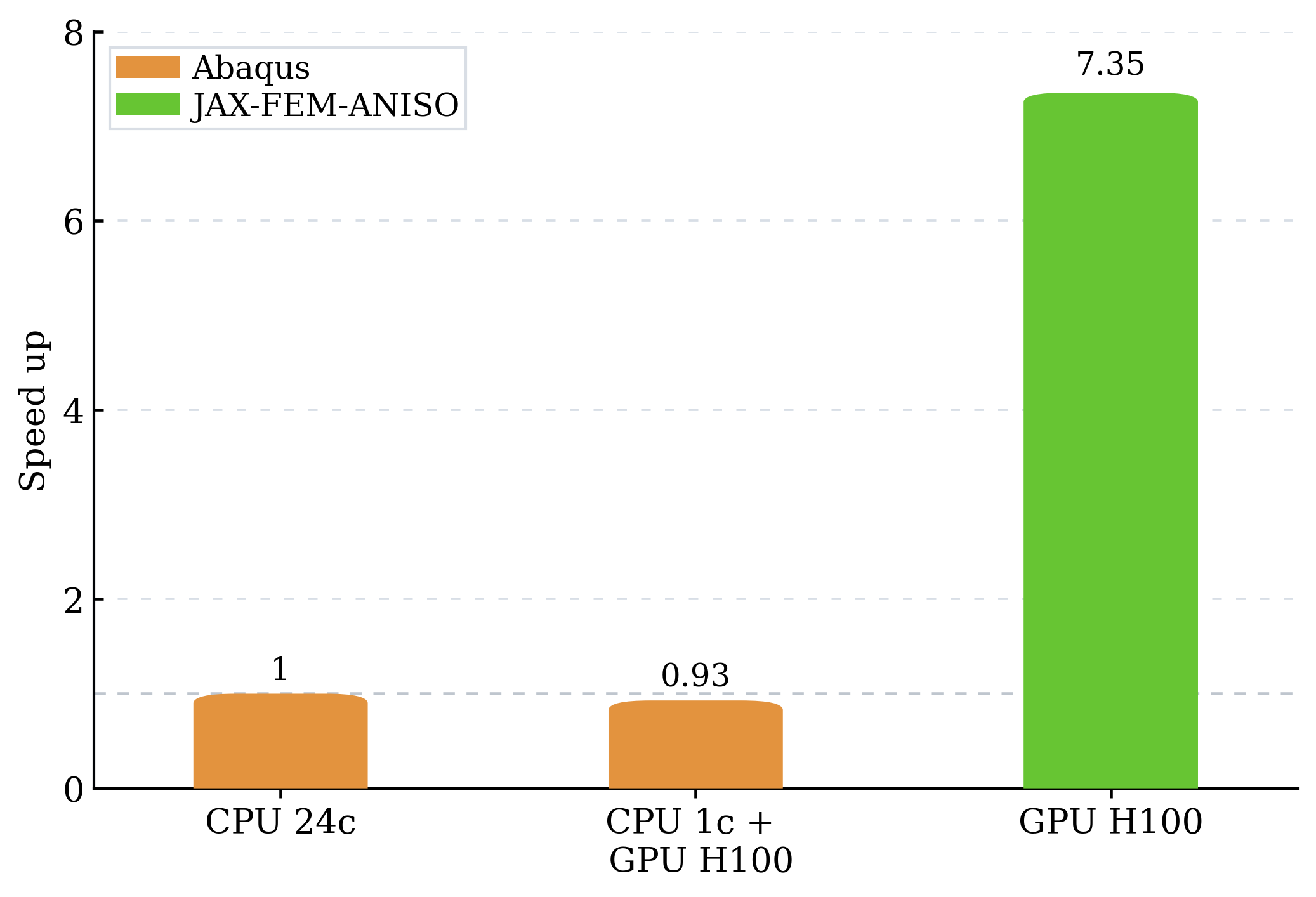}
    \caption[Speed-up comparison between JAX-FEM-ANISO and Abaqus for the larger unit-cube benchmark]{Speed-up comparison between JAX-FEM-ANISO and Abaqus for the larger unit-cube benchmark with approximately 11.2 million degrees of freedom, using the Abaqus CPU (24-core) case as the reference configuration.}
    \label{fig:ForwardGPULarge}
\end{figure}

To examine a larger problem size, the same unit-cube geometry is discretized with $250\times150\times120$ elements, corresponding to approximately 11.2 million degrees of freedom. This case is solved with JAX-FEM-ANISO on an H100 GPU and with Abaqus MPI on a 24-core, 128 GB workstation. The Abaqus simulation takes 123 nonlinear iterations, whereas JAX-FEM-ANISO takes 122 iterations. The corresponding computation times are also summarized in Table~\ref{tab:forward_gpu_time}, while Fig.~\ref{fig:ForwardGPULarge} shows that JAX-FEM-ANISO retains a 7.35$\times$ speed-up even in this large-scale setting. These results indicate that, although large problems are increasingly constrained by memory traffic and host-device data movement, the JAX-FEM-ANISO implementation remains effective for compute- and memory-intensive nonlinear simulations.

\begin{table}[!htbp]
\centering
\begin{threeparttable}
\begin{minipage}{\standardtablewidth}
\caption{Material parameters used in the unit-cube performance simulations.}
\label{tab:performance_material_parameters}
\setlength{\tabcolsep}{2pt}
\centering
\begin{tabular*}{\textwidth}{@{\extracolsep{\fill}}cccccccccc@{}}
\toprule
$E$ & $Q$ & $b$ & $\sigma_0$ & $r_{11}$ & $r_{22}$ & $r_{33}$ & $r_{12}$ & $r_{13}$ & $r_{23}$ \\
\midrule
219 GPa & 410 MPa & 3.8 & 138 MPa & 1.5449 & 1.0666 & 1 & 1.0307 & 1 & 1 \\
\bottomrule
\end{tabular*}
\end{minipage}
\end{threeparttable}
\end{table}

\begin{table}[!htbp]
\centering
\begin{threeparttable}
\begin{minipage}{0.95\textwidth}
\caption[Wall-clock time comparison between JAX-FEM-ANISO and Abaqus for the unit-cube benchmark at different problem sizes]{Wall-clock time comparison between JAX-FEM-ANISO and Abaqus for the unit-cube benchmark at different problem sizes.}
\label{tab:forward_gpu_time}
\centering
\begin{tabularx}{\textwidth}{@{}l l X c c@{}}
\toprule
Problem size & Software & Configuration & Memory (RAM) & Time (h) \\
\midrule
\(\sim\)3M DOFs & Abaqus &  CPU (16 cores) & 166 GB & 4.56 \\
 &  & CPU (24 cores)$^\ast$ & 128 GB & 5.30 \\
 &  & CPU (32 cores) & 166 GB & 4.40 \\
 &  & CPU (64 cores) & 221 GB & 4.55 \\
 &  & CPU (128 cores) & 471 GB & 5.31 \\
 &  & GPU (H100) & 80 GB GPU & 5.40 \\
\cmidrule(lr){2-5}
 & JAX-FEM-ANISO & GPU (A100) & 80 GB GPU & 1.00 \\
 &  & GPU (H100) & 80 GB GPU & 0.56 \\
\midrule
\(\sim\)11.2M DOFs & Abaqus & CPU (24 cores)$^\ast$ & 128 GB & 22.50 \\
 &  & GPU (H100) & 80 GB GPU & 24.24 \\
\cmidrule(lr){2-5}
 & JAX-FEM-ANISO & GPU (H100) & 80 GB GPU & 3.06 \\
\bottomrule
\end{tabularx}

\vspace{0.3em}

\raggedright
\footnotesize $^\ast$Operating System: Windows 11 workstation; Processor: Intel(R) Xeon(R) w7-2495X CPU @ 2.50GHz

All other Abaqus runs use Red Hat Enterprise Linux release 8.10; Processor: Intel(R) Xeon(R) Gold 6338 CPU @ 2.00GHz
\end{minipage}
\end{threeparttable}
\end{table}

\FloatBarrier
\subsection{JAX-AD-Based Sensitivity Analysis and Gradient Verification}
\begin{figure}[!htbp]
\centering
\includegraphics[width=0.5\textwidth]{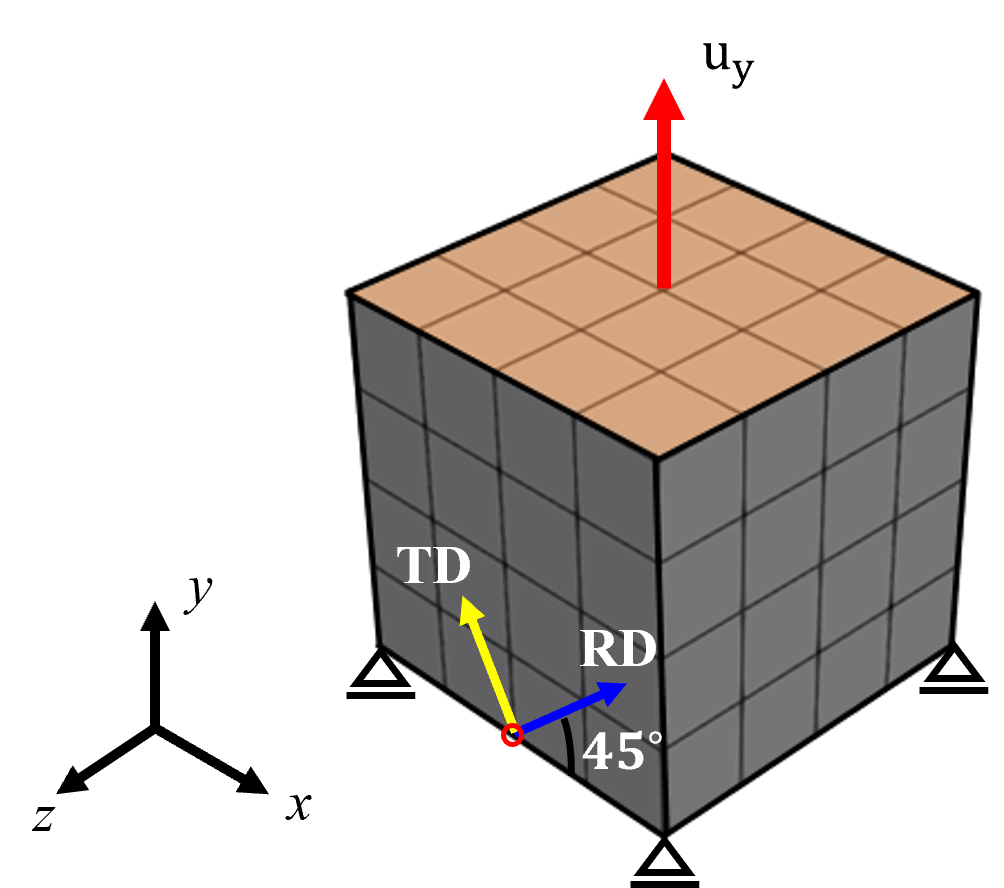}
\caption{Loading and boundary conditions for the unit-cube domain considered for gradient verification.}
\label{fig:cube_domain}
\end{figure}

To verify the AD-based gradients of the objective function, we use the unit-cube geometry. The domain is discretized using $4\times4\times4$ hexahedral elements. A material orientation of 45° is assigned to the domain, as shown in Fig.~\ref{fig:cube_domain}. To simulate the uniaxial response, the cube faces in the $yz$, $xz$, and $xy$ planes are constrained in the $x$ ($u_x = 0$), $y$ ($u_y = 0$), and $z$ ($u_z = 0$) directions, respectively. A total displacement of $u_y = 0.1$~mm is applied to the top face in five equal load steps. The reference displacement data required for inverse calculations are obtained from a forward simulation using the material parameters listed in Table~\ref{tab:material_params}. For the Hill--48 yield model, we set $r_{11} = 1$ so that $\sigma_{11}$ matches the reference yield strength $\sigma_0$. Poisson's ratio ($\nu$) is set to 0.3.

\begin{figure}[!htbp]
\centering
\includegraphics[width=0.6\textwidth]{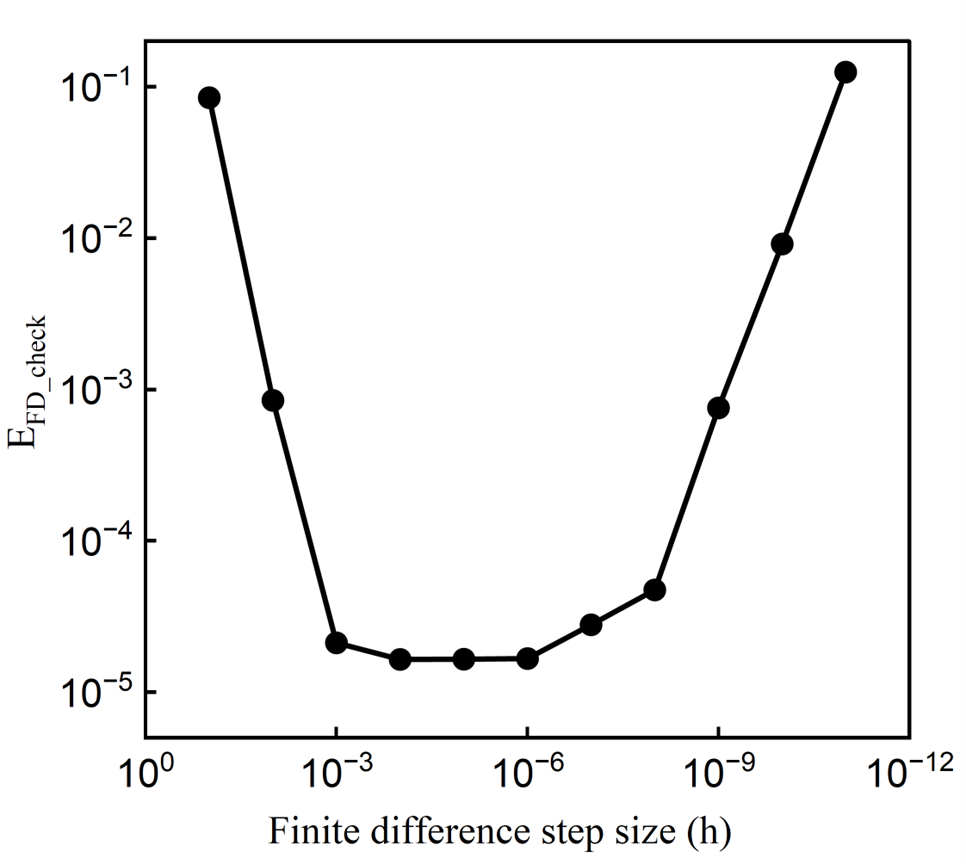}
\caption{Finite-difference error check $E_{FD,check}$ as a function of step size $h$.}
\label{fig:fd_error}
\end{figure}

% QUERY TO AUTHORS: The parameter vector lists nine entries, but the numerical vector below appears to contain eight values. Please confirm whether one value is missing.
For gradient calculations, the objective function is evaluated at $\boldsymbol{\theta} = [\text{E (GPa)}, \sigma_0 \text{ (MPa)}, \text{Q (MPa)}, \text{b}, r_{22}, r_{33}, r_{12}, r_{13}, r_{23}]$ with the following values: $[190, 140, 340, 3.6, 1.06, 0.98, 0.98, 0.98]$. The gradients of the objective function with respect to the elastic, anisotropic plasticity, and hardening material parameters are computed using the adjoint-based JAX-AD approach and listed in Table~\ref{tab:gradient_comparison}. The adjoint-based JAX-AD method provides accurate sensitivities for gradient-based optimization. To verify our implementation, we also calculate gradients using finite-difference approximation (FD), which is a standard approach for checking the correctness of parameter sensitivities. As a sanity check for the JAX-AD gradients, we define the following error metric,
\begin{equation}
E_{FD,check} = \left\| \left[ \frac{\mathrm{d}J}{\mathrm{d}\boldsymbol{\theta}} \right]_{JAX\text{-}AD} \cdot \bm{D}  
- \left[ \frac{\mathrm{d}J}{\mathrm{d}\boldsymbol{\theta}} \right]_{FD} \cdot \bm{D} \right\|
\end{equation}
where $\bm{D}$ represents the direction vector in parameter space. In this paper, each component of the direction vector is set to $D_i = 0.1$. The accuracy of finite-difference gradients is highly sensitive to the chosen step size ($h$). Therefore, FD gradients and $E_{FD,check}$ are calculated at step sizes $h = 10^{-1}, 10^{-2}, \ldots, 10^{-11}$. The finite-difference error-check results are plotted in Fig.~\ref{fig:fd_error}. In finite-difference approximation methods, the gradient error commonly decreases with smaller step sizes up to a critical point. Beyond this threshold, round-off errors dominate and the accuracy degrades. The results show this characteristic behavior: the error ($E_{FD,check}$) initially decreases as $h$ is reduced to $10^{-4}$, remains approximately constant from $10^{-4}$ to $10^{-5}$, and then increases with further step-size reduction to $10^{-11}$ because of round-off accumulation. This behavior supports the accuracy of the JAX-AD gradients. Table~\ref{tab:gradient_comparison} reports the absolute percentage difference between gradients computed using JAX-AD and finite differences at $h = 10^{-6}$ for each material parameter. The gradients computed by the two methods differ by less than 0.2\%, demonstrating accurate end-to-end differentiability for nonlinear anisotropic elastoplastic analysis. This close agreement also verifies the custom Jacobian-vector product implementations for both local and global Newton--Raphson iterations within the JAX automatic differentiation framework.

\begin{table}[!htbp]
\centering
\begin{threeparttable}
\begin{minipage}{\standardtablewidth}
\caption{Elastic modulus, Hill--48 yield, and hardening parameters.}
\label{tab:material_params}
\setlength{\tabcolsep}{2pt}
\centering
\begin{tabular*}{\textwidth}{@{\extracolsep{\fill}}ccccccccc@{}}
\toprule
E (GPa) & $\sigma_0$ (MPa) & Q (MPa) & b & $r_{22}$ & $r_{33}$ & $r_{12}$ & $r_{13}$ & $r_{23}$ \\
\midrule
200 & 150 & 400 & 4 & 1.5 & 1.2 & 1.1 & 1 & 1 \\
\bottomrule
\end{tabular*}
\end{minipage}
\end{threeparttable}
\end{table}

\begin{table}[!htbp]
\centering
\begin{threeparttable}
\begin{minipage}{\standardtablewidth}
\caption{Comparison of material-parameter sensitivities computed using JAX-AD and finite differences at step size $h = 10^{-6}$.}
\label{tab:gradient_comparison}
\setlength{\tabcolsep}{1pt}
\centering
\begin{tabular*}{\textwidth}{@{\extracolsep{\fill}}lccccccccc@{}}
\toprule
 & E & $\sigma_0$ & Q & b & $r_{22}$ & $r_{33}$ & $r_{12}$ & $r_{13}$ & $r_{23}$ \\
\midrule
JAX-AD gradients & $-0.044$ & $-0.015$ & $0.123$ & $0.139$ & $-6.143$ & $-8.640$ & $-9.518$ & $-0.024$ & $0.0039$ \\
\midrule
\multicolumn{10}{c}{Absolute percentage difference between JAX-AD and FD gradients} \\
\midrule
Difference (\%) & 0.004 & 0.0225 & 0.0068 & 0.006 & 0.0025 & 0.0004 & 0.0003 & 0.0570 & 0.1729 \\
\bottomrule
\end{tabular*}
\end{minipage}
\end{threeparttable}
\end{table}

\FloatBarrier

\subsection{Inverse Identification Using Information-Rich Tensile Test}

\begin{figure}[!htbp]
\centering
\includegraphics[width=0.5\textwidth]{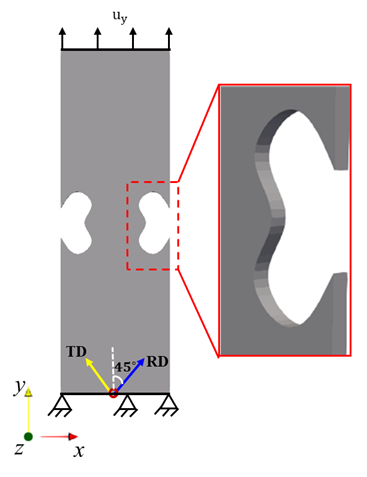}
\caption{Information-rich tensile specimen geometry adapted from Zhang et al.~\cite{Zhang2022}.}
\label{fig:specimen_geometry}
\end{figure}

This section examines the proposed JAX-AD-based framework for identifying material parameters from nonconventional specimen geometries. We consider the tensile specimen geometry proposed by Zhang \textit{et al.}~\cite{Zhang2022}, which was developed through shape optimization to increase information richness for inverse identification of anisotropic plasticity parameters. The geometry is shown in Fig.~\ref{fig:specimen_geometry}. This configuration serves as an inverse-analysis benchmark because it generates highly heterogeneous strain fields that activate a wide range of deformation modes, thereby reducing the need for multiple conventional tests. Previous studies have shown that this geometry, particularly when combined with a 45° material orientation with respect to the tensile direction, improves parameter identifiability relative to standard specimens.

\begin{figure}[!htbp]
\centering
\begin{minipage}[t]{0.30\textwidth}
\raggedright
\hspace*{1pt}%
\begin{tikzpicture}
    \node[anchor=south west, inner sep=0] (image) at (0,0) {\includegraphics[width=0.97\linewidth]{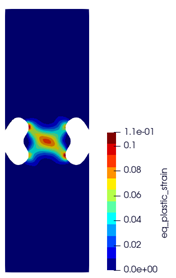}};
    \begin{scope}[x={(image.south east)},y={(image.north west)}]
        \node[anchor=north west, fill=white, inner sep=1pt] at (0.02, 0.98) {(a)};
    \end{scope}
\end{tikzpicture}
\end{minipage}
\hfill
\begin{minipage}[t]{0.60\textwidth}
\centering
\begin{tikzpicture}
    \node[anchor=south west, inner sep=0] (image) at (0,0) {\includegraphics[width=\linewidth]{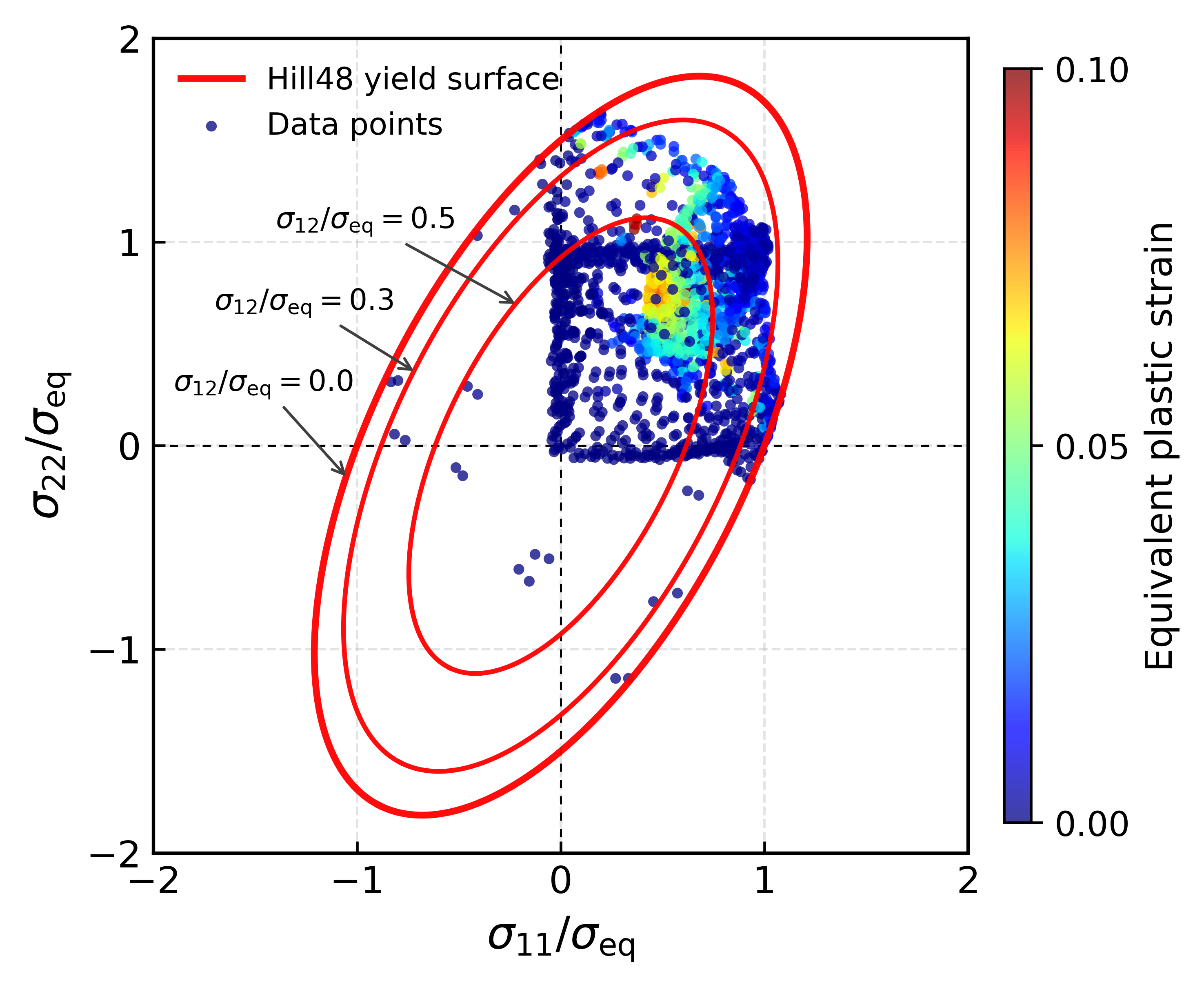}};
    \begin{scope}[x={(image.south east)},y={(image.north west)}]
        \node[anchor=north west, fill=white, inner sep=1pt] at (0.02, 0.98) {(b)};
    \end{scope}
\end{tikzpicture}
\end{minipage}
\caption{(a) Distribution of equivalent plastic strain and (b) stress-state distribution for integration points subjected to plastic deformation. Both axes are normalized by the equivalent Hill stress; $\sigma_{11}$ and $\sigma_{22}$ are aligned with the rolling direction (RD) and transverse direction (TD), respectively.}
\label{fig:plastic_strain}
\end{figure}

Forward-pass simulations are performed on the 3D specimen. The domain is discretized using 5,250 hexahedral elements, resulting in 25,083 degrees of freedom. A material orientation of 45° with respect to the loading direction is considered. The bottom surface is fixed, and a total displacement of 0.5~mm is applied to the top face in the positive $y$-direction over 10 unequal load steps with $t = \{0.005, 0.006, 0.02, 0.045, 0.09, 0.15, 0.3, 0.5, 0.75, 1\}$ to ensure convergence of the nonlinear solver. The ground-truth displacement data needed for the inverse problem are obtained from the forward simulations. We use $r_{11} = 1$, so the yield strength ($\sigma^y_{11}$) equals the reference yield strength ($\sigma_0$). Because relatively small out-of-plane shear stresses are expected, we set $r_{13} = 1$ and $r_{23} = 1$. The elastic modulus ($E$) and Poisson's ratio ($\nu$) are taken as 200~GPa and 0.3, respectively. All other material parameters are listed in Table~\ref{tab:inverse_results}. The equivalent plastic strain distribution at the final time step is plotted in Fig.~\ref{fig:plastic_strain}(a). To analyze stress-state dispersion, Fig.~\ref{fig:plastic_strain}(b) shows the stress distribution for integration points subjected to plastic deformation. The stress components are normalized by the equivalent stress. The generated stress states are widely dispersed, which increases the likelihood of activating different anisotropic plasticity parameters and improves inverse identification.

In the inverse identification, we focus on extracting anisotropic yield-surface and isotropic hardening parameters. The gradients of the objective function with respect to the material parameters are calculated using JAX-AD. The objective function is minimized using the bound-constrained L-BFGS-B algorithm with external gradients available in \texttt{scipy.optimize.minimize}. The optimization is terminated based on predefined convergence criteria involving the maximum number of iterations, maximum number of function evaluations, tolerance for changes in function value, and tolerance on the gradient norm. The convergence parameters used in this paper are listed in Table~\ref{tab:convergence_criteria}, and the same criteria are applied in all examples. Because parameter normalization significantly influences optimizer convergence, we normalize all parameters as follows,
\begin{equation}
\rho_i = \frac{2(\theta_i - \theta_{i,\min})}{\theta_{i,ref}} - 1
\quad \text{where } \rho_i \in [-1, 1]
\end{equation}
\begin{equation}
\theta_i = \frac{(\rho_i + 1) \theta_{i,ref}}{2} + \theta_{i,\min}^i
\end{equation}
where $\rho_i$ denotes the normalized parameter, $\theta_{i,\min}^i$ represents the minimum value, and $\theta_{i,ref}^i$ represents the difference between the maximum and minimum values of the parameter. We perform optimization on the normalized parameters with lower and upper bounds defined as $[-1, 1]$ for each parameter.

With this setup, inverse identification is performed for the tensile specimen using noiseless ground-truth displacement data obtained from the forward simulation described above. Because gradient-based optimization depends on the initial guess, we consider three arbitrarily selected initial guesses:

(C1) $\boldsymbol{\theta}_{min} = [140, 300, 3, 0.9, 0.9, 0.9]$ and $\boldsymbol{\theta}_{ref} = [40, 300, 2, 0.8, 0.8, 0.8]$ with $\rho_i = 0.2$;

(C2) $\boldsymbol{\theta}_{min} = [130, 200, 1, 0.9, 0.9, 0.9]$ and $\boldsymbol{\theta}_{ref} = [70, 200, 4, 0.8, 0.8, 0.8, 0.8]$ with $\rho_i = 0.5$;

\begin{figure}[!htbp]
\centering
\includegraphics[width=0.5\textwidth]{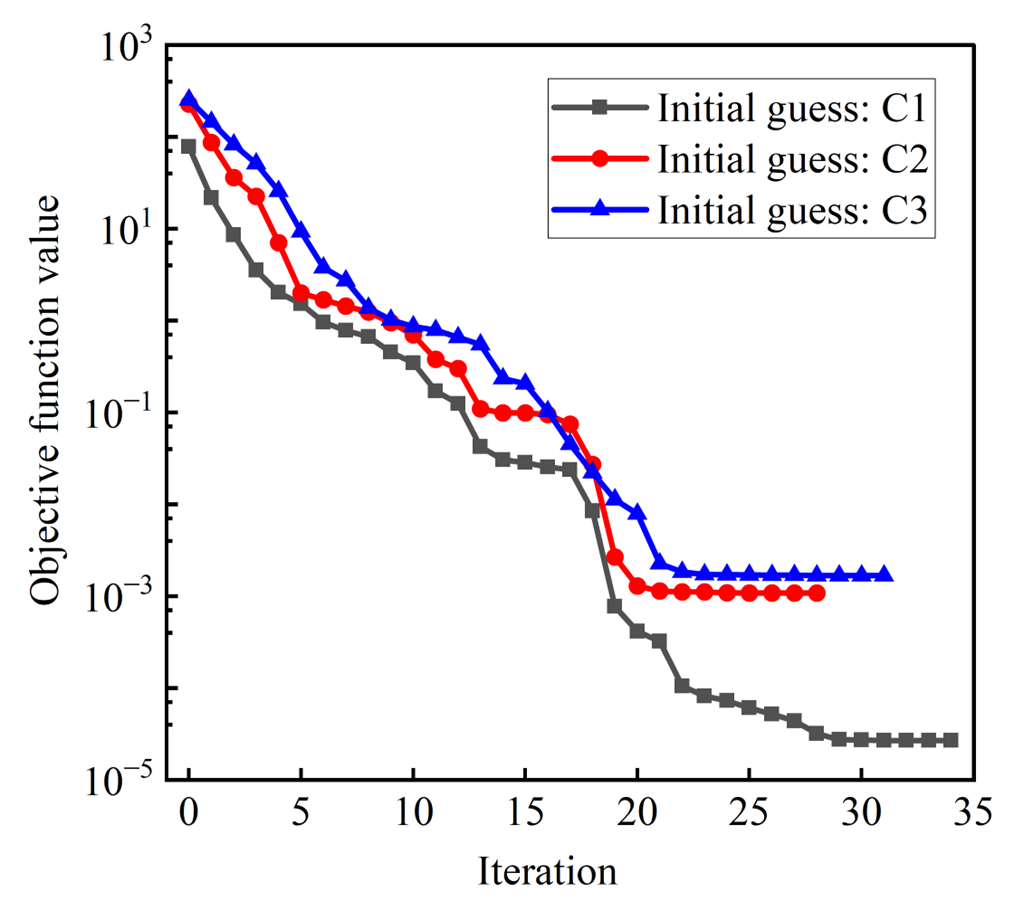}
\caption{Objective-function histories until convergence of the optimization.}
\label{fig:convergence_history}
\end{figure}

\begin{figure}[!htbp]
\centering
\includegraphics[width=0.5\textwidth]{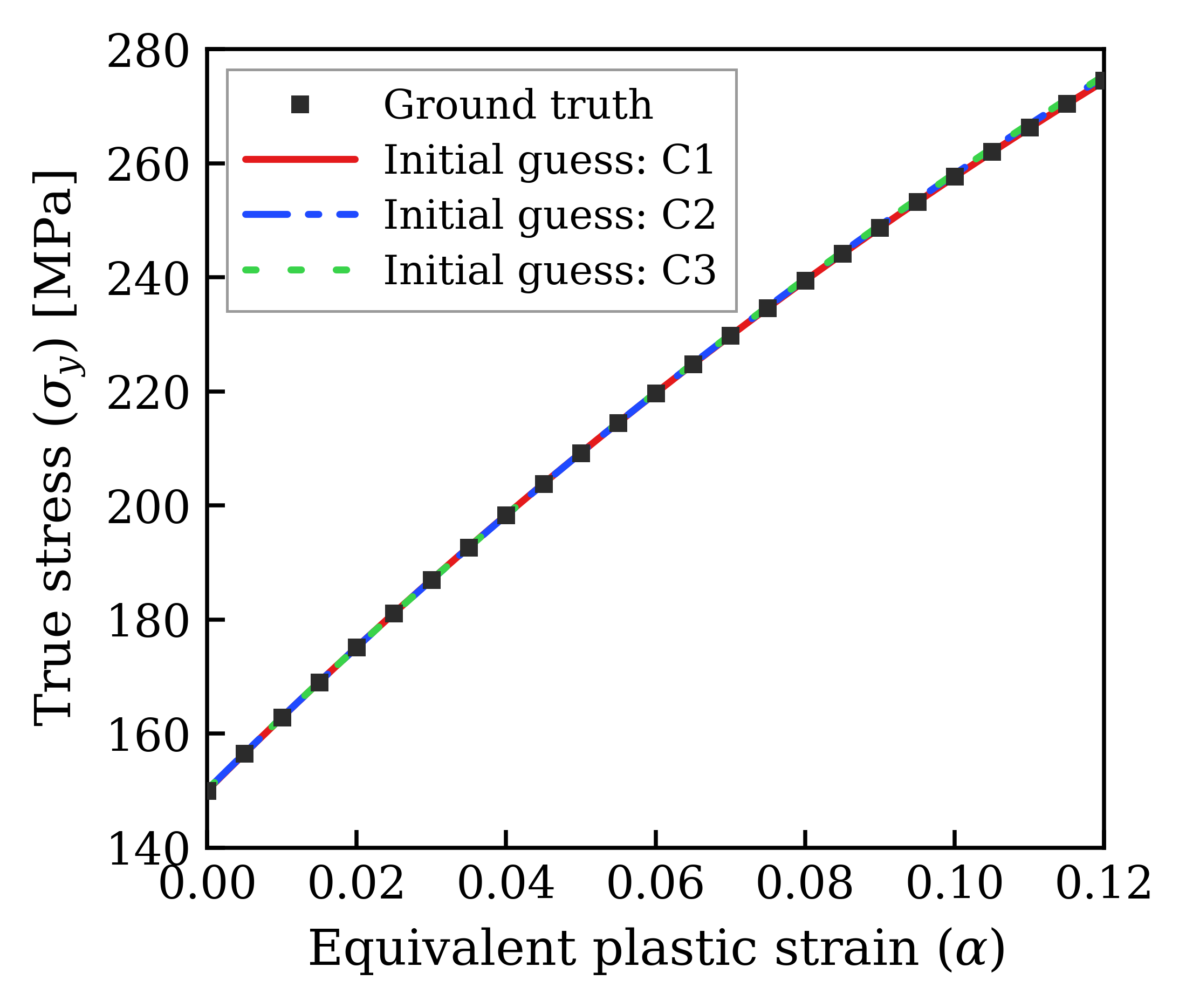}
\caption{Comparison of hardening curves for different material-parameter initializations.}
\label{fig:hardening_comparison}
\end{figure}

(C3) $\boldsymbol{\theta}_{min}$ and $\boldsymbol{\theta}_{ref}$ defined in (C2) with $\rho_i = 0.8$

\noindent where $\boldsymbol{\theta} = [\sigma_0 \text{ (MPa)}, Q \text{ (MPa)}, b, r_{22}, r_{33}, r_{12}]$ denotes the material parameters considered for optimization. Table~\ref{tab:inverse_results} lists the results for all three initial guesses. Table~\ref{tab:convergence_stats} reports the number of iterations, total number of function evaluations, and computation time to convergence, and Fig.~\ref{fig:convergence_history} shows the corresponding convergence histories. The anisotropic plasticity parameters are recovered almost exactly for all three initial guesses, with the largest error below 0.1 percent. The hardening parameters, however, converge to different values for different initial guesses. This behavior is expected because different combinations of $Q$ and $b$ can produce the same hardening curve. Fig.~\ref{fig:hardening_comparison} compares the hardening responses obtained from the recovered parameters with the ground-truth response. The hardening curves for all three cases match the ground-truth result closely.

The initialization of material parameters can significantly affect the total computation time of inverse optimization: poor initialization can require more iterations and function evaluations and may lead to nonconvergence of the optimizer or the forward nonlinear solvers. To compare the computation time of FD and JAX-AD-based approaches, gradients are computed for a single inverse iteration using both approaches. The computation times are listed in Table~\ref{tab:computation_time}. For this example, the JAX-AD-based approach is approximately 9.25 times more efficient than the FD approach.

\begin{table}[!htbp]
\centering
\begin{threeparttable}
\begin{minipage}{\standardtablewidth}
\caption{Inverse parameter identification for the tensile specimen with different initial guesses.}
\label{tab:inverse_results}
\setlength{\tabcolsep}{2pt}
\centering
\begin{tabular*}{\textwidth}{@{\extracolsep{\fill}}lcccccc@{}}
\toprule
 & $\sigma_0$ (MPa) & Q & b & $r_{22}$ & $r_{33}$ & $r_{12}$ \\
\midrule
Ground truth & 150 & 400 & 4 & 1.5 & 1.2 & 1.1 \\
\midrule
\multicolumn{7}{c}{Initial guess: C1} \\
\midrule
Prediction & 149.99 & 397.29 & 4.03 & 1.50 & 1.20 & 1.10 \\
Error (\%) & -0.01 & ** & ** & 0.00 & 0.00 & 0.00 \\
\midrule
\multicolumn{7}{c}{Initial guess: C2} \\
\midrule
Prediction & 150.09 & 418.25 & 3.80 & 1.50 & 1.20 & 1.10 \\
Error (\%) & 0.06 & ** & ** & 0.00 & 0.00 & 0.00 \\
\midrule
\multicolumn{7}{c}{Initial guess: C3} \\
\midrule
Prediction & 150.12 & 423.04 & 3.75 & 1.50 & 1.20 & 1.10 \\
Error (\%) & 0.08 & ** & ** & 0.00 & 0.00 & 0.00 \\
\bottomrule
\end{tabular*}
\vspace{0.3em}

\raggedright
\footnotesize ** The values of parameters $Q$ and $b$ are not unique for the hardening curve in the considered plastic strain range. The hardening curves obtained from the predicted parameters are compared in Fig.~\ref{fig:hardening_comparison}.
\end{minipage}
\end{threeparttable} 
\end{table}

\begin{table}[!htbp]
\centering
\begin{threeparttable}
\begin{minipage}{\standardtablewidth}
\caption{Computation time comparison between JAX-AD and FD approaches for a single iteration of the inverse problem.}
\label{tab:computation_time}
\centering
\begin{tabular*}{\textwidth}{@{\extracolsep{\fill}}lcc@{}}
\toprule
 & Computation time & Speed-up \\
\midrule
JAX-AD & 234 seconds & 1 \\
FD & 2166 seconds & 9.25 \\
\bottomrule
\end{tabular*}
\end{minipage}
\end{threeparttable}
\end{table}

\begin{table}[!htbp]
\centering
\begin{threeparttable}
\begin{minipage}{\standardtablewidth}
\caption{Convergence criteria for the L-BFGS-B optimizer.}
\label{tab:convergence_criteria}
\centering
\begin{tabular*}{\textwidth}{@{\extracolsep{\fill}}cccc@{}}
\toprule
Max. iterations & Max. function & Tolerance function & Tolerance gradient \\
(maxiter) & evaluations & value change (ftol) & norm (gtol) \\
 & (maxfun) &  &  \\
\midrule
45 & 90 & $10^{-10}$ & $10^{-10}$ \\
\bottomrule
\end{tabular*}
\end{minipage}
\end{threeparttable}
\end{table}

\begin{table}[!htbp]
\centering
\begin{threeparttable}
\begin{minipage}{\standardtablewidth}
\caption{Comparison of iterations, function evaluations, and computation time for different cases.}
\label{tab:convergence_stats}
\setlength{\tabcolsep}{4pt}
\centering
\begin{tabular*}{\textwidth}{@{\extracolsep{\fill}}lccc@{}}
\toprule
 & Iterations & Function evaluations & Total computation time \\
\midrule
Initial guess: C1 & 35 & 62 & 204 min \\
Initial guess: C2 & 29 & 50 & 167 min \\
Initial guess: C3 & 32 & 50 & 182 min \\
\bottomrule
\end{tabular*}
\end{minipage}
\end{threeparttable}
\end{table}

\FloatBarrier 

\subsection{Inverse Identification Using Cruciform Specimen}

Cruciform specimens subjected to biaxial loading are effective for extracting in-plane anisotropic properties because they generate diverse strain states, including plane-strain tension and equibiaxial tension, that are difficult to achieve in conventional uniaxial tests~\cite{Zhang2023}. However, inverse identification from full-field digital image correlation (DIC) data is sensitive to measurement noise, which can propagate through the optimization process and degrade parameter estimates, particularly for parameters with lower sensitivity. To assess the robustness of the inverse identification framework, we examine the effect of measurement noise by adding varying levels of synthetic noise to displacement fields obtained from forward simulations. This noise-sensitivity analysis helps quantify the practical limitations of DIC-based calibration.

\begin{figure}[!htbp]
\centering
\includegraphics[width=0.5\textwidth]{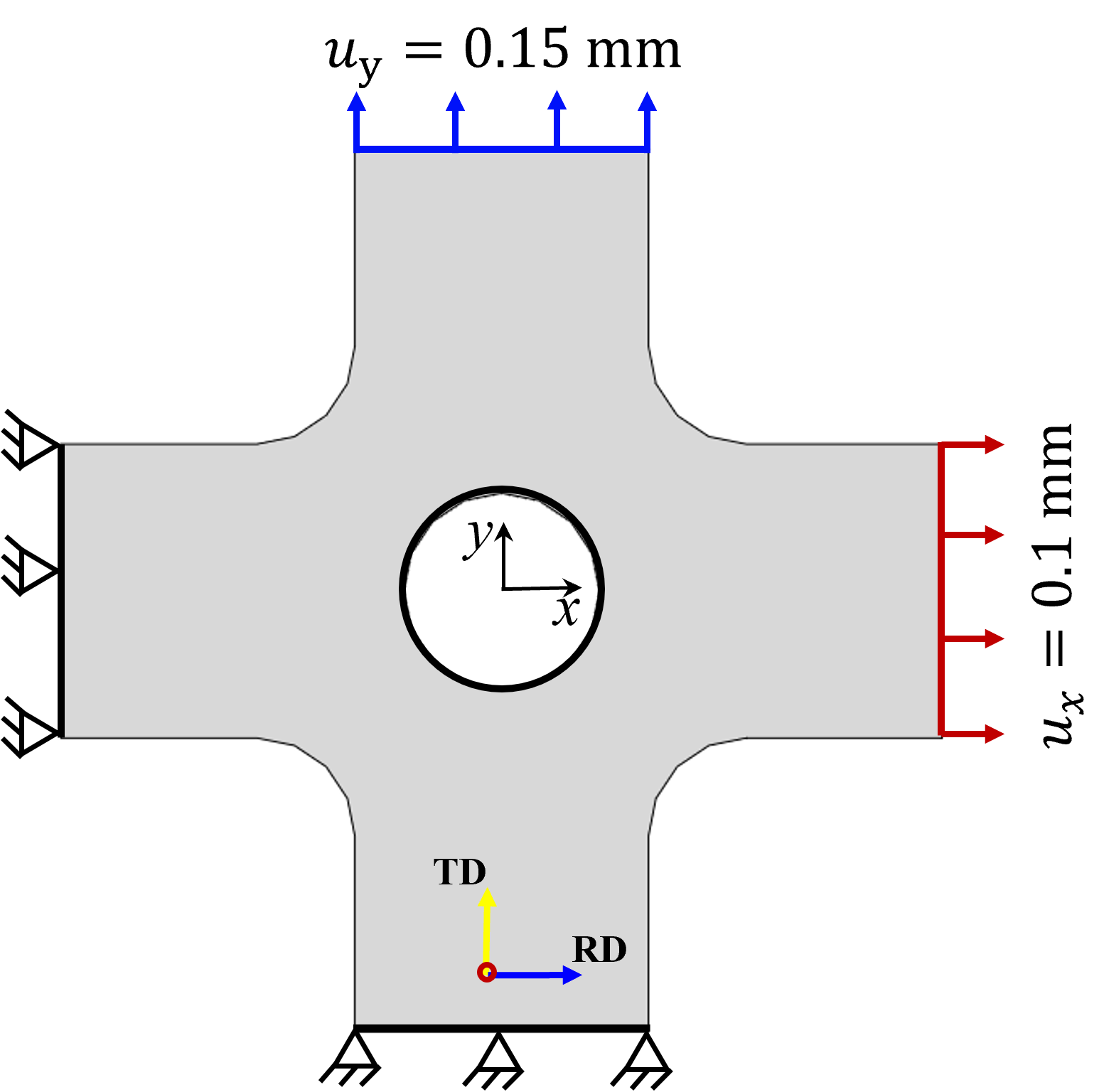}
\caption{Cruciform specimen geometry with the considered loading and boundary conditions.}
\label{fig:cruciform_geometry}
\end{figure}

\begin{figure}[!htbp]
\centering
\includegraphics[width=0.5\textwidth]{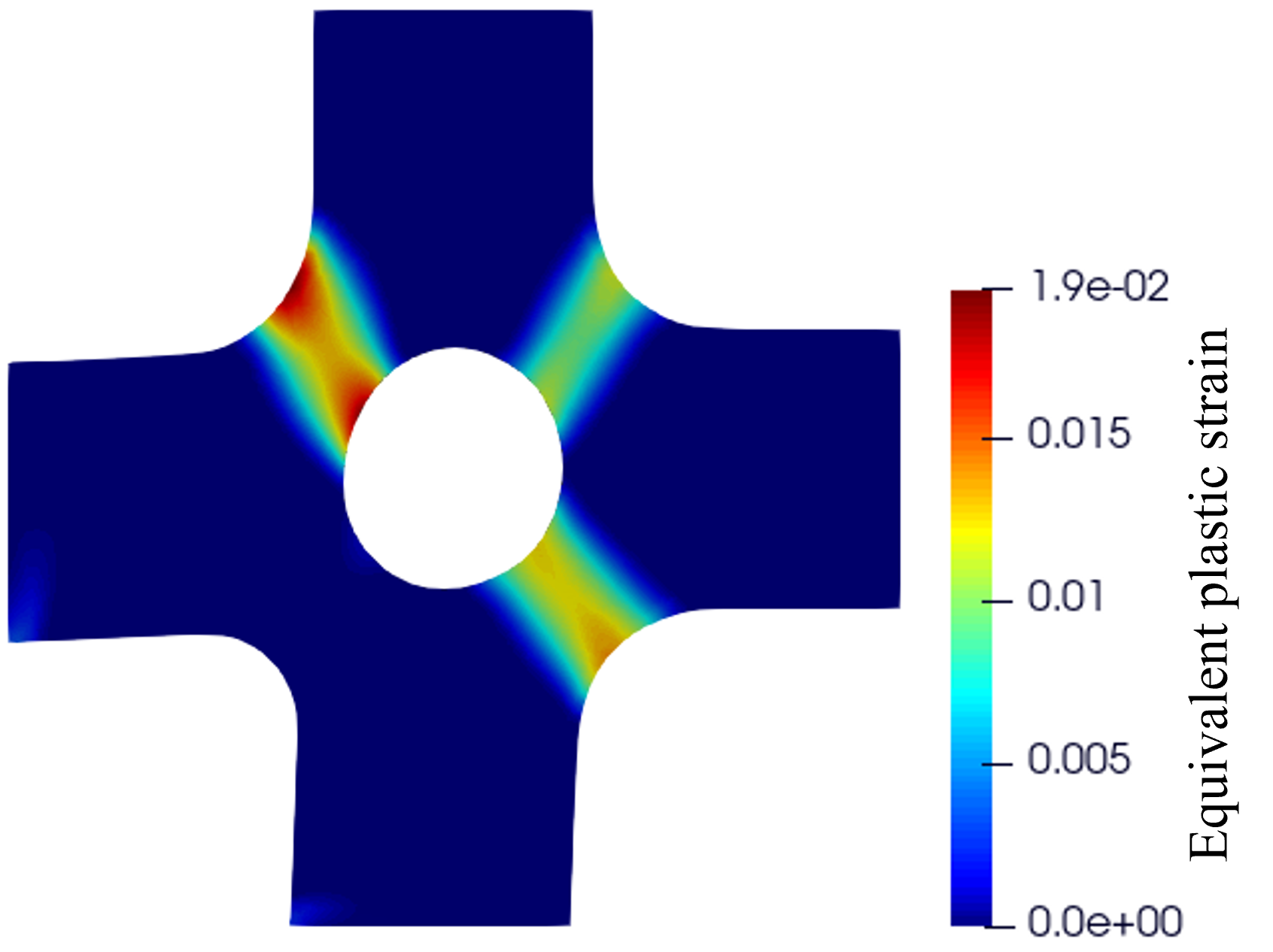}
\caption{Plastic-strain distribution at the final load step.}
\label{fig:cruciform_plastic_strain}
\end{figure}

We consider the cruciform geometry described by Zhang et al.~\cite{Zhang2023}. The specimen has four arms extending symmetrically from a central gauge region, with each arm spanning from -45~mm to +45~mm in both the horizontal ($x$) and vertical ($y$) directions, defining a total bounding box of $[-45,45] \times [-45,45]$ in the plane. The central gauge section contains a circular perforation with a radius of 10~mm at the origin of the specimen coordinate system. The arm junctions also have 10~mm fillet radii, and the specimen thickness is 1.0~mm in the out-of-plane direction ($z$-axis). The domain is discretized with 7,656 hexahedral elements, resulting in 47,346 degrees of freedom. The material coordinate system is defined with the rolling direction aligned with the $x$-axis, as indicated in Fig.~\ref{fig:cruciform_geometry}. The left and bottom edges are fully constrained, and non-proportional displacements of 0.1~mm (right edge) and 0.15~mm (top edge) are prescribed as shown in Fig.~\ref{fig:cruciform_geometry}. The displacements are applied in 9 unequal steps as $u_x = 0.1t$ and $u_y = 0.15t$ with $t = \{0.05, 0.1, 0.17, 0.28, 0.45, 0.62, 0.8, 0.96, 1\}$. As in the previous example, we set the elastic modulus $E = 200$~GPa, Poisson's ratio $\nu = 0.3$, and $r_{11} = r_{13} = r_{23} = 1$. All other anisotropic plasticity and hardening parameters required to generate synthetic noiseless displacement fields via forward simulations are provided in Table~\ref{tab:cruciform_results}. Fig.~\ref{fig:cruciform_plastic_strain} shows the equivalent plastic strain distribution at the final time step.

\begin{figure}[!htbp]
\centering
\includegraphics[width=0.9\textwidth]{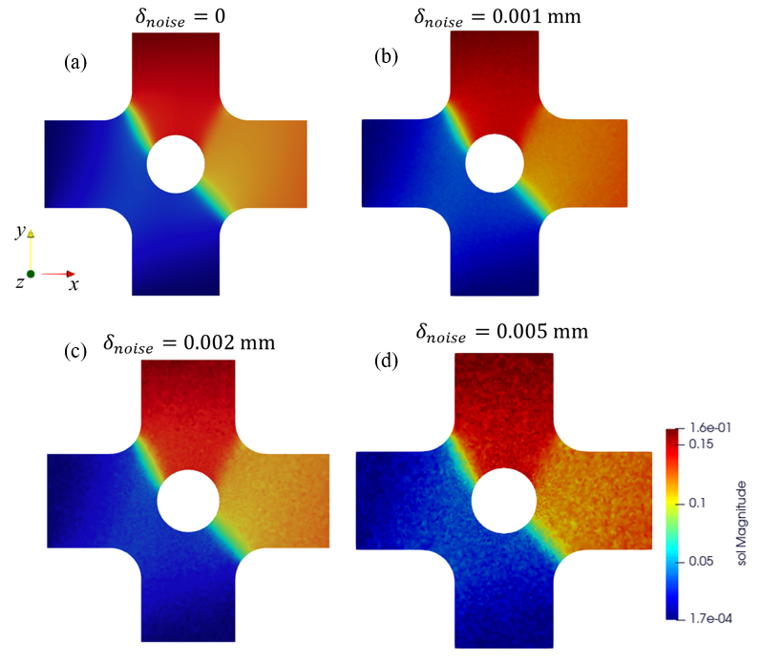}
\caption{Visualization of noisy synthetic data used for inverse identification at different noise levels.}
\label{fig:noisy_data_visualization}
\end{figure}

To examine the effect of noisy data on inverse identification, we generate synthetic noisy data by adding varying levels of random noise to the ground-truth displacement data obtained from the forward simulation. A random value drawn from a normal distribution with zero mean and unit variance is added to both the $u_x$ and $u_y$ components of the displacement field. The random value is scaled by a factor $\delta_{noise}$ such that $\bm{u}_{noisy} = \bm{u} + n\delta_{noise}$, where $n \sim \mathcal{N}(0,1)$. When $\delta_{noise} = 0$, the noiseless ground-truth displacement data are recovered. Using this method, we create noisy datasets with four noise levels: (a) $\delta_{noise} = 0$, (b) $\delta_{noise} = 0.001$~mm, (c) $\delta_{noise} = 0.002$~mm, and (d) $\delta_{noise} = 0.005$~mm, as shown in Fig.~\ref{fig:noisy_data_visualization}. The considered noise levels are analogous to those in experimental DIC measurements.

\begin{figure}[!htbp]
\centering
\includegraphics[width=0.5\textwidth]{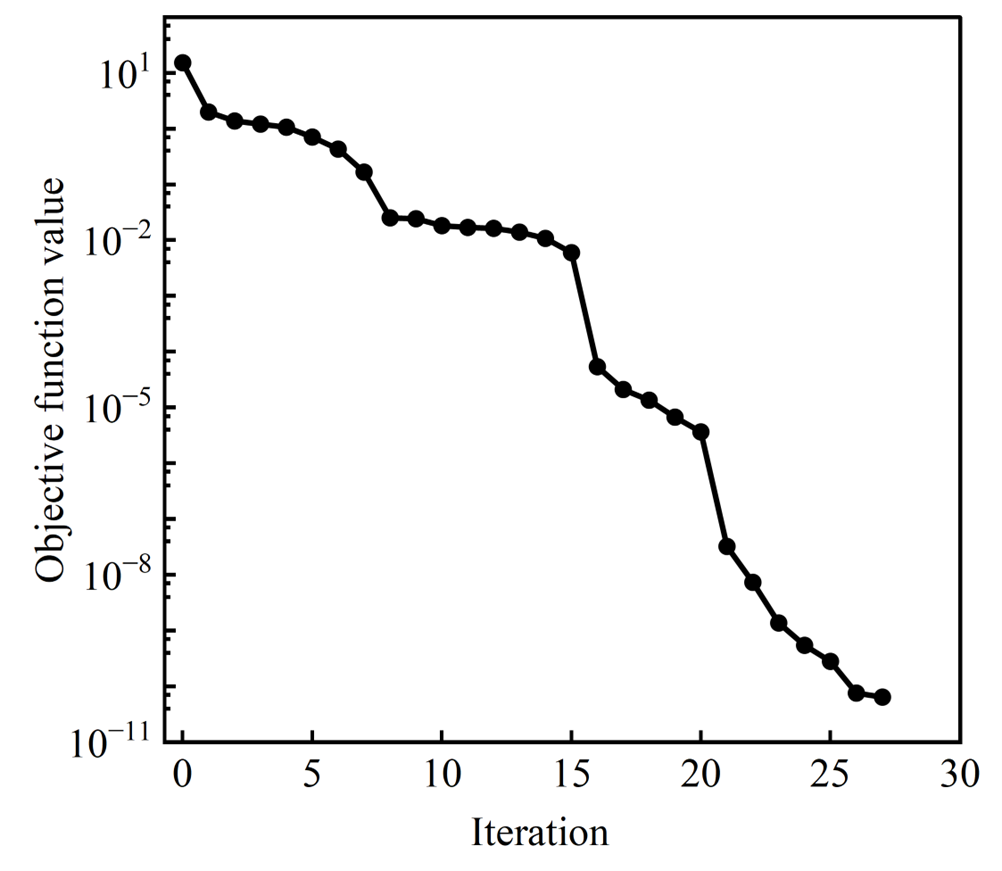}
\caption{Convergence history for inverse identification using noise-free synthetic data.}
\label{fig:cruciform_convergence_noiseless}
\end{figure}

\begin{figure}[!htbp]
\centering
\begin{subfigure}[b]{0.48\textwidth}
    \centering
    \includegraphics[width=\textwidth]{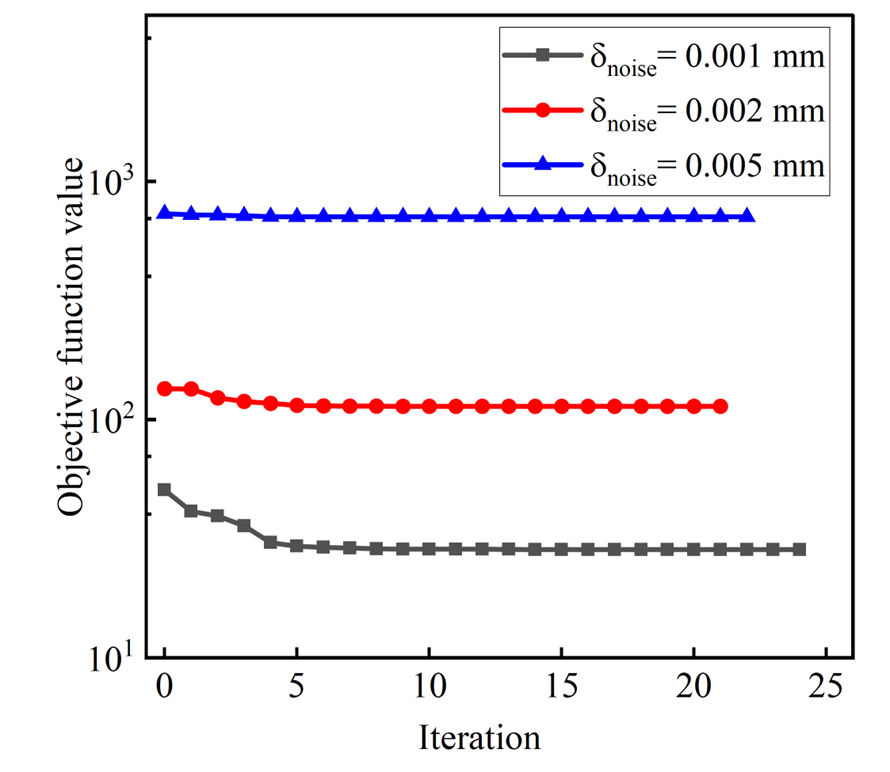}
    \caption{}
    \label{fig:cruciform_convergence_noisy_objective}
\end{subfigure}
\hfill
\begin{subfigure}[b]{0.48\textwidth}
    \centering
    \includegraphics[width=\textwidth]{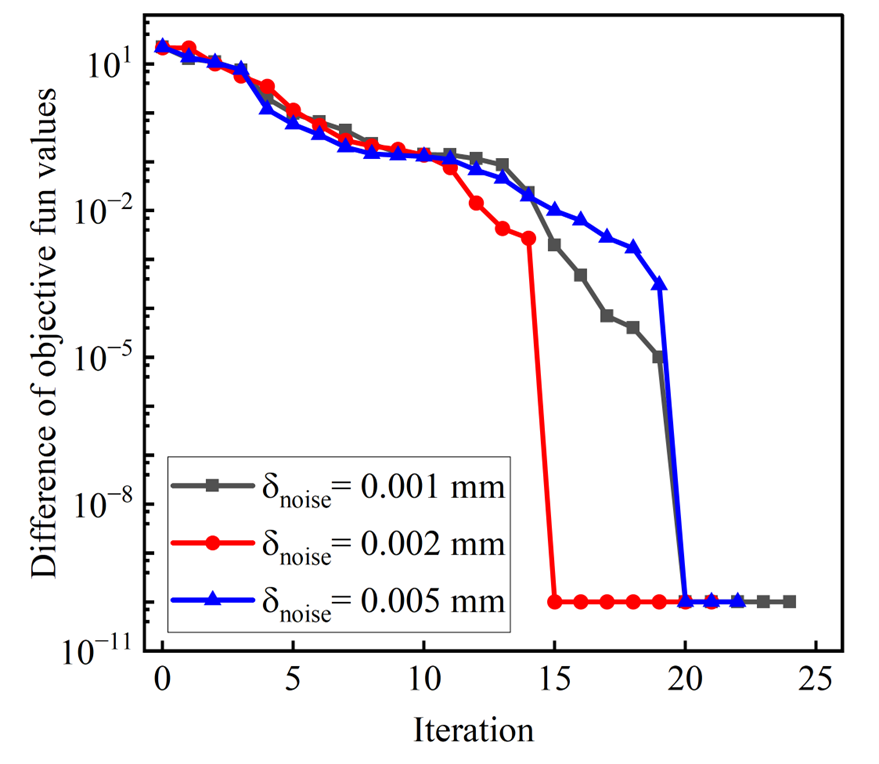}
    \caption{}
    \label{fig:cruciform_convergence_noisy_difference}
\end{subfigure}
\caption{Convergence histories for inverse identification using noisy synthetic data: (a) objective function values and (b) reduction in objective function values measured as the difference between the initial and current objective function values during optimization.}
\label{fig:cruciform_convergence_noisy}
\end{figure}

For the optimization, the material parameters $\boldsymbol{\theta} = [\sigma_0 \text{ (MPa)}, Q \text{ (MPa)}, b, r_{22}, r_{33}, r_{12}]$ are initialized for all cases as follows: $\boldsymbol{\theta}_{min} = [140, 300, 3, 0.85, 0.85, 0.85]$ and $\boldsymbol{\theta}_{ref} = [40, 200, 2, 0.5, 0.5, 0.5]$ with $\rho_i = 0.2$. The native JAX BiCGSTAB linear solver is used for both the forward and inverse solves. The simulations are performed on an RTX-8000 GPU with 48 GB of memory. Table~\ref{tab:cruciform_results} summarizes the material-parameter identification results obtained using JAX-AD gradients with the LBFGS-B optimizer for both noiseless and noisy datasets. Table~\ref{tab:cruciform_convergence} reports the iterations, function evaluations, and total computation time. The convergence histories for noiseless and noisy data are shown in Fig.~\ref{fig:cruciform_convergence_noiseless} and Fig.~\ref{fig:cruciform_convergence_noisy}, respectively. Fig.~\ref{fig:cruciform_convergence_noisy}(a) shows the objective-function decay for each noisy dataset, while Fig.~\ref{fig:cruciform_convergence_noisy}(b) shows the reduction in the objective function relative to its initial value. The results indicate that noise can significantly affect the overall optimization time. For the noiseless ground-truth data, the objective function decreases to a very small value at the end of the optimization, yielding an almost exact recovery of the material parameters. For the noisy datasets, the objective function converges to higher values, as expected. Nevertheless, the recovered material parameters remain accurate, with errors below 0.5\% for all parameters across all considered noise levels.

These findings show that the proposed JAX-AD-based differentiable framework provides accurate gradients for the considered noise levels in a computationally efficient and automatic manner. This combination of accuracy and efficiency supports the identification of spatially uniform mechanical parameters. The next section extends the same framework to spatially varying material properties.

\begin{table}[!htbp]
\centering
\begin{threeparttable}
\begin{minipage}{\standardtablewidth}
\caption{Inverse parameter identification for the cruciform specimen with noisy ground-truth data.}
\label{tab:cruciform_results}
\setlength{\tabcolsep}{2pt}
\centering
\begin{tabular*}{\textwidth}{@{\extracolsep{\fill}}lcccccc@{}}
\toprule
 & $\sigma_0$ (MPa) & Q (MPa) & b & $r_{22}$ & $r_{33}$ & $r_{12}$ \\
\midrule
Ground truth & 150.00 & 400.00 & 4.00 & 1.25 & 0.95 & 0.90 \\
\midrule
\multicolumn{7}{c}{$\delta_{noise} = 0$} \\
\midrule
Prediction & 150.00 & 400.00 & 4.00 & 1.25 & 0.95 & 0.90 \\
Error (\%) & 0.00 & 0.00 & 0.00 & 0.00 & 0.00 & 0.00 \\
\midrule
\multicolumn{7}{c}{$\delta_{noise} = 0.001$ mm} \\
\midrule
Prediction & 150.15 & 400.78 & 4.00 & 1.25 & 0.95 & 0.90 \\
Error (\%) & 0.10 & 0.20 & 0.00 & 0.00 & 0.00 & 0.00 \\
\midrule
\multicolumn{7}{c}{$\delta_{noise} = 0.002$ mm} \\
\midrule
Prediction & 150.06 & 400.76 & 4.00 & 1.25 & 0.95 & 0.90 \\
Error (\%) & 0.04 & 0.19 & 0.00 & 0.00 & 0.00 & 0.00 \\
\midrule
\multicolumn{7}{c}{$\delta_{noise} = 0.005$ mm} \\
\midrule
Prediction & 150.45 & 399.52 & 3.99 & 1.25 & 0.95 & 0.90 \\
Error (\%) & 0.30 & -0.12 & -0.25 & 0.00 & 0.00 & 0.00 \\
\bottomrule
\end{tabular*}
\end{minipage}
\end{threeparttable}
\end{table}

\begin{table}[!htbp]
\centering
\begin{threeparttable}
\begin{minipage}{\standardtablewidth}
\caption{Comparison of iterations, function evaluations, and computation time for different cases.}
\label{tab:cruciform_convergence}
\setlength{\tabcolsep}{4pt}
\centering
\begin{tabular*}{\textwidth}{@{\extracolsep{\fill}}lccc@{}}
\toprule
Noise level & Iterations & Function evaluations & Total computation time \\
\midrule
$\delta_{noise} = 0$ mm & 29 & 35 & 144 min \\
$\delta_{noise} = 0.001$ mm & 25 & 44 & 174 min \\
$\delta_{noise} = 0.002$ mm & 22 & 38 & 148 min \\
$\delta_{noise} = 0.005$ mm & 23 & 47 & 199 min \\
\bottomrule
\end{tabular*}
\end{minipage}
\end{threeparttable}
\end{table}

\FloatBarrier

\subsection{Inverse Identification of Spatially Varying Material Properties}

Engineering structures often exhibit spatial heterogeneity in material properties due to manufacturing processes, welding, additive manufacturing, or localized degradation. In such cases, assuming uniform properties may be inadequate for accurate structural predictions~\cite{tammas2017}. Identifying spatially varying anisotropic plasticity parameters is important for high-fidelity simulations of these components, but the resulting high-dimensional parameter space creates significant computational challenges. Traditional finite-difference-based gradient methods become prohibitively expensive as the number of spatially varying parameters increases. The JAX-AD-based differentiable framework addresses this limitation by providing gradients through automatic differentiation, enabling inverse identification of field-distributed material properties. This section demonstrates recovery of spatially heterogeneous anisotropic plasticity parameter distributions.

\begin{figure}[!htbp]
\centering
\begin{subfigure}{0.48\textwidth}
    \centering
    \includegraphics[width=\textwidth]{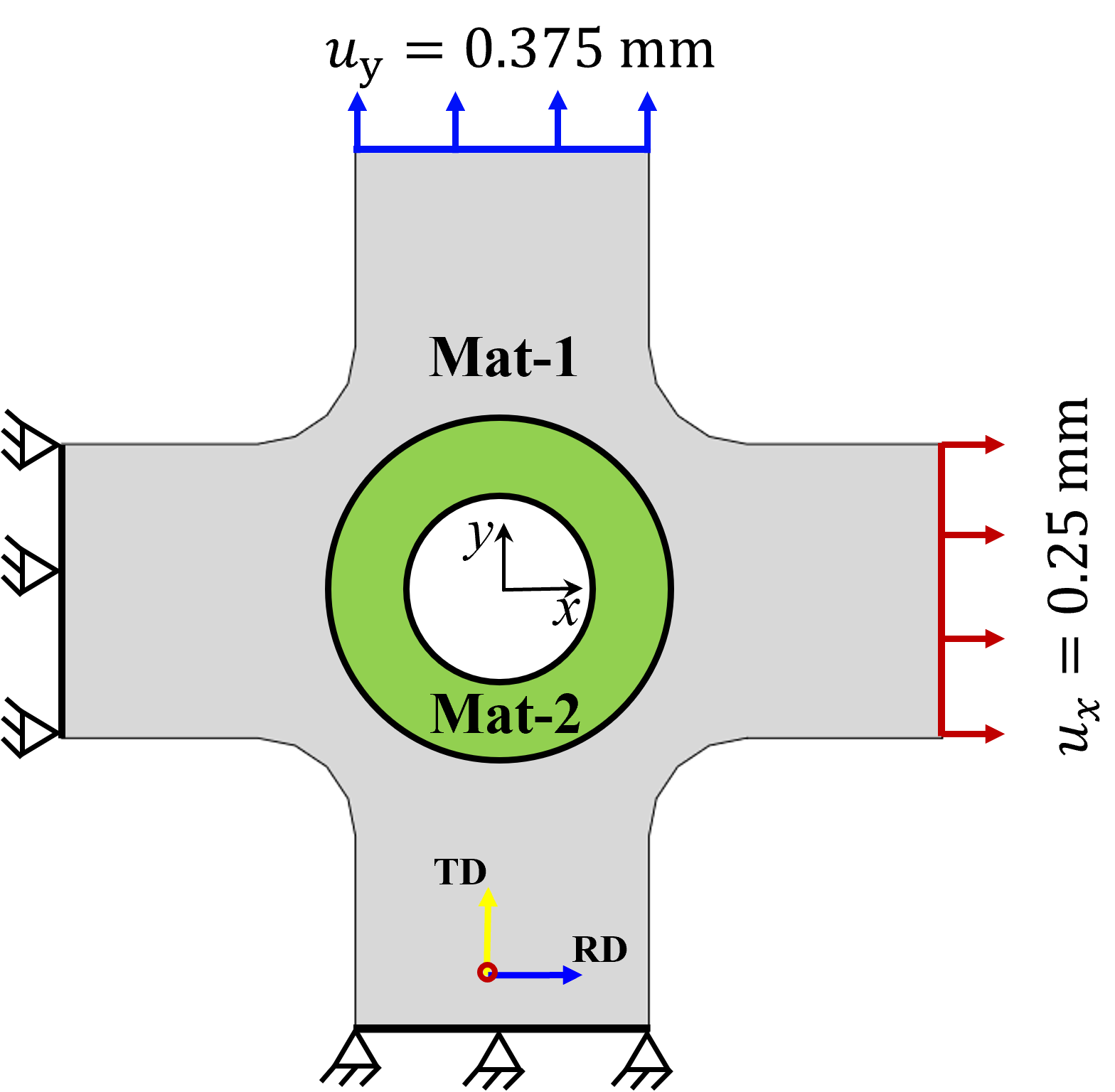}
    \caption{}
    \label{fig:spatially_varying_a}
\end{subfigure}
\hfill
\begin{subfigure}{0.48\textwidth}
    \centering
    \includegraphics[width=\textwidth]{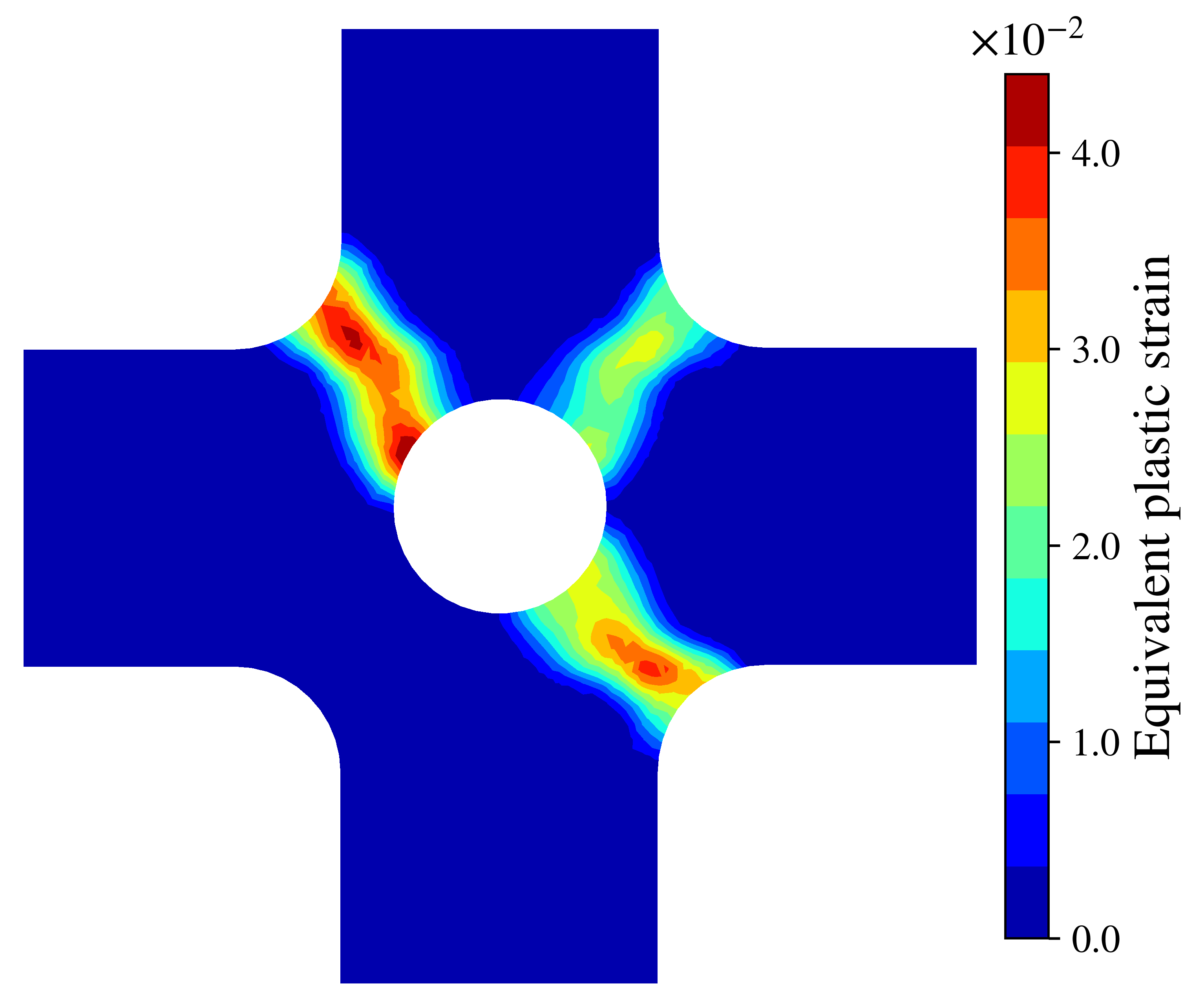}
    \caption{}
    \label{fig:spatially_varying_b}
\end{subfigure}
\caption{(a) Cruciform geometry with spatially varying material properties; (b) distribution of plastic strain at the final loading step.}
\label{fig:spatially_varying_geometry}
\end{figure}

We again use the cruciform specimen geometry from the previous section. The geometry is divided into two regions, and different material properties are assigned to each region, as illustrated in Fig.~\ref{fig:spatially_varying_geometry}. The radius of the outer circular interface is 17~mm. The material parameters adopted for Material-1 and Material-2 are summarized in Table~\ref{tab:spatially_varying_results}, while the remaining parameters ($E = 200$~GPa, $\nu = 0.3$ and $r_{11} = r_{13} = r_{23} = 1$) are identical for both materials. The boundary conditions and displacement loading are shown in Fig.~\ref{fig:spatially_varying_geometry}(a); the prescribed total displacement is applied in nine unequal increments. The corresponding plastic strain distribution from the forward simulation is shown in Fig.~\ref{fig:spatially_varying_geometry}(b).

\begin{figure}[!htbp]
\centering
\includegraphics[width=0.5\textwidth]{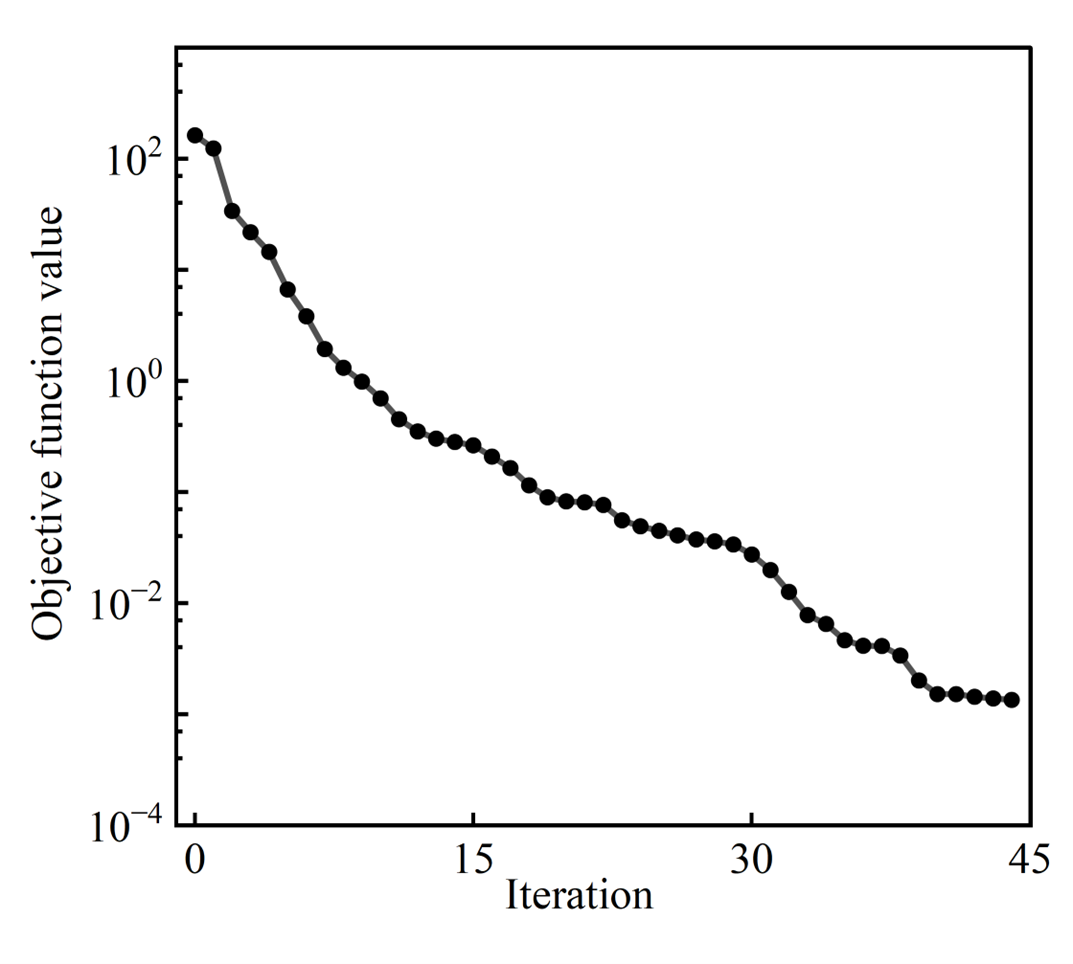}
\caption{Convergence history for identifying spatially varying material properties.}
\label{fig:spatially_varying_convergence}
\end{figure}

\begin{figure}[!htbp]
\centering
\includegraphics[width=0.5\textwidth]{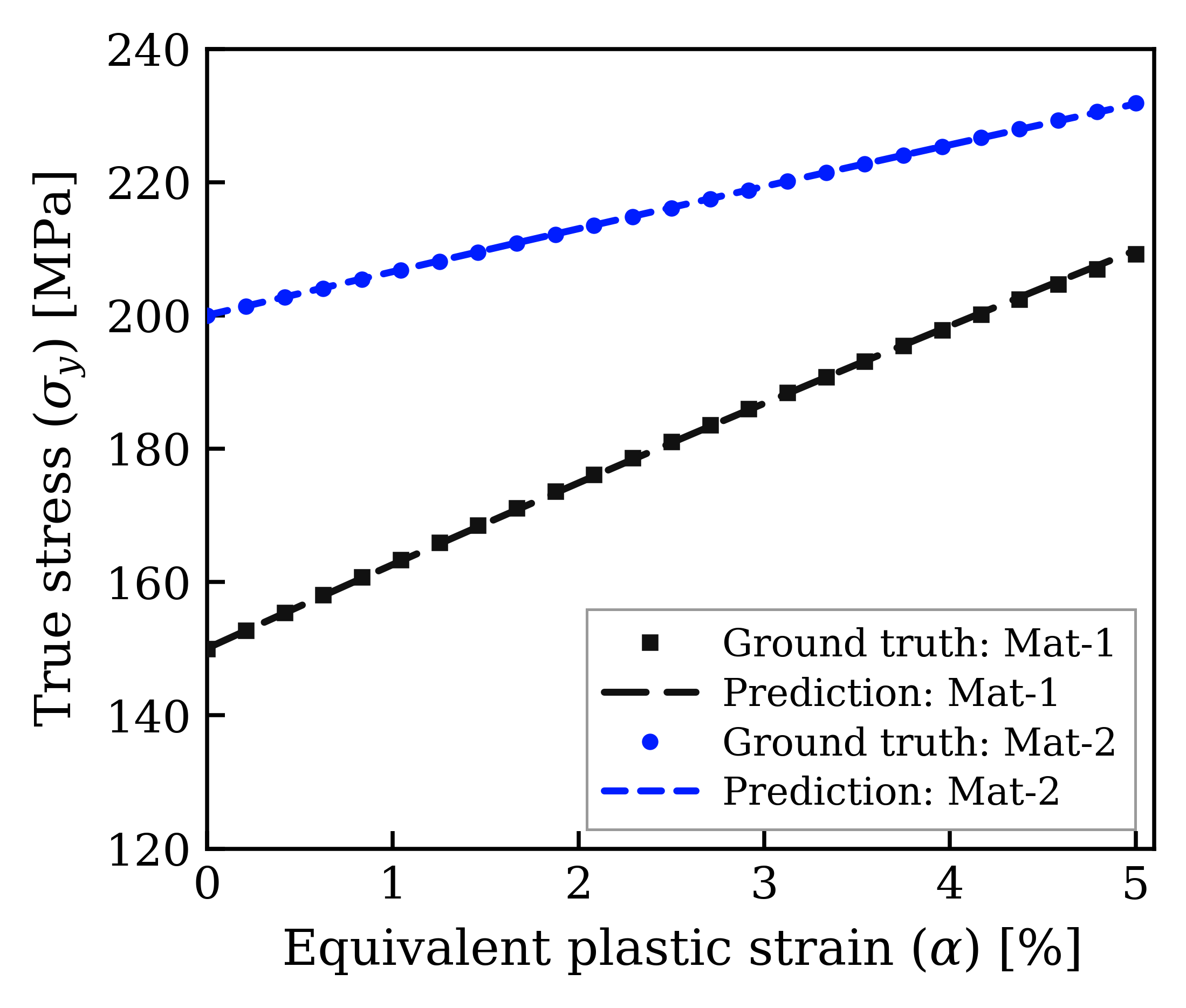}
\caption{Comparison of inverse-identified hardening curves with ground-truth results.}
\label{fig:spatially_varying_hardening}
\end{figure}

For the inverse identification problem, the unknown material parameters are taken as 
$\boldsymbol{\theta} = [\sigma_0^1 \text{ (MPa)}, Q^1 \text{ (MPa)}, b^1, r_{22}^1, r_{33}^1, r_{12}^1, \sigma_0^2 \text{ (MPa)}, Q^2 \text{ (MPa)}, b^2, r_{22}^2, r_{33}^2, r_{12}^2]$, 
with $\boldsymbol{\theta}_{min} = [130, 300, 2, 0.85, 0.85, 0.85, 140, 300, 3, 0.85, 0.85, 0.85]$ and 
$\boldsymbol{\theta}_{ref} = [100, 700, 5, 0.5, 0.5, 0.5, 100, 700, 5, 0.5, 0.5, 0.5]$ with $\rho_i = 0.2$. Here superscripts 1 and 2 refer to material-1 and material-2, respectively. Identical initial guesses, lower bounds, and upper bounds are assigned to the parameters of both materials. Introducing a second material significantly increases the total number of optimization parameters. The inverse identification analyses are performed on an RTX 8000 GPU (48 GB RAM) using a JAX-based BiCGSTAB linear solver. Table~\ref{tab:spatially_varying_results} reports the optimized parameters obtained from JAX-AD gradients together with their percentage errors. Fig.~\ref{fig:spatially_varying_convergence} shows the convergence history. The optimization converges in 146 minutes, requiring 45 iterations and 63 function evaluations. Despite starting from spatially uniform initial parameters, the framework recovers location-specific parameters for both materials, with errors below 0.2\%. Fig.~\ref{fig:spatially_varying_hardening} compares the hardening responses obtained from the recovered hardening parameters ($Q$ and $b$) with the ground-truth hardening curves for both materials. The recovered hardening curves show excellent agreement with the ground-truth results. sd

\begin{table}[!htbp]
\centering
\begin{threeparttable}
\begin{minipage}{\standardtablewidth}
\caption{Inverse parameter identification for the cruciform specimen with spatially varying material properties.}
\label{tab:spatially_varying_results}
\setlength{\tabcolsep}{2pt}
\centering
\begin{tabular*}{\textwidth}{@{\extracolsep{\fill}}lcccccc@{}}
\toprule
 & $\sigma_0$ (MPa) & Q (MPa) & b & $r_{22}$ & $r_{33}$ & $r_{12}$ \\
\midrule
\multicolumn{7}{c}{Material-1} \\
\midrule
Ground truth & 150.00 & 400.00 & 4.00 & 1.25 & 0.95 & 0.90 \\
Prediction & 150.06 & 600.99 & 2.59 & 1.25 & 0.95 & 0.90 \\
Error (\%) & 0.04 & ** & ** & 0.00 & 0.00 & 0.00 \\
\midrule
\multicolumn{7}{c}{Material-2} \\
\midrule
Ground truth & 200.00 & 800.00 & 1.00 & 0.90 & 1.10 & 1.20 \\
Prediction & 200.02 & 503.61 & 1.61 & 0.90 & 1.10 & 1.20 \\
Error (\%) & 0.01 & ** & ** & 0.00 & 0.00 & 0.00 \\
\bottomrule
\end{tabular*}
\vspace{0.3em}

\raggedright
\footnotesize ** The values of parameters $Q$ and $b$ are not unique for the hardening curve in the considered plastic strain range. The hardening curves obtained from the predicted parameters are compared in Fig.~\ref{fig:spatially_varying_hardening}.
\end{minipage}
\end{threeparttable}
\end{table}

\FloatBarrier
\subsection{Inverse Identification of Barlat's Yld2004-18p Model Using a Topology-Optimized Specimen}

Inverse identification of Barlat models is challenging because these models contain many parameters and require highly informative experimental data. Conventional inverse characterization approaches rely on multiple standardized tests, such as uniaxial tension and simple shear tests in different orientations, to generate data for parameter identification. This process is time-consuming and resource-intensive. Recently, Goncalves et al.~\cite{goncalves2023, goncalves2024} used topology optimization to design specimen geometries that generate highly heterogeneous strain fields, thereby enhancing the information content of experimental data for inverse identification. Using the proposed JAX-AD-based framework, the parameters of complex constitutive models such as Barlat's Yld2004-18p can be identified from data generated with topology-optimized specimen geometries. This section applies the framework to this challenging inverse problem.

\begin{figure}[!htbp]
\centering
\includegraphics[width=0.75\textwidth]{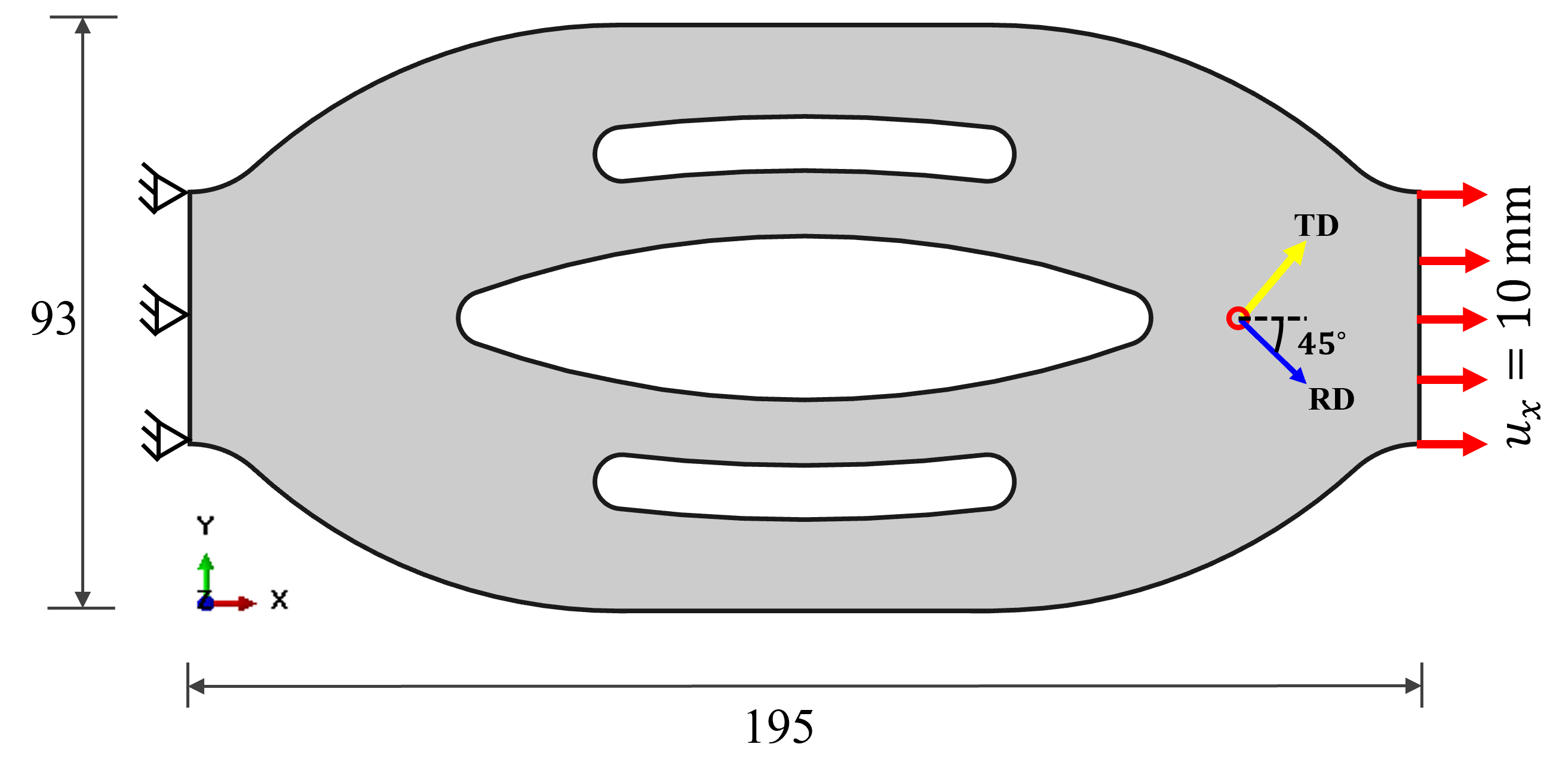}
\caption{Topology-optimized specimen geometry used for inverse identification of Barlat's Yld2004-18p model, including the applied boundary conditions and loading direction. The specimen thickness is set to 1.0 mm. All dimensions are in mm.}
\label{fig:topology_optimized_geometry}
\end{figure}

\begin{figure}[!htbp]
\centering
\begin{minipage}[t]{0.7\textwidth}
\centering
\includegraphics[width=\textwidth]{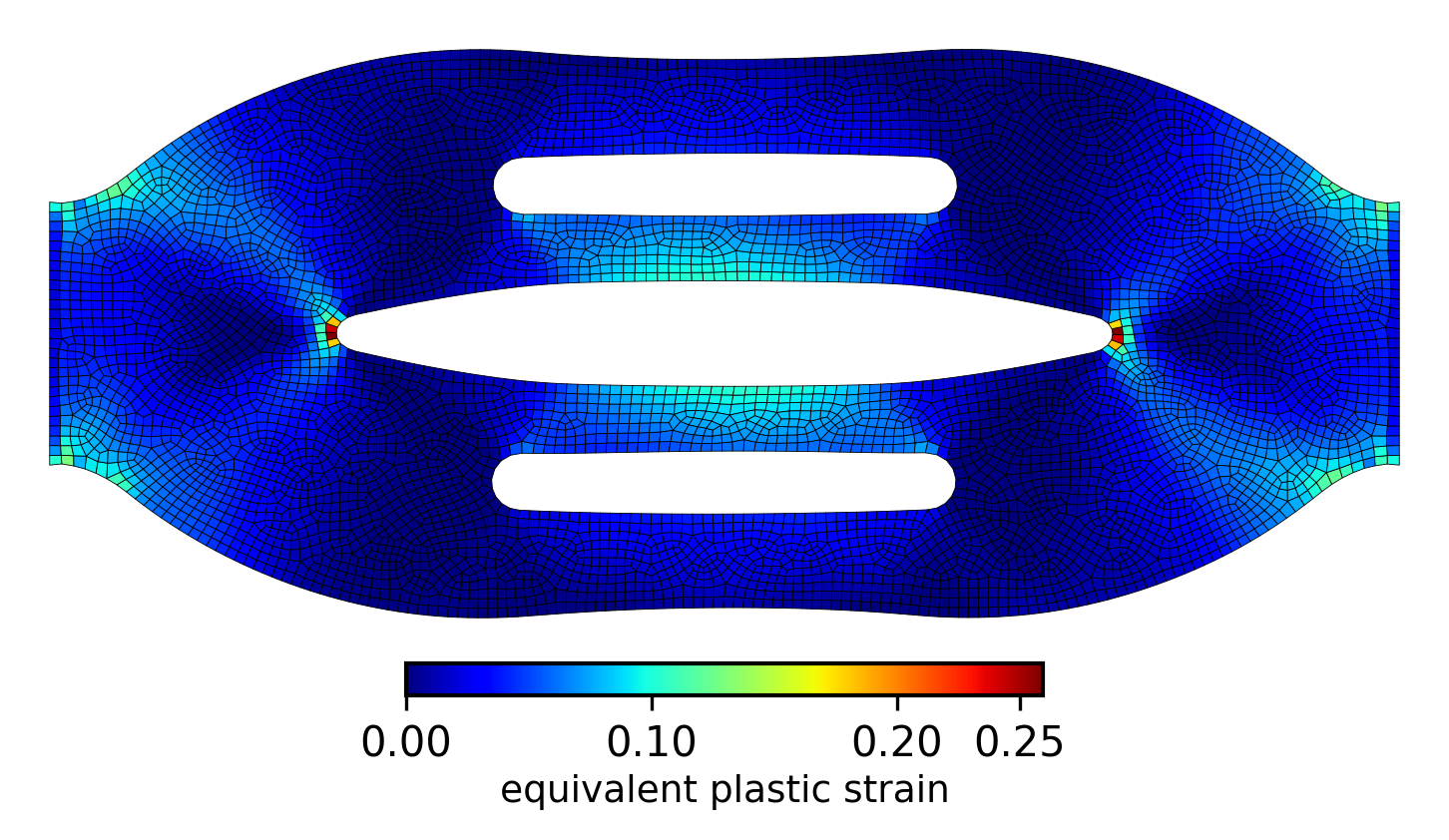}

\small (a)
\end{minipage}

\vspace{0.5em}

\begin{minipage}[t]{0.48\textwidth}
\centering
\includegraphics[width=\textwidth]{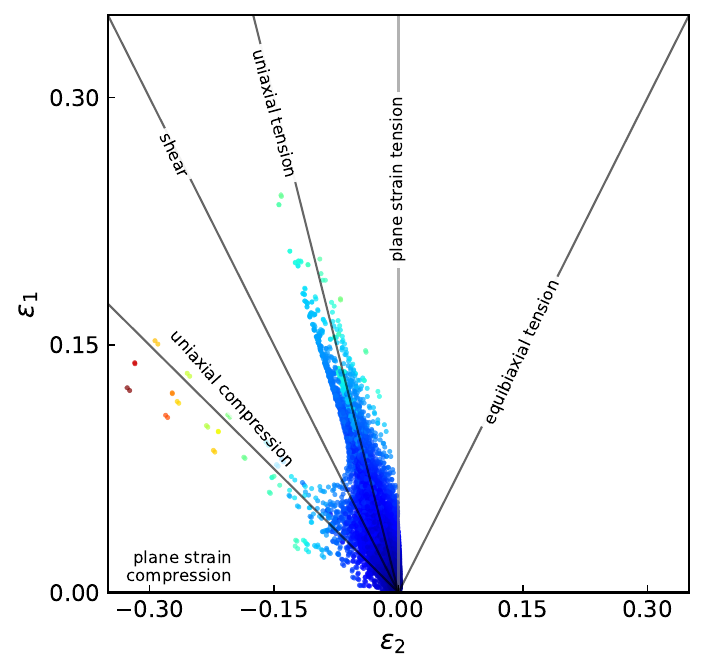}

\small (b)
\end{minipage}
\hfill
\begin{minipage}[t]{0.48\textwidth}
\centering
\includegraphics[width=\textwidth]{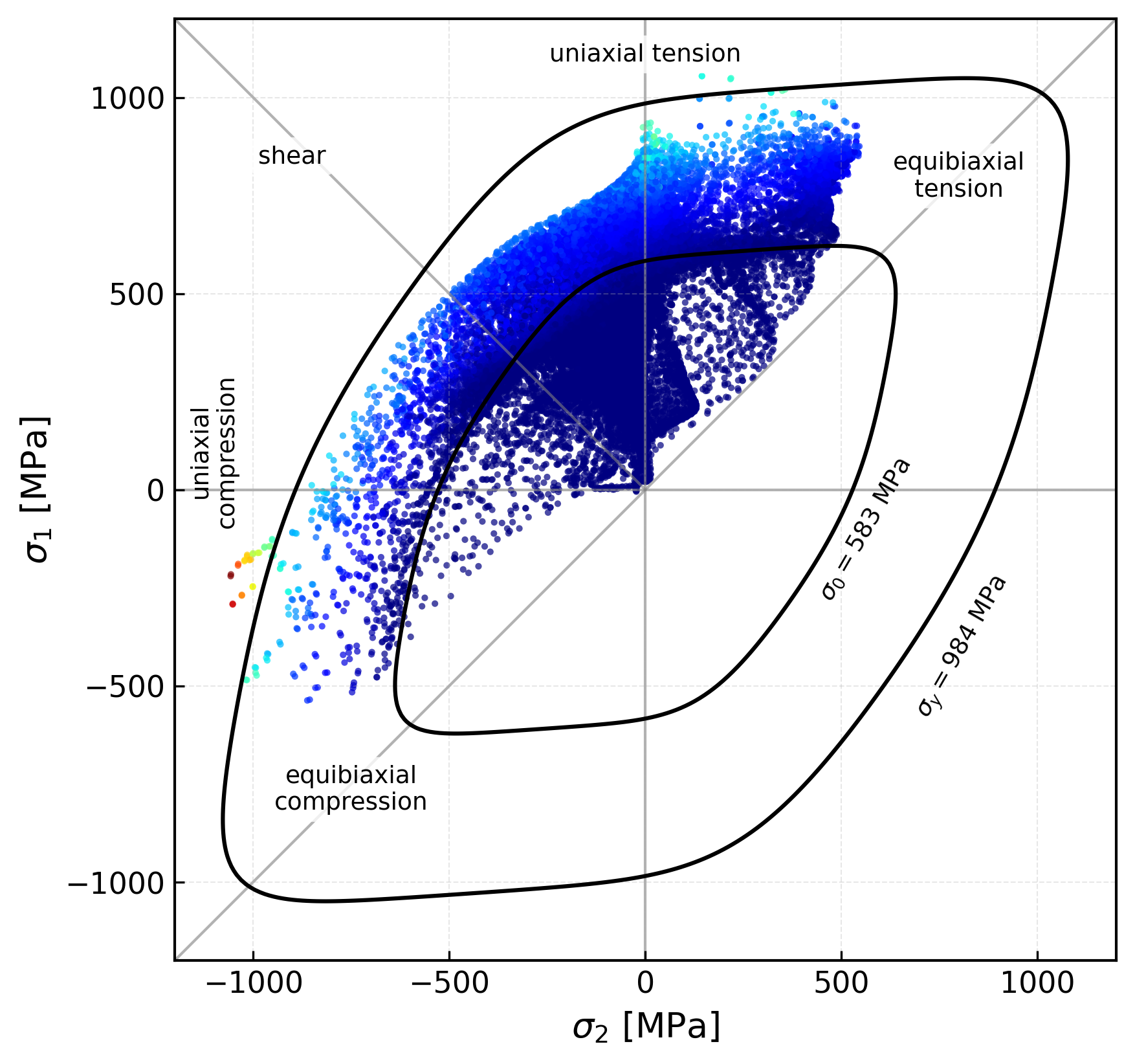}

\small (c)
\end{minipage}
\caption{(a) Equivalent plastic-strain distribution with the deformed geometry at the final time step, (b) principal strain-state diagram at the final time step, and (c) the corresponding principal stress-state diagram with yield surfaces overlaid. Only material points undergoing plastic deformation are shown in the state-space plots.}
\label{fig:topology_optimized_plastic_strain}
\label{fig:topology_optimized_stress_strain}
\end{figure}

We consider the specimen geometry designed by Goncalves et al.~\cite{goncalves2023} using topology optimization for inverse identification of Barlat's Yld2004-18p model. The geometry is shown in Fig.~\ref{fig:topology_optimized_geometry}. The specimen is discretized with 6,575 hexahedral elements, resulting in 41,472 degrees of freedom. A material orientation of 45 degrees is assigned to the specimen with respect to the loading direction. The material parameters used to generate the ground-truth displacement data from forward simulations are listed in Table~\ref{tab:barlat_forward_parameters}. The boundary and loading conditions are applied as shown in Fig.~\ref{fig:topology_optimized_geometry}. A total displacement of 10 mm is applied in several steps using the automatic time stepper. Fig.~\ref{fig:topology_optimized_plastic_strain}(a) shows the equivalent plastic strain distribution at the final time step. Fig.~\ref{fig:topology_optimized_stress_strain}(b) and Fig.~\ref{fig:topology_optimized_stress_strain}(c) show the principal strain and stress-state diagrams at the final time step, respectively. Only material points subjected to plastic deformation are plotted in the stress- and strain-state diagrams to show the diversity of states generated by the topology-optimized specimen geometry. The specimen generates a wide range of stress and strain states, from plane-strain tension to uniaxial compression, which are highly informative for inverse identification of anisotropic plasticity parameters.

We also implement a warm-start initialization strategy to reduce the computational cost of forward simulations during inverse optimization. In this approach, the displacement history at every step from the previous forward pass is stored and used as the initial guess for the corresponding time steps in the current forward pass. This strategy allows smaller time steps to be skipped and accelerates convergence, especially in the later stages of optimization when both the objective function and parameter changes are small. Warm-start initialization is enabled when the change in each parameter between the current and previous forward passes is less than 10\%.

\begin{table}[!htbp]
    \centering
\begin{threeparttable}
\begin{minipage}{\standardtablewidth}
    \caption{Ground-truth material parameters used in the Barlat Yld2004-18p forward simulation and the corresponding inverse identification results.}
    \label{tab:barlat_forward_parameters}
    \label{tab:barlat_results}
    \setlength{\tabcolsep}{2pt}
\centering
    \begin{tabular*}{\textwidth}{@{\extracolsep{\fill}}ccccccccc@{}}
        \toprule
        \multicolumn{9}{c}{Ground-truth parameters (forward simulation)} \\
        \midrule
        $c'_{12}$ & $c'_{13}$ & $c'_{21}$ & $c'_{23}$ & $c'_{31}$ & $c'_{32}$ & $c'_{44}$ & $c'_{55}$ & $c'_{66}$ \\
        \midrule
        -0.069888 & 0.936408 & 0.079143 & 1.003060 & 0.524741 & 1.363180 & 1.023770 & 1.069060 & 0.954322 \\
        \midrule
        $c''_{12}$ & $c''_{13}$ & $c''_{21}$ & $c''_{23}$ & $c''_{31}$ & $c''_{32}$ & $c''_{44}$ & $c''_{55}$ & $c''_{66}$ \\
        \midrule
        0.981171 & 0.476741 & 0.575316 & 0.866827 & 1.145010 & -0.079294 & 1.051660 & 1.147100 & 1.404620 \\
        \midrule
        $m$ & $E$ & $\nu$ & $\sigma_0$ & Q & b &  &  &  \\
        \midrule
        8.0 & 200 GPa & 0.3 & 583 MPa & 467.5 MPa & 13.0 &  &  &  \\
        \midrule
        \multicolumn{9}{c}{Identified parameters (inverse optimization)} \\
        \midrule
        $c'_{12}$ & $c'_{13}$ & $c'_{21}$ & $c'_{23}$ & $c'_{31}$ & $c'_{32}$ &  &  &  \\
        \midrule
        -0.048525 & 0.944853 & 0.082811 & 1.008399 & 0.528159 & 1.371080 &  &  &  \\
        \midrule
        $c''_{12}$ & $c''_{13}$ & $c''_{21}$ & $c''_{23}$ & $c''_{31}$ & $c''_{32}$ & $\sigma_0$ & $Q$ &  \\
        \midrule
        0.993085 & 0.487031 & 0.566246 & 0.856798 & 1.139342 & -0.075833 & 582.27 & 464.97 &  \\
        \bottomrule
    \end{tabular*}
\end{minipage}
\end{threeparttable}
\end{table}

\begin{figure}[!htbp]
\centering
\begin{subfigure}[b]{0.48\textwidth}
    \centering
    \includegraphics[width=\textwidth]{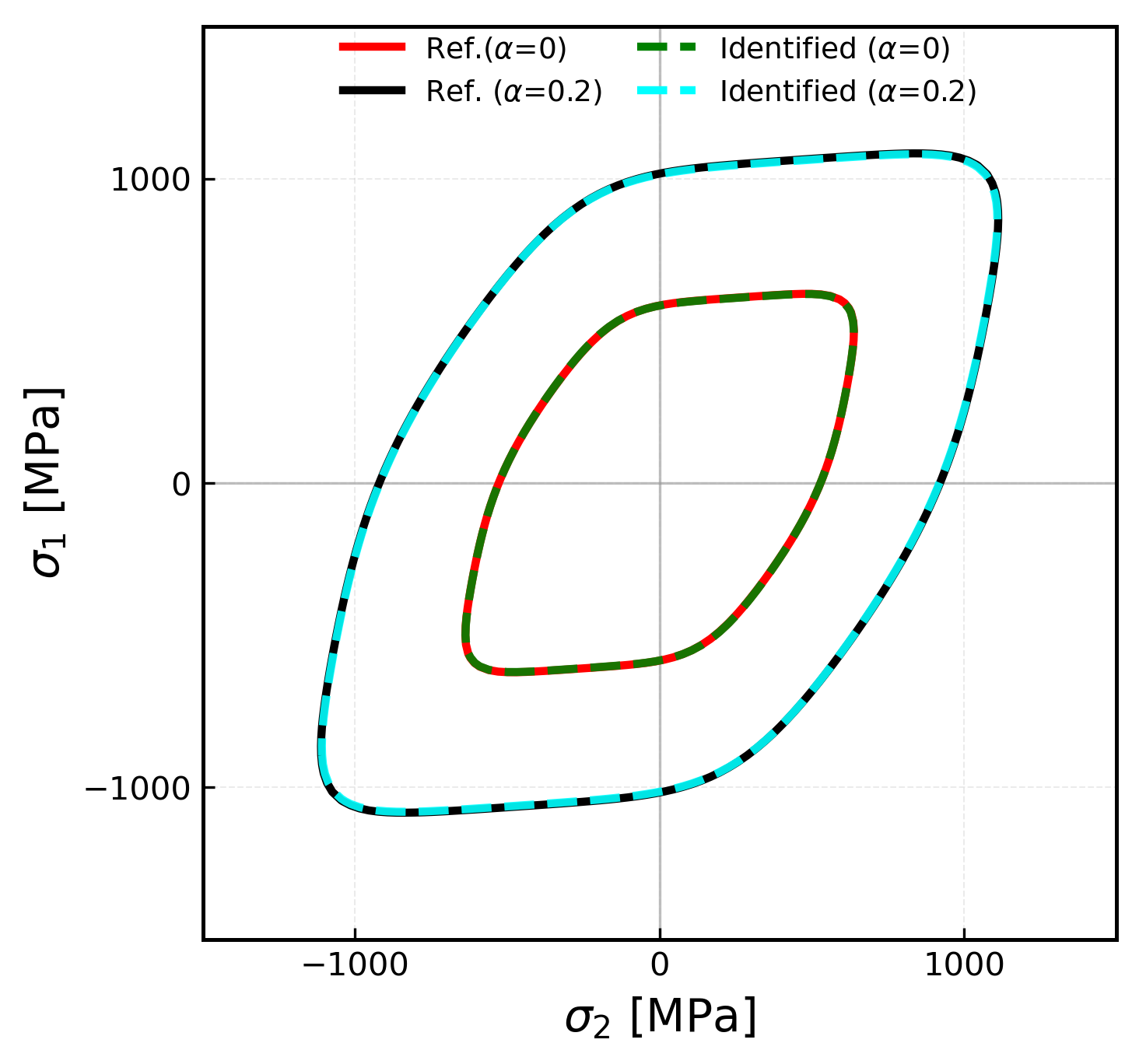}
    \caption{}
    \label{fig:barlat_yield_surface}
\end{subfigure}
\hfill
\begin{subfigure}[b]{0.5\textwidth}
    \centering
    \includegraphics[width=\textwidth]{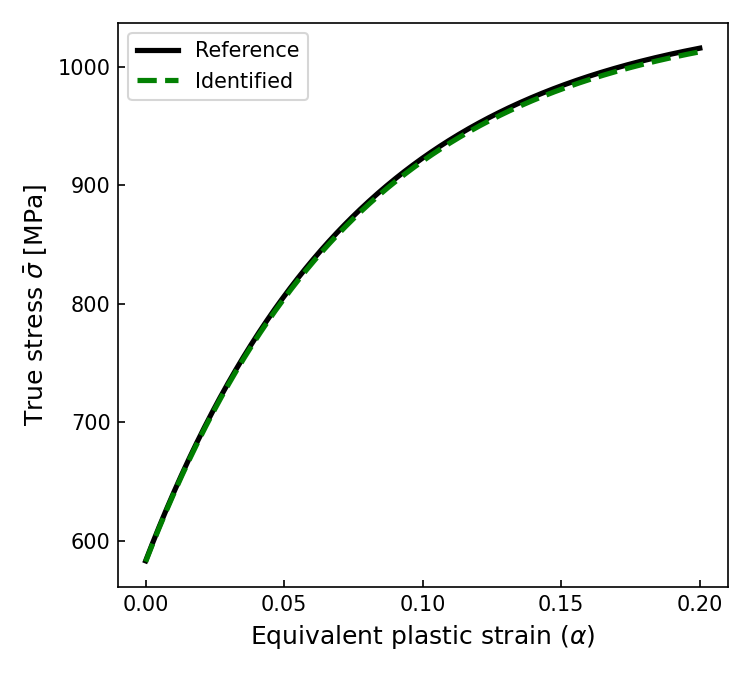}
    \caption{}
    \label{fig:barlat_hardening_curve}
\end{subfigure}
\caption{Inverse identification of Barlat's model: (a) comparison of reference and identified yield surfaces in the principal-stress plane for two hardening states ($\alpha=0$ and $\alpha=0.2$), and (b) comparison of reference and identified hardening curves.}
\label{fig:barlat_hardening_comparison}
\end{figure}

\begin{figure}[!htbp]
\centering
\begin{subfigure}[b]{0.48\textwidth}
    \centering
    \includegraphics[width=\textwidth]{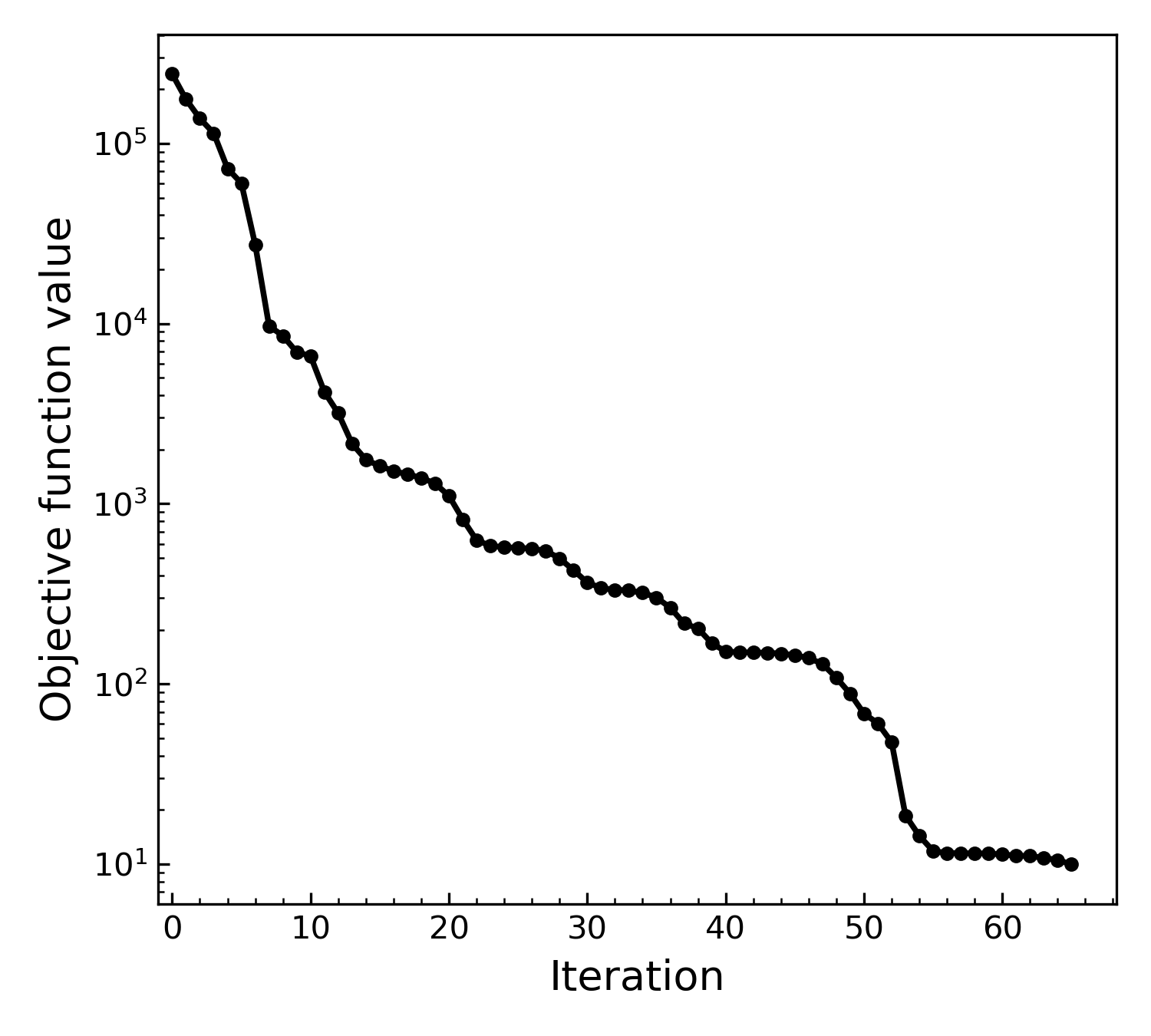}
    \caption{}
    \label{fig:barlat_convergence}
\end{subfigure}
\hfill
\begin{subfigure}[b]{0.48\textwidth}
    \centering
    \includegraphics[width=\textwidth]{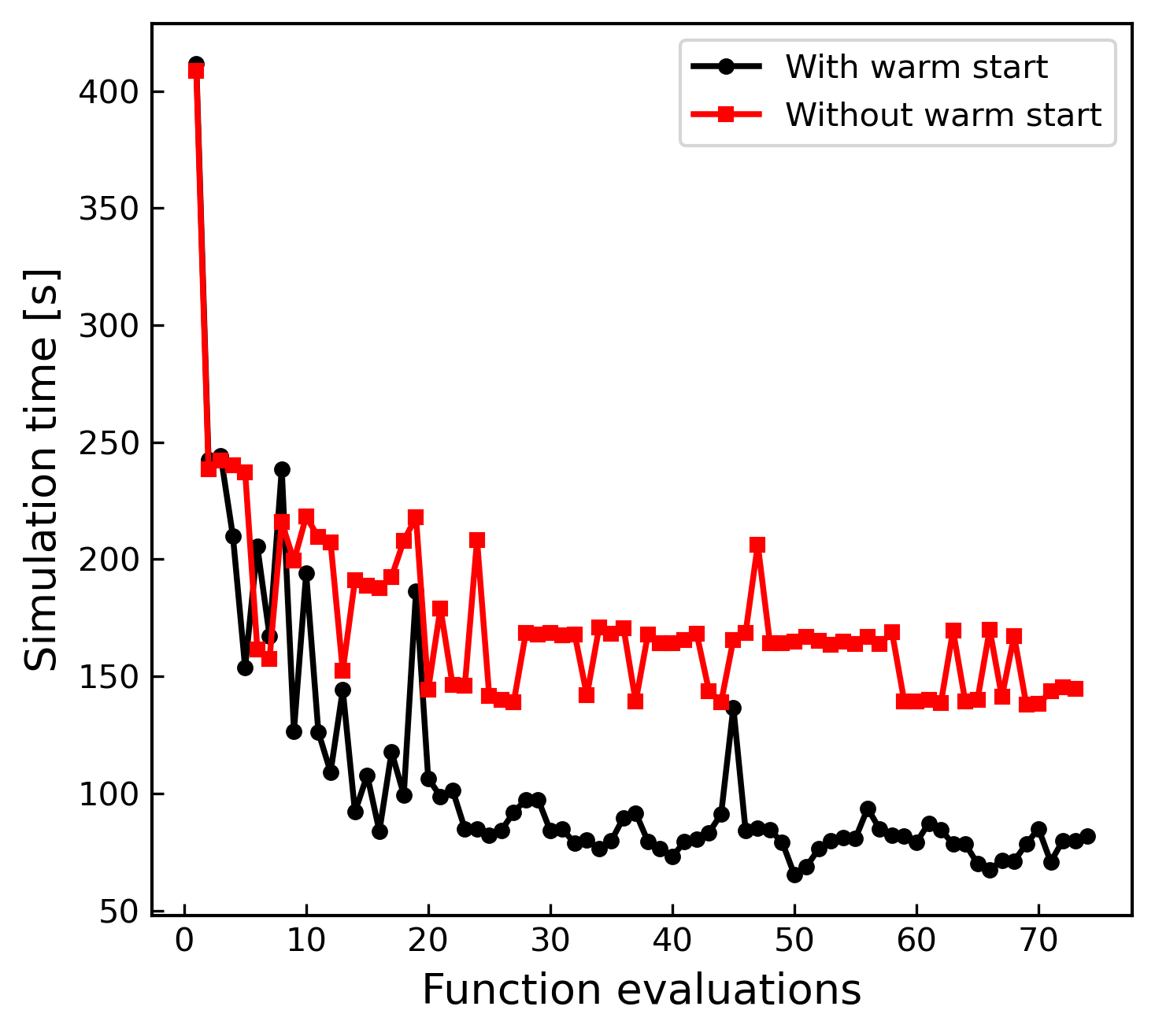}
    \caption{}
    \label{fig:barlat_inv_sim_time}
\end{subfigure}
\caption{Inverse optimization for Barlat material parameter identification: (a) convergence history of the optimization, and (b) comparison of forward-pass simulation times with and without warm-start initialization.}
\label{fig:barlat_convergence_simtime}
\end{figure}

Because negligible out-of-plane shear stresses are expected for the current specimen with a thickness of 1 mm, we consider only the following parameter set for inverse identification: $\boldsymbol{\theta} = [\sigma_0, Q, c'_{12}, c'_{13}, c'_{21}, c'_{23}, c'_{31}, c'_{32}, c''_{12}, c''_{13}, c''_{21}, c''_{23}, c''_{31}, c''_{32}]$. For this demonstration, the upper and lower bounds of the parameters are defined using a $\pm 50\%$ variation from the ground-truth parameters. In the normalized parameter space $\rho_i \in [-1, 1]$, all parameters are initialized with $\rho_i = 0.5$. For this problem, the maximum numbers of iterations and function evaluations are set to 65 and 100, respectively.
The optimization is performed using an H100 GPU with 80 GB of memory and a CuDSS-based direct sparse linear solver. Because the complex constitutive plasticity model can lead to a nonsymmetric tangent stiffness matrix, direct sparse linear solvers are more robust and efficient than iterative solvers for moderate-size forward and inverse problems. Table~\ref{tab:barlat_results} summarizes the inverse-identification results. Fig.~\ref{fig:barlat_hardening_comparison}(a) and Fig.~\ref{fig:barlat_hardening_comparison}(b) compare the identified yield surfaces and hardening curves with the reference data, respectively. Fig.~\ref{fig:barlat_convergence_simtime}(a) shows the convergence history. The optimization converges in 65 iterations and 73 function evaluations, with a total computation time of 131 min with warm-start initialization and 172 min without it. Fig.~\ref{fig:barlat_convergence_simtime}(b) compares forward-pass simulation times during inverse optimization with and without the warm-start initialization strategy. The warm-start strategy reduces the overall inverse optimization time by approximately a factor of 1.6. For complex yield surfaces, different parameter combinations can lead to the same yield surface, which explains the non-uniqueness of the predicted parameters. Nevertheless, the results demonstrate that the proposed framework can identify complex constitutive models using data generated from highly informative specimen geometries, which can significantly reduce the experimental effort required for material characterization.

\section{Conclusions}
This work presented an end-to-end differentiable, GPU-accelerated finite element framework for inverse characterization of finite-strain anisotropic plasticity. The formulation combines rate-consistent anisotropic plasticity, JAX-based automatic differentiation, and adjoint sensitivity analysis to enable efficient gradient-based optimization from full-field displacement data. By exploiting automatic vectorization, JIT compilation, and GPU-resident computations, the framework provides a scalable route for forward simulation and inverse material identification in large nonlinear elastoplastic problems.

\begin{itemize}
    \item The proposed JAX-FEM-ANISO implementation provides automatic and accurate sensitivities for PDE-constrained inverse optimization. Gradient verification against finite differences confirmed the correctness of the adjoint-based derivatives while avoiding the poor scaling and step-size sensitivity of finite-difference gradients.

    \item GPU acceleration was integrated across the main nonlinear FEM bottlenecks: constitutive updates, stiffness calculations, sparse assembly, row elimination, and linear solves. The memory-aware implementation manages GPU resources effectively, enabling fast large-scale forward and inverse finite-strain plasticity simulations.

    \item The inverse identification examples demonstrated accurate recovery of anisotropic yield and hardening behavior from information-rich displacement fields. The framework was effective for Hill--48 plasticity in tensile and cruciform specimens, including noisy synthetic displacement data representative of DIC measurements.

    \item The same differentiable workflow was extended to more challenging parameter spaces, including spatially varying parameters and the Barlat Yld2004-18p model. These examples show that AD-based adjoint gradients make high-dimensional inverse problems feasible where finite-difference-based approaches would be computationally prohibitive.
\end{itemize}

Future work will focus on applying the framework to real experimental DIC data for inverse parameter identification. Experimental measurements introduce noise, imperfect boundary conditions, incomplete observations, and possible model discrepancy, which can affect optimization robustness and parameter uniqueness. Addressing these issues through regularization, noise-aware objective functions, and uncertainty quantification will be an important step toward practical deployment in experimental material characterization.

\section*{Acknowledgment}
The authors would like to acknowledge support from the Defense Advanced Research Projects Agency (DARPA), Multiobjective Engineering and Testing of Alloy Structures (METALS) program, “RADICAL: Rapid Array Dimple based Co-design of gradient materiaL and geometry” project (No. HR0011-24-2-0302), monitored by Dr. Andrew Detor, General Motors LLC, and the DoD Vannevar Bush Faculty Fellowship (N00014-19-1-2642) to JC.

\section*{Declaration on the Use of Generative AI}
The authors used ChatGPT, a large language model developed by OpenAI, to assist with English language editing and grammar refinement. The scientific content, technical interpretations, and conclusions are solely the responsibility of the authors.

\appendix
\renewcommand{\theequation}{\thesection\arabic{equation}}
\renewcommand{\thealgorithm}{\thesection\arabic{algorithm}}
\renewcommand{\thefigure}{\thesection\arabic{figure}}
\renewcommand{\thetable}{\thesection\arabic{table}}
\renewcommand{\thesection}{\Alph{section}}
\makeatletter
\newcommand{\appendixsectionformat}[1]{Appendix~\csname the#1\endcsname.\quad}
\renewcommand{\@seccntformat}[1]{%
  \ifcsname appendix#1format\endcsname
    \csname appendix#1format\endcsname{#1}%
  \else
    \csname the#1\endcsname\quad
  \fi
}
\expandafter\let\csname appendixsectionformat\endcsname\appendixsectionformat
\@addtoreset{equation}{section}
\@addtoreset{algorithm}{section}
\@addtoreset{figure}{section}
\@addtoreset{table}{section}
\makeatother
\section{Constitutive Model Details} \label{sec:appendix_A}

This Appendix collects the yield-function parameterizations summarized in Section~\ref{subsubsec:constitutive_model}.

\subsection{Hill--48 model}
The tensor $\mathbb{H}$ is constructed in a local orthotropic basis and rotated into the global coordinate system using the material orientation $\tens{\Theta}$,
\begin{equation}
    \mathbb{H}_{\text{global}} = \tens{\Theta} \,\tens{\Theta} : \mathbb{H}_{\text{local}} : \tens{\Theta}^\top \tens{\Theta}^\top
\end{equation}
with the local Hill tensor parameterized as,
\begin{equation}
\mathbb{H}_{\text{local}} = \begin{bmatrix}
    p_1 & p_4 & p_6 & 0 & 0 & 0 \\
    & p_2 & p_5 & 0 & 0 & 0 \\
    & & p_3 & 0 & 0 & 0 \\
    \text{sym.} & & & \frac{1}{2}p_7 & 0 & 0 \\
    & & & & \frac{1}{2}p_8 & 0 \\
    & & & & & \frac{1}{2}p_9
\end{bmatrix}
\end{equation}
where $\{p_i\}_{i=1}^9$ are material parameters.

The effective stress $\phi(\bm{T})$ is defined in terms of a reference yield stress $\sigma_0$ and six independent material parameters, namely the initial axial yield stresses $\sigma^y_{ii}$ and shear yield stresses $\tau^y_{ij}$ with respect to the principal axes of orthotropy. Introducing the normalized yield ratios,
\begin{equation}
\begin{aligned}
r_{11} &= \frac{\sigma^y_{11}}{\sigma_0}, &
r_{22} &= \frac{\sigma^y_{22}}{\sigma_0}, &
r_{33} &= \frac{\sigma^y_{33}}{\sigma_0} \\
r_{12} &= \frac{\sqrt{3}\tau^y_{12}}{\sigma_0}, &
r_{13} &= \frac{\sqrt{3}\tau^y_{13}}{\sigma_0}, &
r_{23} &= \frac{\sqrt{3}\tau^y_{23}}{\sigma_0}
\end{aligned}
\end{equation}
The coefficients $p_i$ are expressed as,
\begin{equation}
\begin{aligned}
p_1 = \frac{2}{3\, r_{11}^2}, \qquad
p_2 = \frac{2}{3\, r_{22}^2}, \qquad
p_3 = \frac{2}{3\, r_{33}^2} \\
p_7 = \frac{1}{r_{12}^2}, \qquad
p_8 = \frac{1}{r_{23}^2}, \qquad
p_9 = \frac{1}{r_{13}^2}
\end{aligned}
\label{eq:p_coefficients}
\end{equation}
The remaining dependent coefficients are,
\begin{equation}
p_4 = \frac{1}{2}(p_3-p_1-p_2), \qquad 
p_5 = \frac{1}{2}(p_1-p_2-p_3), \qquad 
p_6 = \frac{1}{2}(p_2-p_1-p_3)
\end{equation}
The isotropic case is recovered when,
\begin{equation}
r_{11} = r_{22} = r_{33} = r_{12} = r_{23} = r_{13} = 1
\end{equation}

\subsection{Barlat Yld2004-18p model}

For the Barlat model, the transformed stresses appearing in Eq.~\eqref{eq:Phi_barlat} are defined by two linear transformations of the deviatoric stress $\bm{s} = \mathbb{P} : \bm{T}$,
\begin{equation}
    \bm{s}' = \mathbb{C}' : \bm{s} = \mathbb{C}' : \mathbb{P} : \bm{T}, \quad
    \bm{s}'' = \mathbb{C}'' : \bm{s} = \mathbb{C}'' : \mathbb{P} : \bm{T}
\end{equation}
In Voigt notation,
\begin{equation}
    \mathbb{P} = \frac{1}{3}
    \begin{bmatrix}
        2 & -1 & -1 & 0 & 0 & 0 \\
        -1 & 2 & -1 & 0 & 0 & 0 \\
        -1 & -1 & 2 & 0 & 0 & 0 \\
        0 & 0 & 0 & 3 & 0 & 0 \\
        0 & 0 & 0 & 0 & 3 & 0 \\
        0 & 0 & 0 & 0 & 0 & 3
    \end{bmatrix}
\end{equation}

\begin{figure}[!htbp]
    \centering
    \includegraphics[width=0.5\textwidth]{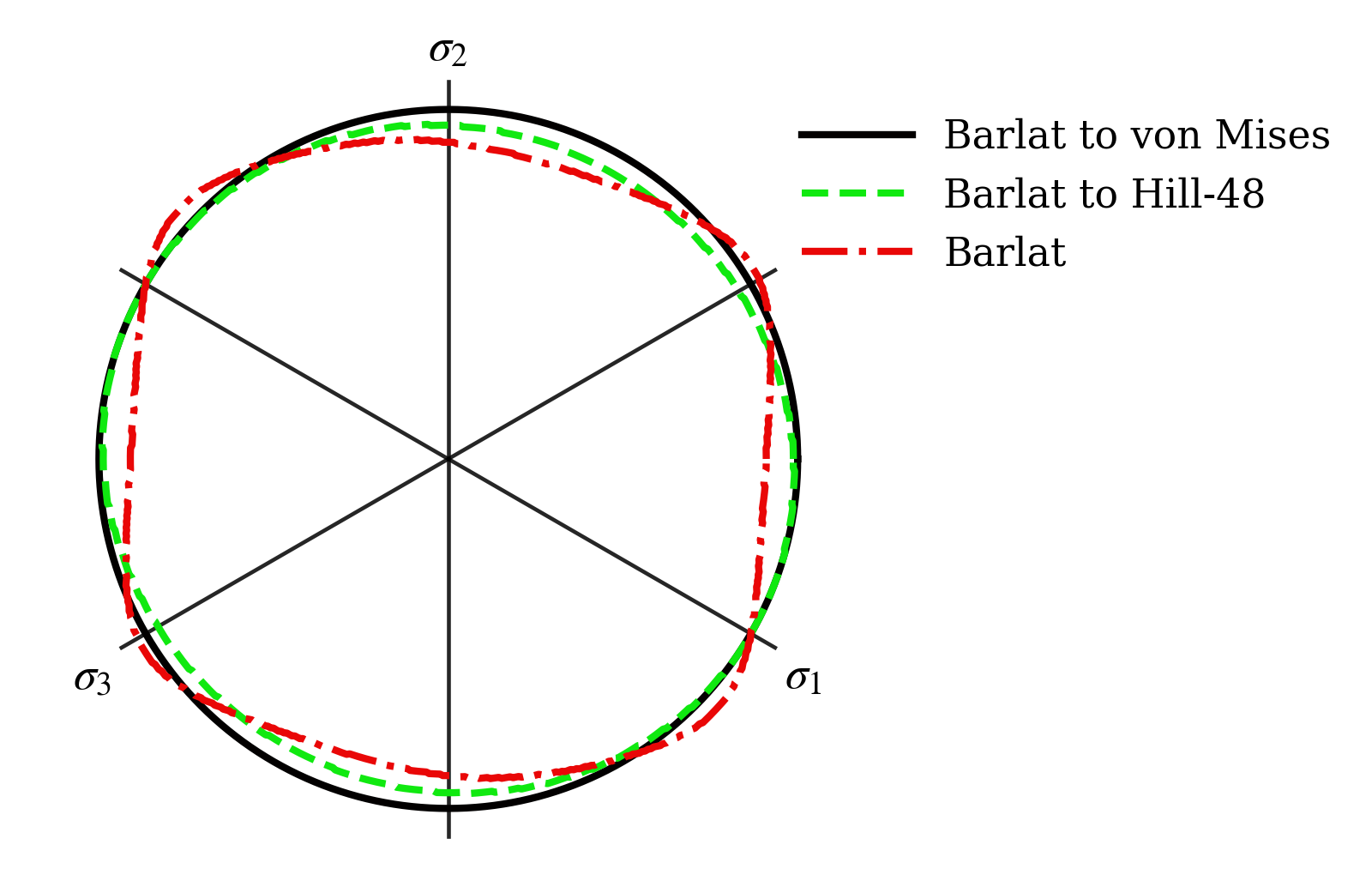}
    \caption{Reduction of the anisotropic Barlat yield criterion to the von Mises and Hill--48 models in the $\pi$-plane under the corresponding parameter choices. The models are normalized such that the yield stress in uniaxial tension along the first principal axis is $\sigma^y_{11} = 1$. The material parameters used in this figure are reported in Table~\ref{tab:barlat_verification_material_parameters}.}
    \label{fig:pi_plane_barlat_reductions}
\end{figure}

In the material coordinate system, the tensors $\mathbb{C}'$ and $\mathbb{C}''$ are represented in matrix notation as,
\begin{equation}
\begin{gathered}
[\mathbb{C}'] = \begin{bmatrix}
    0 & -c'_{12} & -c'_{13} & 0 & 0 & 0 \\
    -c'_{21} & 0 & -c'_{23} & 0 & 0 & 0 \\
    -c'_{31} & -c'_{32} & 0 & 0 & 0 & 0 \\
    0 & 0 & 0 & c'_{44} & 0 & 0 \\
    0 & 0 & 0 & 0 & c'_{55} & 0 \\
    0 & 0 & 0 & 0 & 0 & c'_{66}
\end{bmatrix} \\[10pt]
[\mathbb{C}''] = \begin{bmatrix}
    0 & -c''_{12} & -c''_{13} & 0 & 0 & 0 \\
    -c''_{21} & 0 & -c''_{23} & 0 & 0 & 0 \\
    -c''_{31} & -c''_{32} & 0 & 0 & 0 & 0 \\
    0 & 0 & 0 & c''_{44} & 0 & 0 \\
    0 & 0 & 0 & 0 & c''_{55} & 0 \\
    0 & 0 & 0 & 0 & 0 & c''_{66}
\end{bmatrix}
\end{gathered}
\end{equation}
where the 18 coefficients $\{c'_{ij}, c''_{ij}\}$ are material parameters that characterize the anisotropic yielding behavior. These local tensors are then rotated into the global coordinate system using the material orientation tensor $\tens{\Theta}$ in the same manner as for the Hill model. As shown in Fig.~\ref{fig:pi_plane_barlat_reductions}, when all 18 parameters are set to unity and $m=2$ or $m=4$, the Barlat model reduces to the von Mises criterion. With $\mathbb{C}' = \mathbb{C}''$ and $m=2$ or $m=4$, it reduces to the Hill--48 model.

\section{Line-Search Implementation Analysis} \label{sec:appendix_B}

This Appendix gives the line-search update used in the local return-mapping algorithm summarized in Section~\ref{subsubsec:implicit_time_discretization} and then analyzes its effect on convergence.

To improve robustness when the full Newton correction is too aggressive for highly anisotropic nonquadratic yield surfaces, we augment the local update with a line search along the Newton direction, following Scherzinger~\cite{Scherzinger2017}. Instead of taking the full Newton step $\bm{y}^{k+1} = \bm{y}^k + \delta\bm{y}^k$, a scalar step length $\zeta^k \in (0,1]$ is introduced as,
\begin{equation}
    \bm{y}^{k+1} = \bm{y}^k + \zeta^k\,\delta\bm{y}^k
\end{equation}
The step length $\zeta^k$ is chosen to sufficiently reduce the scalar merit function,
\begin{equation}
    \psi^k(\zeta^k) = \tfrac{1}{2}\,\bm{G}\!\bigl(\bm{y}^k + \zeta^k\delta\bm{y}^k\bigr)^\top \bm{G}\!\bigl(\bm{y}^k + \zeta^k\delta\bm{y}^k\bigr)
    \label{eq:merit}
\end{equation}
which combines all components of the local residual (Eq.~\eqref{eq:local_cpp}) into a single scalar measure of convergence quality. Because $\delta\bm{y}^k$ is the Newton direction, the directional derivative at $\zeta^k = 0$ satisfies $(\psi^k)'(0) = -2\psi^k(0)$, which guarantees that $\psi^k$ is locally decreasing at the full Newton step.

To find a suitable $\zeta^k$, the algorithm evaluates both $\psi^k(0)$ and $\psi^k(1)$. The full step is accepted, setting $\zeta^k = 1$, if it satisfies Goldstein's sufficient decrease condition,
\begin{equation}
    \psi^k(1) < \bigl(1 - 2\beta \zeta^k\bigr)\,\psi^k(0)
    \label{eq:goldstein_full}
\end{equation}
where $\beta$ is an algorithmic constant. Otherwise, a quadratic approximation $\hat{\psi}^k(\zeta^k)$ of the merit function ${\psi}^k(\zeta^k)$ is constructed using the three known values $\psi^k(0)$, $(\psi^k)'(0)$, and $\psi^k(1)$,
\begin{equation}
    \hat{\psi}^k\!\left(\zeta^k\right) = \bigl(1 - 2\zeta^k + (\zeta^k)^2\bigr)\,\psi^k(0) + (\zeta^k)^2\,\psi^k(1)
    \label{eq:quadratic_approx}
\end{equation}
Minimizing $\hat{\psi}^k(\zeta^k)$ provides the initial trial step length,
\begin{equation}
    \zeta^k = \frac{\psi^k(0)}{\psi^k(0) + \psi^k(1)}
    \label{eq:alpha_init}
\end{equation}
If the Goldstein condition is not yet satisfied, the quadratic approximation is updated iteratively. At inner iteration $j$, the merit function is re-evaluated at the current trial $\zeta^{(j)}$ and the minimization of the updated quadratic yields the next candidate,
\begin{equation}
    \zeta^{(j+1)} = \frac{\psi^k(0)}{\psi^k(0) + \psi^k\!\left(\zeta^{(j)}\right)}
    \label{eq:alpha_update_ls}
\end{equation}
To prevent the step length from becoming excessively small, a lower bound is enforced at each iteration through a parameter $\eta \in (0,1)$, which controls the minimum fraction of the step length that can be accepted. The updated step length is,
\begin{equation}
    \zeta^{(j+1)} \leftarrow \max\!\left\{\eta\,\zeta^{(j)},\;\zeta^{(j+1)}\right\}
    \label{eq:alpha_safeguard}
\end{equation}
The combined effect of the quadratic approximation and the Goldstein condition ensures that the line search terminates with a step length that provides a meaningful reduction in $\psi^k$ while preserving the quadratic convergence of the Newton scheme when $\zeta^k = 1$ is admissible.

For the numerical comparison below, we consider a deviatoric trial stress state with principal stresses $\sigma_1=8$, $\sigma_2=-5.2$, and $\sigma_3=-2.8$. All material parameters are the same as in the previous Barlat verification example (Table~\ref{tab:barlat_verification_material_parameters}), with perfect plasticity and the anisotropic yield surface defined by the Barlat Yld2004-18p model. The constitutive update is solved using two approaches: (i) a standard Newton method and (ii) a Newton method with the line-search algorithm. Following Scherzinger~\cite{Scherzinger2017}, the line-search parameters are $\beta = 10^{-4}$ and $\eta = 0.1$.

\begin{figure}[!htbp]
    \centering
    \begin{tikzpicture}
        \node[anchor=south west, inner sep=0] (image) at (0,0) {\includegraphics[width=0.8\textwidth]{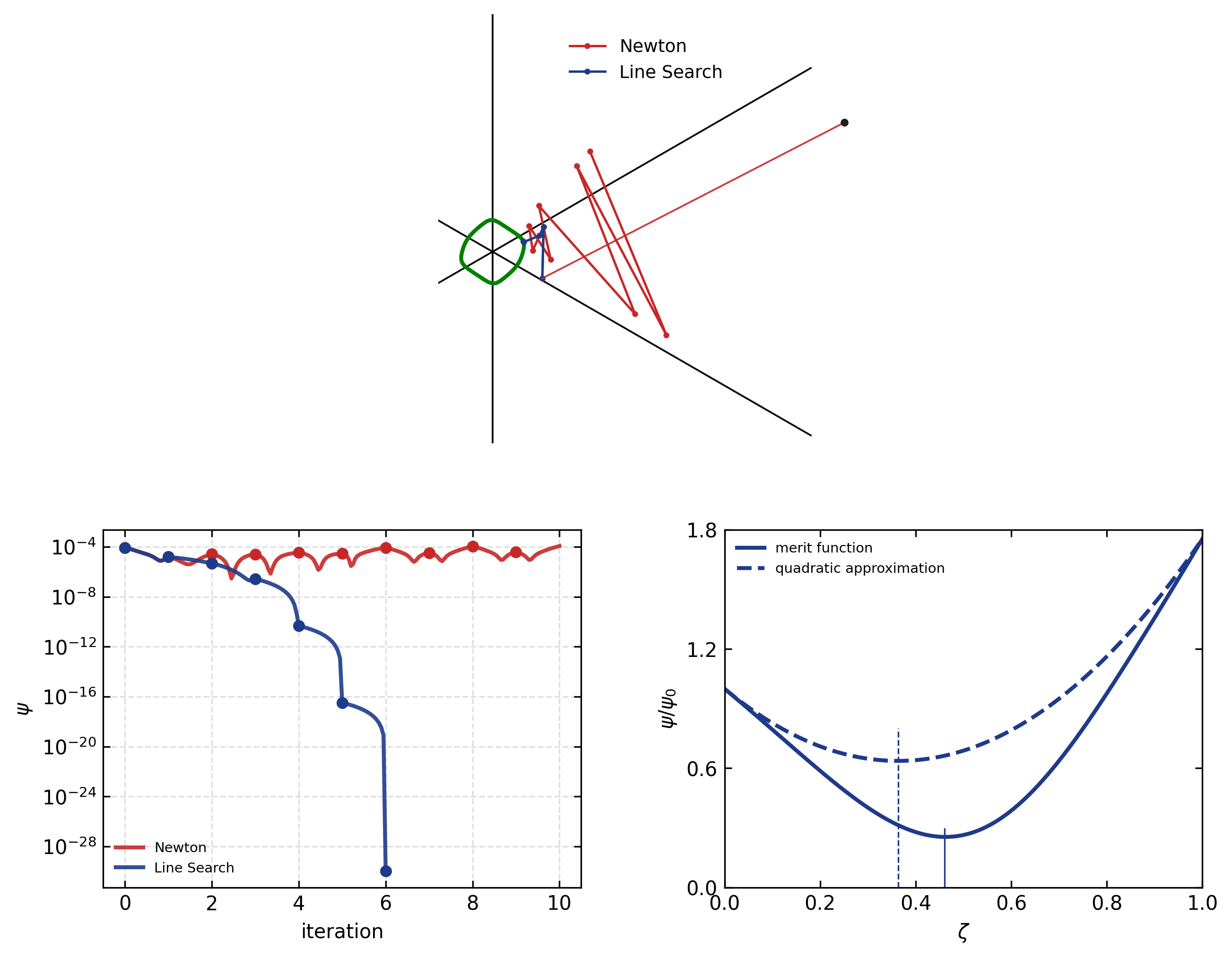}};
        \begin{scope}[x={(image.south east)},y={(image.north west)}]
            \node[anchor=north west] at (0.25,0.98) {(a)};
            \node[anchor=north west] at (0.55,0.6) {$\sigma_1$};
            \node[anchor=north west] at (0.35,0.98) {$\sigma_2$};
            \node[anchor=north west] at (0.32,0.7) {$\sigma_3$};
            \node[anchor=north west] at (0.02,0.52) {(b)};
            \node[anchor=north west] at (0.53,0.52) {(c)};
        \end{scope}
    \end{tikzpicture}
    \caption{Line-search analysis for the single-point return-mapping problem with the Barlat Yld2004-18p yield function: (a) iteration trajectories of the standard Newton method and the Newton method with line search on the $\pi$-plane; (b) normalized merit function $\psi/\psi_0$ versus iteration count; and (c) normalized merit function and its quadratic approximation as functions of the step size $\zeta$ during the third iteration.}
    \label{fig:LineSearchAnalysis}
\end{figure}

\begin{figure}[!htbp]
    \centering
    \begin{tikzpicture}
        \node[anchor=south west, inner sep=0] (image) at (0,0) {\includegraphics[width=0.8\textwidth]{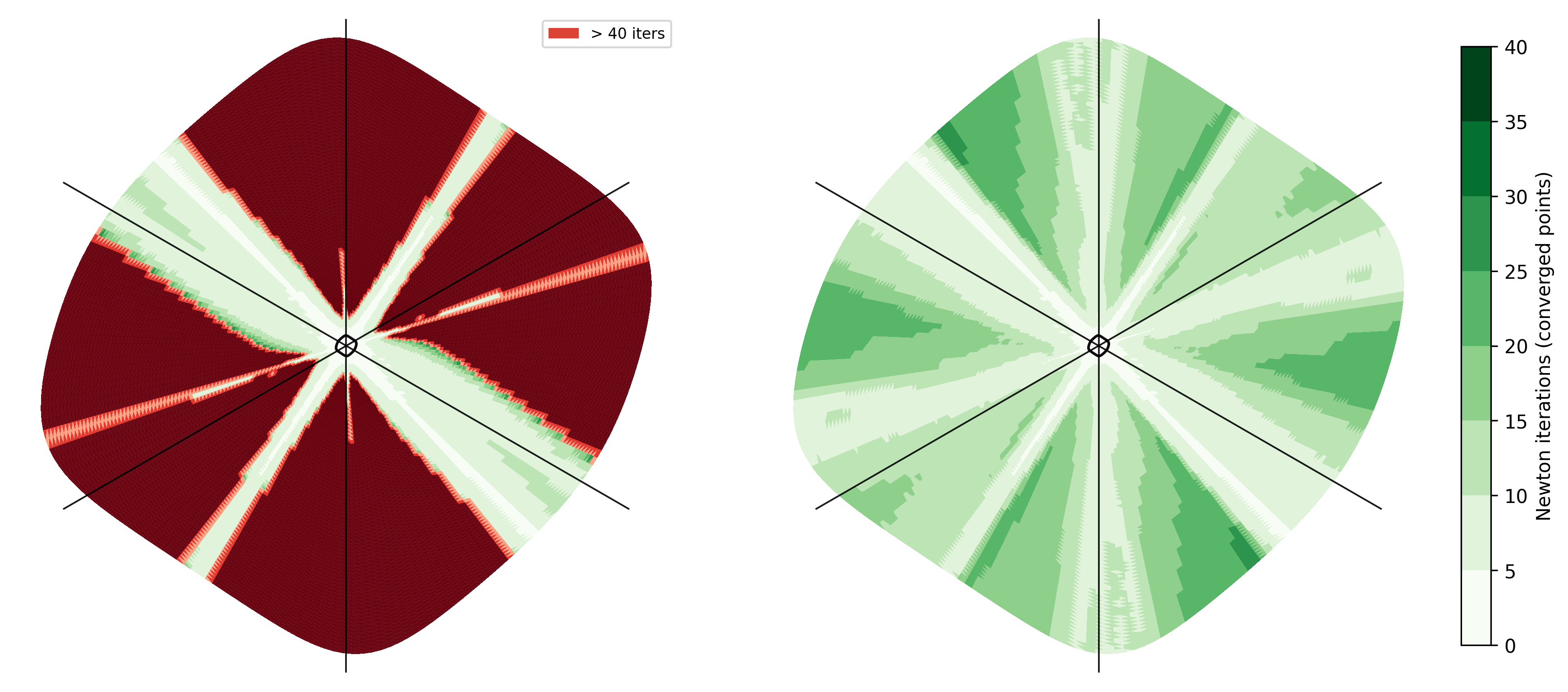}};
        \begin{scope}[x={(image.south east)},y={(image.north west)}]
            \node[anchor=north west] at (0.02,0.98) {(a)};
            \node[anchor=north west] at (0.38,0.25) {$\sigma_1$};
            \node[anchor=north west] at (0.22,0.99) {$\sigma_2$};
            \node[anchor=north west] at (0.02,0.25) {$\sigma_3$};
            \node[anchor=north west] at (0.33,0.9) {Newton};
            \node[anchor=north west] at (0.48,0.98) {(b)};
            \node[anchor=north west] at (0.85,0.25) {$\sigma_1$};
            \node[anchor=north west] at (0.7,0.99) {$\sigma_2$};
            \node[anchor=north west] at (0.5,0.25) {$\sigma_3$};
            \node[anchor=north west] at (0.78,0.92) {Line search};
        \end{scope}
    \end{tikzpicture}
    \caption{Comparison of convergence behavior for a wide range of trial stress states in the principal-stress plane: (a) standard Newton method and (b) Newton method with line search. Red indicates trial states that did not converge within 40 iterations, while the green color scale denotes the number of iterations required for converged points.}
    \label{fig:LineSearchPerformance}
\end{figure}

Fig.~\ref{fig:LineSearchAnalysis}(a) compares the convergence paths of the two methods on the $\pi$-plane. The first two iterations are similar, but from the third iteration onward the standard Newton method begins to oscillate and fails to approach the yield surface in a stable manner. This trend is also evident in Fig.~\ref{fig:LineSearchAnalysis}(b), which shows the evolution of the normalized merit function. The standard Newton method oscillates without converging even after 10 iterations, whereas the line-search method reduces the merit function sharply at each step by selecting a step length $\zeta$ that yields sufficient decrease.

Fig.~\ref{fig:LineSearchAnalysis}(c) shows the normalized merit function and its quadratic approximation with respect to the step size during the third iteration. The full Newton step, $\zeta=1$, increases the merit function relative to its value at $\zeta=0$, whereas the true minimum of the merit function occurs at $\zeta=0.46$. The quadratic model used in the line-search algorithm selects $\zeta=0.363$, which still provides a substantial reduction of the merit function and leads to robust convergence. These results demonstrate that the line-search strategy significantly improves the robustness of the return-mapping algorithm for highly anisotropic yield surfaces and difficult trial stress states.

Fig.~\ref{fig:LineSearchPerformance}(a) and Fig.~\ref{fig:LineSearchPerformance}(b) further compare the convergence behavior of the two algorithms for a wide range of trial stress states sampled in the principal-stress plane. Approximately 16,000 trial stress states are sampled between the actual yield surface, shown in black, and a geometrically similar surface scaled by a factor of 30. For both algorithms, the maximum iteration count is set to 40. Trial states that do not converge within 40 iterations are shown in red, while the color scale for converged points indicates the number of iterations required to reach convergence. The standard Newton method fails to converge for a large portion of the sampled trial states, approximately 9,600 points, whereas the line-search method converges for all sampled points. Moreover, the line-search method converges within 20 iterations for most of the sampled states, further demonstrating its robust convergence properties.

\FloatBarrier

\section{Closest-Point Projection Stress-Update Algorithms} \label{sec:appendix_C}

\begin{figure}[!htbp]
\centering
\includegraphics[width=0.5\textwidth]{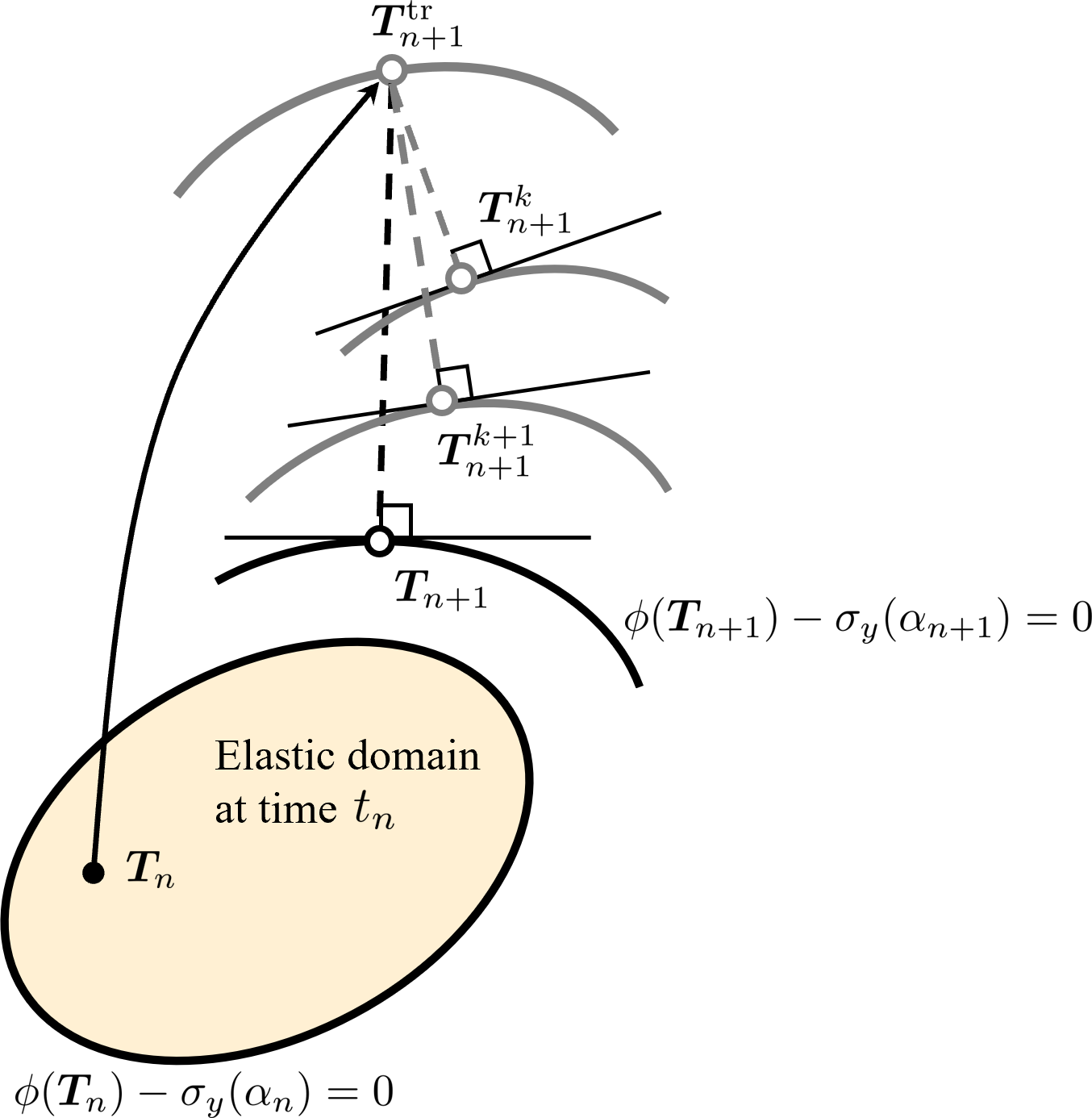}
\caption{Geometric interpretation of the closest-point projection algorithm.}
\label{fig:CPP_algorithm}
\end{figure}

As illustrated in Fig.~\ref{fig:CPP_algorithm}, the closest-point projection algorithm projects the inadmissible trial stress state $\bm{T}_{n+1}^{\text{tr}}$ back onto the updated yield surface along the direction normal to the yield surface, while satisfying both the consistency condition and the flow rule. The yield surface expands due to isotropic hardening as the internal variable $\alpha$ increases from $\alpha_n$ to $\alpha_{n+1}$. The corresponding stress-update and local return-mapping procedures are listed in Algorithms~\ref{alg:global_stress_update} and~\ref{alg:local_cpp}.

\begin{algorithm}[!htbp]
\caption{Global stress-update algorithm}
\label{alg:global_stress_update}
\begin{algorithmic}[1]
\Require $\bm{F}_{n+1}$, $\bm{\varepsilon}^p_n$, $\alpha_n$, $\boldsymbol{\theta}_{\mathcal{M}}$ (material parameters)
\Ensure $\bm{T}_{n+1}$, $\bm{\varepsilon}^p_{n+1}$, $\alpha_{n+1}$, $\bm{S}_{n+1}$, $\bm{P}_{n+1}$

\State Compute kinematics:
\Statex \hspace{\algorithmicindent}
$\bm{C}_{n+1} = \bm{F}_{n+1}^\mathsf{T}\bm{F}_{n+1}$,\quad
$J_{n+1} = \det \bm{F}_{n+1}$,\quad
$\bm{\varepsilon}_{n+1} = \tfrac{1}{2}\ln\bm{C}_{n+1}$

\State Elastic predictor:
$\bm{T}^{\mathrm{tr}}_{n+1} = \mathbb{E}:(\bm{\varepsilon}_{n+1} - \bm{\varepsilon}^p_n)$ \Comment{Eq.~\eqref{eq:T}}

\State Evaluate trial yield function:
$f^{\mathrm{tr}} = \phi^{\mathrm{tr}}(\bm{T}^{\mathrm{tr}}_{n+1}) - \sigma_{y,n}$ \Comment{Eq.~\eqref{eq:yield_final}}

\If{$f^{\mathrm{tr}} \le 0$} \Comment{Elastic step}
    \State $\bm{T}_{n+1} \gets \bm{T}^{\mathrm{tr}}_{n+1}$, 
           $\bm{\varepsilon}^p_{n+1} \gets \bm{\varepsilon}^p_n$,
           $\alpha_{n+1} \gets \alpha_n$,
           $\Delta\gamma \gets 0$
\Else \Comment{Plastic step}
    \State $(\bm{T}_{n+1}, \bm{\varepsilon}^p_{n+1}, \alpha_{n+1}, \Delta\gamma)$
           $\gets \mathrm{LocalReturnMapping}(\bm{\varepsilon}_{n+1}, \bm{\varepsilon}^p_n, \alpha_n, \boldsymbol{\theta}_{\mathcal{M}})$ \Comment{Alg.~\ref{alg:local_cpp}}
\EndIf

\State Pull back logarithmic stress to Lagrangian measures:
\Statex \hspace{\algorithmicindent}
$\bm{S}_{n+1} = \mathbb{P}_{L}(\bm{C}_{n+1}) : \bm{T}_{n+1}$
\Comment{Eq.~\eqref{eq:proj_lagrange}}
\Statex \hspace{\algorithmicindent}
$\bm{P}_{n+1} = \bm{F}_{n+1}\bm{S}_{n+1}$ \Comment{Eq.~\eqref{eq:pk_stress}}

\State \Return $\bm{T}_{n+1}$, $\bm{\varepsilon}^p_{n+1}$, $\alpha_{n+1}$, $\bm{S}_{n+1}$, $\bm{P}_{n+1}$
\end{algorithmic}
\end{algorithm}

\begin{algorithm}[!htbp]
\caption{Local return-mapping algorithm}
\label{alg:local_cpp}
\begin{algorithmic}[1]
\Require $\bm{\varepsilon}_{n+1}$, $\bm{\varepsilon}^p_n$, $\alpha_n$, $\beta$, $\eta$, $\boldsymbol{\theta}_{\mathcal{M}}$
\Ensure $\bm{T}_{n+1}$, $\bm{\varepsilon}^p_{n+1}$, $\alpha_{n+1}$, $\Delta\gamma$
\State Compute trial stress: $\bm{T}^{\mathrm{tr}}_{n+1} = \mathbb{E}:(\bm{\varepsilon}_{n+1} - \bm{\varepsilon}^p_n)$; $\bm{\xi}^0 \gets \bm{T}^{\mathrm{tr}}_{n+1}$
\State Set $\bm{y}^0 = [\bm{\xi}^0 ,\;0,\;\alpha_n]^\mathsf{T}$
\For{$k = 0,1,\dots,k_{\max}$}
    \State Compute $\phi^k$, $\bm{N}^k$ \Comment{Eq.~\eqref{eq:Phi_hill} or Eq.~\eqref{eq:Phi_barlat} and Eq.~\eqref{eq:ep_evol}}
    \State Evaluate residual: 
    $\bm{G}(\bm{y}^k)$ \Comment{Eq.~\eqref{eq:local_cpp}}
    \If{$\|\bm{G}(\bm{y}^k)\| \le \mathrm{tol}$}
        \State \textbf{break}
    \EndIf
    \State Compute Jacobian: $\bm{J}^G(\bm{y}^k) = \partial\bm{G}/\partial\bm{y}\big|_{\bm{y}^k}$ \Comment{via JAX-AD}
    \State Solve Newton direction: $\bm{J}^G(\bm{y}^k)\,\delta\bm{y}^k = -\bm{G}(\bm{y}^k)$.
    \State Evaluate merit: $\psi^k(\zeta)=\tfrac{1}{2}\|\bm{G}(\bm{y}^k+\zeta\,\delta\bm{y}^k)\|^2$ at $\zeta=0$ and $\zeta=1$ \Comment{Eq.~\eqref{eq:merit}}
    \If{$\psi^k(1) < (1-2\beta)\,\psi^k(0)$} \Comment{Eq.~\eqref{eq:goldstein_full}}
        \State $\zeta^k \gets 1$
    \Else
        \State $\zeta^{(0)} \gets \psi^k(0)\big/\bigl(\psi^k(0)+\psi^k(1)\bigr)$ \Comment{Eq.~\eqref{eq:alpha_init}}
        \For{$j = 0,1,\dots j_{\max}$}
            \State $\zeta^{(j+1)} \gets \psi^k(0)\big/\bigl(\psi^k(0)+\psi^k(\zeta^{(j)})\bigr)$ \Comment{Eq.~\eqref{eq:alpha_update_ls}}
            \State $\zeta^{(j+1)} \gets \max\!\bigl\{\eta\,\zeta^{(j)},\;\zeta^{(j+1)}\bigr\}$ \Comment{Eq.~\eqref{eq:alpha_safeguard}}
            \If{$\psi^k(\zeta^{(j+1)}) < (1-2\beta\,\zeta^{(j+1)})\,\psi^k(0)$} \Comment{Eq.~\eqref{eq:goldstein_full}}
                \State $\zeta^k \gets \zeta^{(j+1)}$; \textbf{break}
            \EndIf
        \EndFor
    \EndIf
    \State $\bm{y}^{k+1} \gets \bm{y}^k + \zeta^k\,\delta\bm{y}^k$
\EndFor
\State Extract: $\bm{\xi}_{n+1} \gets \bm{\xi}^k$, $\Delta\gamma \gets \Delta\gamma^k$, $\alpha_{n+1} \gets \alpha^k$
\State Recover: $\bm{\varepsilon}^p_{n+1} = \bm{\varepsilon}^p_n + \Delta\gamma\,\bm{N}_{n+1}$, $\quad\bm{T}_{n+1} = \mathbb{E}:(\bm{\varepsilon}_{n+1} - \bm{\varepsilon}^p_{n+1})$
\State \Return $\bm{T}_{n+1}$, $\bm{\varepsilon}^p_{n+1}$, $\alpha_{n+1}$, $\Delta\gamma$
\end{algorithmic}
\end{algorithm}

\section{Implicit Differentiation of the Local Constitutive Update} \label{sec:appendix_D}

This Appendix summarizes the implicit-differentiation procedure used to evaluate consistent directional derivatives of the local constitutive update. Following Section~\ref{subsubsec:implicit_time_discretization}, the discrete evolution equations in Eq.~\eqref{eq:Ep_update}--Eq.~\eqref{eq:yield_final} are collected into the local nonlinear system in Eq.~\eqref{eq:local_cpp}, written abstractly as,
\begin{equation}
\bm{G}(\bm{y};\bm{p})=\bm{0}
\end{equation}
where $\bm{y}$ denotes the vector of local unknowns appearing in Eq.~\eqref{eq:local_cpp} and $\bm{p}$ denotes the collection of quantities treated as inputs to the local constitutive update. We write the converged local solution as,
\begin{equation}
\bm{y}^{\star}=\bm{y}^{\star}(\bm{p})
\end{equation}
The corresponding updated stress and internal variables are recovered from the constitutive update formulas associated with Eq.~\eqref{eq:local_cpp}. In abstract form, the local output vector may be written as,
\begin{equation}
\bm{z}_{n+1}=\bm{z}(\bm{y}^{\star},\bm{p})
\end{equation}
where $\bm{z}_{n+1}$ collects the updated constitutive variables.

For a prescribed perturbation $\dot{\bm{p}}$ of the input variables, the local Newton iterations are not differentiated directly. Instead, the converged state is treated as an implicitly defined function of $\bm{p}$, following the general framework of automatic implicit differentiation for solver-defined mappings \cite{blondel2022ImplicitDifferentiation}. Differentiating Eq.~(D1) in the direction $\dot{\bm{p}}$ gives,
\begin{equation}
\frac{\partial \bm{G}}{\partial \bm{y}}(\bm{y}^{\star};\bm{p})\,\dot{\bm{y}}
\;+\;
\frac{\partial \bm{G}}{\partial \bm{p}}(\bm{y}^{\star};\bm{p})\,\dot{\bm{p}}
=
\bm{0}
\end{equation}
or, equivalently,
\begin{equation}
\bm{J}_{y}\dot{\bm{y}}=-\dot{\bm{G}}_{p},
\qquad
\bm{J}_{y}:=\frac{\partial \bm{G}}{\partial \bm{y}}(\bm{y}^{\star};\bm{p}),
\qquad
\dot{\bm{G}}_{p}:=\frac{\partial \bm{G}}{\partial \bm{p}}(\bm{y}^{\star};\bm{p})\,\dot{\bm{p}}
\end{equation}
with $\dot{\bm{y}}$ denoting the directional derivative of the converged local solution. Thus, the sensitivity of the constitutive state is obtained from the dense linear solve,
\begin{equation}
\dot{\bm{y}}=-\bm{J}_{y}^{-1}\dot{\bm{G}}_{p}
\end{equation}
In the implementation, $\bm{J}_{y}$ is obtained by automatic differentiation of the local residual with respect to the local unknowns, while $\dot{\bm{G}}_{p}$ is evaluated by applying a directional derivative with respect to the full set of constitutive inputs at fixed $\bm{y}^{\star}$. This avoids tracing the iterations of the local Newton solver while preserving the exact linearization of the converged constitutive state.

Once $\dot{\bm{y}}$ has been computed, the tangent of the local constitutive output follows from the chain rule applied to Eq.~(D3),
\begin{equation}
\dot{\bm{z}}_{n+1}
=
\frac{\partial \bm{z}}{\partial \bm{y}}(\bm{y}^{\star},\bm{p})\,\dot{\bm{y}}
\;+\;
\frac{\partial \bm{z}}{\partial \bm{p}}(\bm{y}^{\star},\bm{p})\,\dot{\bm{p}}
\end{equation}
Eq.~(D5) provides the consistent directional derivative of the local constitutive equations. The derivation is independent of any particular constitutive specialization and therefore applies to a broader class of implicit material updates. This construction yields a consistent constitutive linearization while avoiding explicit differentiation through the iterative return-mapping process.

\section{Sparse Assembly Performance Comparison} \label{sec:appendix_E}

This Appendix reports representative wall-clock times for constructing the global stiffness matrix in CSR format and performing row elimination using a conventional CPU-based SciPy implementation and the GPU-resident CuPy implementation described in Section~\ref{subsec:Implementation_sec}. The comparison highlights the benefit of performing sparse assembly directly on the accelerator for large systems. The reported values are given in seconds as a function of the total number of degrees of freedom (DOFs) in millions. 
The GPU sparse assembly procedure is listed in Algorithm~\ref{alg:gpu_chunked_assembly} and computation times are compared in Table~\ref{tab:assembly_computation_times}.

\begin{algorithm}[!htbp]
\caption{GPU based sparse assembly of the global tangent matrix}
\label{alg:gpu_chunked_assembly}
\begin{algorithmic}[1]
\Require Triplets $(\bm{V},\bm{I},\bm{J})$, constrained dofs, chunk count $N_c$, row batch size $N_b$
\Ensure Global tangent matrix $\bm{K}$ in GPU CSR format
\State Initialize empty GPU CSR matrix $\bm{K}$
\State Partition $(\bm{V},\bm{I},\bm{J})$ into $N_c$ chunks
\For{each triplet batch $(\bm{V}^{(k)}, \bm{I}^{(k)}, \bm{J}^{(k)})$}
\State $\bm{K} \gets \bm{K} + \mathrm{CSR}(\mathrm{COO}(\bm{V}^{(k)}, \bm{I}^{(k)}, \bm{J}^{(k)}))$
\EndFor
\State Build constrained-row index set $\mathcal{Q}$ on the GPU
\State Expose CSR arrays \texttt{indptr}, \texttt{indices}, and \texttt{data}
\State Partition $\mathcal{Q}$ into batches of size at most $N_b$
\For{each row batch $\mathcal{Q}^{(m)}$}
\State Enforce identity rows with one CUDA thread per row
\EndFor
\State \Return $\bm{K}$
\end{algorithmic}
\end{algorithm}

\begin{table}[!htbp]
\centering
\begin{threeparttable}
\begin{minipage}{\standardtablewidth}
\caption{Computation times in seconds for CSR sparse matrix assembly and row elimination using SciPy on CPU and CuPy on H100 GPU.}
\label{tab:assembly_computation_times}
\setlength{\tabcolsep}{6pt}
\centering
\begin{tabular*}{\textwidth}{@{\extracolsep{\fill}}lcc@{}}
\toprule
DOFs & SciPy CSR & CuPy CSR \\
\midrule
3.0 M & 18 sec & 4 sec \\
4.6 M & 25 sec & 6 sec \\
5.5 M & 30 sec & 7 sec \\
7.3 M & 52 sec & 7 sec \\
9.2 M & 55 sec & 8 sec \\
\bottomrule
\end{tabular*}
\end{minipage}
\end{threeparttable}
\end{table}

\section{Linear Solver Performance Comparison} \label{sec:appendix_F}

This Appendix summarizes representative wall-clock times for the linear solvers discussed in Section~\ref{subsec:Implementation_sec}. The comparison includes iterative GPU-based solvers implemented through JAX and AMGX, CPU-based direct solvers through UMFPACK and Pardiso, and GPU-based direct solvers through CuDSS. The reported values are given in seconds for problems of increasing size, measured by the total number of degrees of freedom (DOFs). The linear systems are obtained from a linear elastic problem on a unit-cube domain. Blank entries indicate cases that were not computed.

\begin{table}[!htbp]
\centering
\begin{threeparttable}
\begin{minipage}{0.95\textwidth}
\caption{Computation times in seconds for different linear solvers}
\label{tab:solver_computation_times}
\setlength{\tabcolsep}{2pt}
\centering
\begin{tabular*}{\textwidth}{@{\extracolsep{\fill}}lccccc@{}}
\toprule
DOFs & \shortstack{JAX-BICGSTAB \\ (H100 GPU)}  & \shortstack{AMGX-BICGSTAB \\ (H100 GPU)} & \shortstack{Direct CuDSS \\ (H100 GPU)} & \shortstack{Direct UMFPACK \\ (CPU)} & \shortstack{Direct Pardiso \\ (CPU)} \\
\midrule
27,783   & 0.53 & 0.034 & 0.35  & 8.5  & 0.29  \\
60,543   & 0.49 & 0.080 & 0.68  & 52   & 0.64  \\
118,203  & 0.74 & 0.12  & 1.44  & --   & 1.53  \\
475,983  & 1.86 & 0.37  & 6.51  & --   & 14.17 \\
1,987,983  & 9.87 & 1.99  & --  & --   & -- \\
3,090,903  & 11.21 & 1.96  & --  & --   & -- \\
4,621,053  & 31.16 & 7.41  & --  & --   & -- \\
5,536,113  & 39.83 & 9.24  & --  & --   & -- \\
\bottomrule
\end{tabular*}
\end{minipage}
\end{threeparttable}
\end{table}

\bibliographystyle{unsrt}
\bibliography{references}

\end{document}